%% file: main.tex
\documentclass[11pt]{article}

\usepackage[colorlinks,bookmarksopen,bookmarksnumbered,citecolor=blue,urlcolor=red]{hyperref}
\usepackage{setspace}
\usepackage[margin=1in,right=1in,top=1in,bottom=1in]{geometry} 
\usepackage{amsmath,amsthm,amssymb,scrextend, enumerate, mathrsfs, float}
\usepackage{pgf,tikz}
\usetikzlibrary{arrows,shapes.arrows,shapes.geometric,shapes.multipart,decorations.pathmorphing,positioning,swigs}
\usepackage{fancyhdr}
\usepackage{tikz}
\usepackage{pgfplots}
\usepackage{natbib}
\usepackage{amsmath, threeparttable}
\usepackage[labelfont=bf]{caption}
\usepackage[mathscr]{euscript}
 \let\mathscr\relax
\usepackage[scr]{rsfso}
\usepackage{amsthm}
\usepackage{amssymb}
\usepackage{multicol}
\usepackage{multirow}
\usepackage{booktabs}
\usepackage{lscape}
\usepackage{algorithm}
\usepackage{algpseudocode}

\usepackage{authblk}

\newtheorem{theorem}{Theorem}
\newtheorem{corollary}{Corollary}
\newtheorem{proposition}{Proposition}
\newtheorem{lemma}{Lemma}

\newtheorem{assumption}{Assumption}

\newtheorem{remark}{Remark}

\newcommand{\E}{\mathbb{E}}
\newcommand{\mc}{\mathcal}
\newcommand{\Var}{\text{var}}
\renewcommand{\subsectionmark}[1]{}

\newcommand{\indep}{\perp\!\!\!\perp}

\title{Sharp Instrumental Variable Bounds under The Marginal Sensitivity Model}

\author[1]{Antonio Olivas-Martinez}
\author[1,2]{Eric Tchetgen Tchetgen}

\affil[1]{Department of Biostatistics, Epidemiology and Informatics,
Perelman School of Medicine, University of Pennsylvania,
Philadelphia, USA}

\affil[2]{Department of Statistics and Data Science,
The Wharton School, University of Pennsylvania,
Philadelphia, USA}
\date{}
\begin{document}

\maketitle



\begin{abstract}
  Marginal sensitivity models bound the degree of unmeasured confounding but do not incorporate instrumental-variable restrictions, even when an instrument is available. We develop a framework combining these models with instrumental-variable assumptions to sharpen inference on population treatment effects without requiring treatment monotonicity or treatment-effect homogeneity. For continuous and finite ordered outcomes, we characterize sharp identified sets for potential-outcome means and the average treatment effect, together with observable restrictions implied jointly by the instrumental-variable and sensitivity assumptions. The bounds admit explicit formulas for finite ordered outcomes, while optimization over prespecified threshold classes provides tractable outer bounds for continuous outcomes. We develop cross-fitted one-step estimators that accommodate flexible nuisance estimation. Under suitable conditions, the resulting outward confidence sets asymptotically cover the sharp identified interval without requiring optimal threshold estimation. Stronger conditions yield semiparametric efficient estimation and inference for restricted-class bounds. Numerical experiments show narrower confidence sets when incorporating an instrument and compare the performance of outward inference and inference for individual bounds. We apply the framework to the Oregon Health Insurance Experiment to illustrate how combining instrumental-variable assumptions with restrictions on the degree of unmeasured confounding can strengthen conclusions about the effect of Medicaid coverage on emergency-department use.
\end{abstract}

\section{Introduction}
\label{sec:introduction}

Analyses of observational data commonly estimate treatment effects by adjusting for observed baseline covariates. Their causal interpretation, however, depends on the assumption that these covariates are sufficient to control confounding. Because this assumption cannot generally be verified from the observed data, investigators are encouraged to examine how their conclusions change under interpretable departures from unconfoundedness.

The marginal sensitivity model (MSM) has become a prominent framework for sensitivity analysis in observational studies, motivating a growing literature on sharp identification and robust estimation \citep{tan2006distributional,zhao2019sensitivity,dorn2023sharp,
tan2024model,dorn2025doubly}. Its appeal lies partly in its direct interpretation: on the odds-ratio scale, a sensitivity parameter $\Lambda\geq1$ bounds how strongly treatment assignment may depend on an individual's potential outcome beyond what is explained by the observed covariates. Thus, $\Lambda=1$ recovers unconfoundedness, whereas larger values
permit progressively stronger treatment--outcome confounding. For each $\Lambda$, the model yields an interval of causal effects compatible with the observed data and the allowed degree of confounding. In its conventional form, however, the MSM does not use structural information that may be available from an instrumental variable.

An instrument offers a different response to unmeasured confounding. Definitions vary across the IV literature, but typically a valid instrument is associated with treatment, is conditionally independent of the potential outcomes, and does not affect the outcome except through treatment \citep{swanson2018partial}. These assumptions can impose testable restrictions on the observed-data distribution \citep{bonet2001instrumentality}, but they do not generally point identify the population average treatment effect (ATE)
\citep{robins1989analysis,manski1990nonparametric,balke1997bounds}. Under monotonicity, they may instead identify an average effect among compliers \citep{angrist1996identification}; restrictions on treatment-effect modification by unobserved factors can identify the population ATE \citep{wang2018bounded}. The former changes the target population, while the latter may be scientifically restrictive. Without such assumptions, IV analyses instead yield partial identification, and the resulting bounds may be too wide to determine even the direction of the effect. For continuous outcomes, finite IV-only bounds on mean effects generally require credible limits on the potential-outcome support \citep{manski1990nonparametric,swanson2018partial}.

The MSM and instrumental-variable assumptions provide complementary restrictions on unmeasured confounding.
This raises the central question of this paper: can an instrument strengthen an MSM sensitivity analysis of the population ATE without replacing the target by a complier effect or imposing treatment-effect homogeneity?

We answer this question by proposing an instrumental-variable restricted marginal sensitivity model (IV-MSM). The model imposes MSM restrictions within instrument strata and requires them to be jointly compatible with the IV assumptions. Sensitivity parameters may differ across strata, allowing asymmetric restrictions on confounding and, if desired, leaving selected strata unrestricted.
We develop the main results for a binary instrument and provide
extensions of identification, estimation, and inference to finite
discrete instruments in Appendix~\ref{sec:supp_finite_extensions}.

\paragraph{Illustrative example.}
Imagine a randomized clinical trial with nonadherence, no baseline covariates, and a binary outcome.
Let $Z$ denote randomized treatment allocation, with $\Pr(Z=1)=1/2$; let $T$ denote treatment received; and let $Y$ denote the
observed outcome. Write $Y^t$ for the potential outcome if treatment received
were set to $t$. Randomization makes $Z$ independent of the potential
outcomes, and we suppose that allocation affects the outcome only through
treatment received. Suppose the observed probabilities are:
\begin{center}
\begin{tabular}{c@{\qquad}ccc}
\toprule
$z$ & $\Pr(T=1\mid Z=z)$ & $\Pr(Y=1\mid T=1,Z=z)$
& $\Pr(Y=1\mid T=0,Z=z)$\\
\midrule
$0$ & $0.30$ & $0.45$ & $0.34$\\
$1$ & $0.70$ & $0.60$ & $0.40$\\
\bottomrule
\end{tabular}
\end{center}
Thus, the instrument increases the probability of treatment from $0.30$
to $0.70$. This observed law is compatible with the IV, consistency,
and positivity assumptions introduced formally in Section~\ref{sec:setup}.

For $t,z\in\{0,1\}$, consistency and the law of total
expectation give
\[
\begin{aligned}
\E(Y^t\mid Z=z)
={}&\Pr(T=t\mid Z=z)\Pr(Y=1\mid T=t,Z=z)\\
&+\Pr(T=1-t\mid Z=z)
  \Pr(Y^t=1\mid T=1-t,Z=z).
\end{aligned}
\]
The treatment probabilities and the observed success probability
are identified from the data. The missing success probability
$\Pr(Y^t=1\mid T=1-t,Z=z)$ concerns individuals who did not
receive treatment $t$ and is not identified without additional
assumptions.

A conventional MSM conditioning on $Z$ constrains this missing
probability by restricting treatment selection within each
allocation stratum:
\[
\Lambda_z^{-1}
\leq
\frac{\operatorname{odds}\{\Pr(T=t\mid Y^t=y,Z=z)\}}
     {\operatorname{odds}\{\Pr(T=t\mid Z=z)\}}
\leq
\Lambda_z,
\qquad t,z,y\in\{0,1\},
\]
where $\operatorname{odds}(p)=p/(1-p)$.
Thus, within each stratum, knowing an individual's potential
outcome may multiply the treatment odds by at most $\Lambda_z$
or divide them by at most $\Lambda_z$.
By Bayes' rule, the restrictions for the two outcome values,
together with their probabilities summing to one, bound the
missing success probability relative to its observed counterpart.
Substitution into the preceding decomposition gives an interval
$I_z^t(\Lambda)=[\ell_z^t(\Lambda),u_z^t(\Lambda)]$
for $\E(Y^t\mid Z=z)$, where we set
$\Lambda_0=\Lambda_1=\Lambda$ for this illustration and use
the same sensitivity parameter for both potential outcomes.
At $\Lambda=1$, the model assumes $T\indep Y^t\mid Z$;
larger values allow progressively stronger departures from
this no-unmeasured-confounding assumption.

The conventional MSM allows the potential-outcome means
to differ across allocation strata and marginalizes over $Z$
to obtain the population bounds
\[
\left[
\sum_{z=0}^1\Pr(Z=z)\ell_z^t(\Lambda),
\;
\sum_{z=0}^1\Pr(Z=z)u_z^t(\Lambda)
\right].
\]

For comparison, a conventional MSM ignoring $Z$ applies
the same calculation to the pooled population, replacing
the conditional treatment odds above by
$\operatorname{odds}\{\Pr(T=t\mid Y^t=y)\}$ and
$\operatorname{odds}\{\Pr(T=t)\}$.
The same numerical $\Lambda$ then encodes a different
restriction because treatment selection is no longer
conditioned on allocation.

The IV-MSM adds $Y^t\indep Z$ to the within-stratum sensitivity restrictions, requiring a common potential-outcome mean across strata.
For binary outcomes, its feasible values are
precisely those in
\[
I_{\mathrm{IV\text{-}MSM}}^t(\Lambda)
=
I_0^t(\Lambda)\cap I_1^t(\Lambda)=
\left[
\max_{z\in\{0,1\}}\ell_z^t(\Lambda),
\;
\min_{z\in\{0,1\}}u_z^t(\Lambda)
\right],
\qquad t\in\{0,1\}.
\]
Thus, instead of averaging the stratum-specific bounds, the IV-MSM intersects their intervals.
When nonempty, this intersection is strictly narrower unless $\ell_0^t(\Lambda)=\ell_1^t(\Lambda)$ and $u_0^t(\Lambda)=u_1^t(\Lambda)$.
An empty intersection reveals incompatibility of the
sensitivity and IV restrictions with the observed distribution.

Without the sensitivity restrictions, the missing probability $\Pr(Y^t=1\mid T=1-t,Z=z)$ can take any value in $[0,1]$.
Setting it to zero or one in the preceding decomposition gives the stratum-specific lower and upper bounds:
\[
\begin{aligned}
\ell_{z,\mathrm{IV}}^t
&=\Pr(T=t\mid Z=z)\Pr(Y=1\mid T=t,Z=z),\\
u_{z,\mathrm{IV}}^t
&=\ell_{z,\mathrm{IV}}^t+\Pr(T=1-t\mid Z=z).
\end{aligned}
\]
Under the IV requirement, and without further restrictions,
intersecting these stratum-specific intervals gives the
IV-only interval, corresponding to the natural Robins--Manski bounds \citep{robins1989analysis,manski1990nonparametric}:
\[
I_{\mathrm{IV}}^t
=
\left[
\max_{z\in\{0,1\}}\ell_{z,\mathrm{IV}}^t,
\;
\min_{z\in\{0,1\}}u_{z,\mathrm{IV}}^t
\right].
\]

When $\Lambda_0=\Lambda_1=\Lambda$, the sharp IV-MSM interval is contained in, but need not equal, the intersection of the Robins--Manski interval and the conventional-MSM interval at the same $\Lambda$, whether conditioning on $Z$ or not.
Thus, merely intersecting the separately derived IV-only and MSM bounds does not generally recover the sharp IV-MSM bounds; see Remark~\ref{rem:combining_iv_msm} in Section~\ref{sec:sharp_bounds} for further discussion.

\begin{figure}[!ht]
\centering
\includegraphics[width=\textwidth]{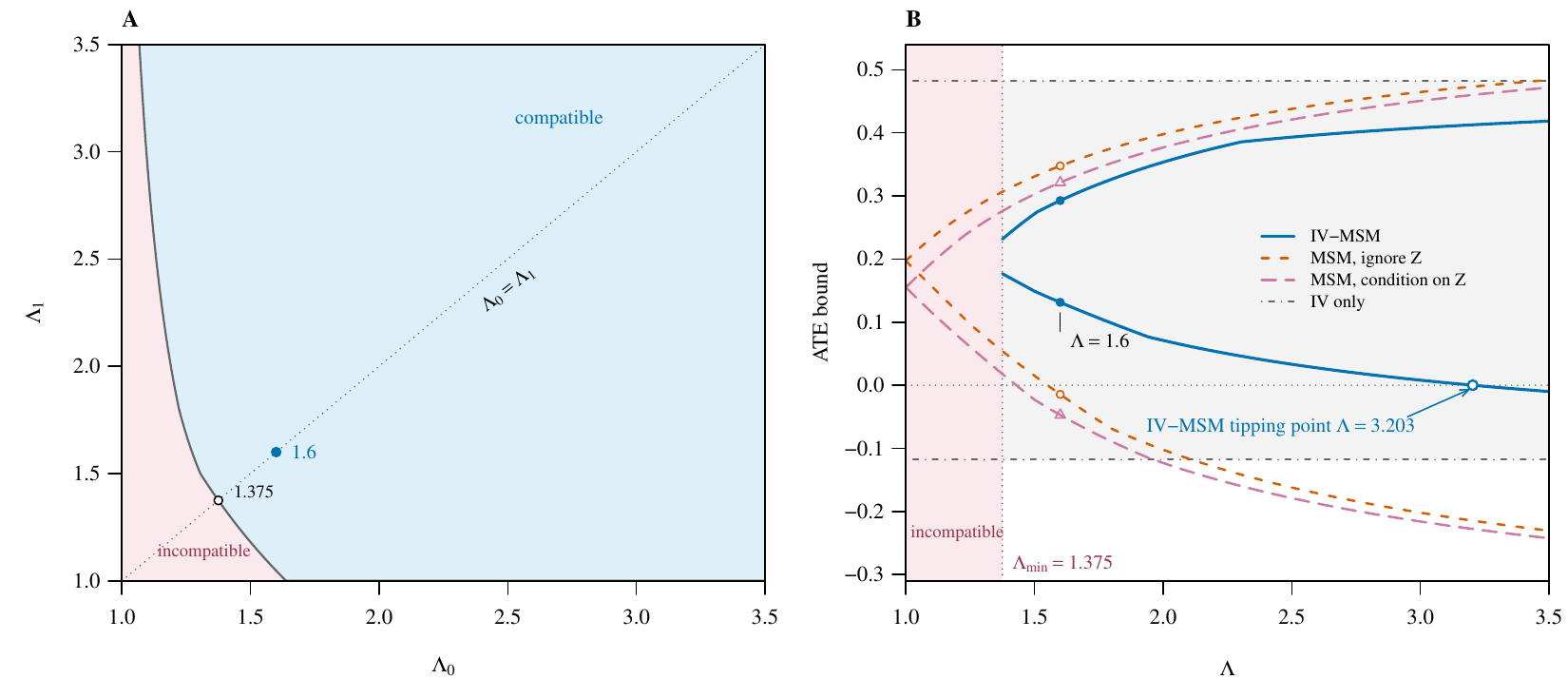}
\caption{How the instrument strengthens the MSM in the illustrative binary example. (a) Sensitivity-parameter pairs compatible with the observed distribution and the IV assumptions. The diagonal is the common-parameter choice $\Lambda_0=\Lambda_1=\Lambda$, and the marked point is the value $\Lambda=1.6$. (b) Population ATE bounds along this diagonal under the IV-MSM, the conventional MSM with and without conditioning on $Z$, and the IV assumptions alone. The shaded region is the IV-only identified interval. Values $\Lambda<1.375$ are incompatible with the observed distribution and the IV assumptions. The IV-MSM lower bound reaches zero at $\Lambda=3.203$.}
\label{fig:intro_binary_example_revised}
\end{figure}

Figure~\ref{fig:intro_binary_example_revised} illustrates the bounds and compatibility restrictions.
Panel (a) shows which pairs $(\Lambda_0,\Lambda_1)$ are compatible with the observed distribution and the IV assumptions.
The diagonal corresponds to the common-parameter choice $\Lambda_0=\Lambda_1=\Lambda$ and enters the compatible region at $\Lambda=1.375$.
Panel (b) shows the population ATE bounds along this diagonal.
At $\Lambda=1.6$, the IV-MSM gives $[0.1313,0.2926]$,
which excludes zero and is approximately 56\% shorter than
the conventional MSM interval conditioning on $Z$,
$[-0.0472,0.3213]$, and 55\% shorter than the interval
ignoring $Z$, $[-0.0145,0.3476]$.
The first comparison isolates the additional IV restrictions;
the second compares models in which the same numerical
sensitivity parameter encodes different restrictions.
The IV-only interval, $[-0.1180,0.4820]$, contains both
conventional-MSM intervals, so intersecting either with it
leaves that interval unchanged and still containing zero.

Our contributions are threefold. First, we derive observable compatibility restrictions and characterize the sharp identified sets for the potential-outcome means and population ATE under the IV-MSM.
For continuous outcomes, the sharp bounds representations involve conditional weighted quantile functions that can be challenging to estimate. We therefore also consider optimization over prespecified classes of threshold functions, yielding tractable outer bounds on the sharp identified set.
Second, we characterize conditions under which the restricted-class bound functionals are pathwise differentiable, derive their efficient influence functions, and construct cross-fitted one-step estimators.
Third, we establish valid outward inference under weaker conditions than those required for inference on individual
restricted-class bounds. We also use the observable compatibility restrictions to construct a level-controlled test of candidate sensitivity parameters under the IV assumptions.

Section~\ref{sec:setup} introduces the causal framework and the IV-MSM. Sections~\ref{sec:compatibility_dual} and~\ref{sec:sharp_bounds} derive the compatibility restrictions and sharp bounds. Section~\ref{sec:estimation} develops estimation, inference, and compatibility testing, and Section~\ref{sec:binary_bounds} specializes the method to binary outcomes. Section~\ref{sec:numerical_experiments} presents numerical experiments, and Section~\ref{sec:oregon_application} applies our method to the Oregon Health Insurance Experiment \citep{finkelstein2012oregon}. Section~\ref{sec:discussion} discusses related work and extensions of our method.

\section{Setup and the IV-MSM}\label{sec:setup}

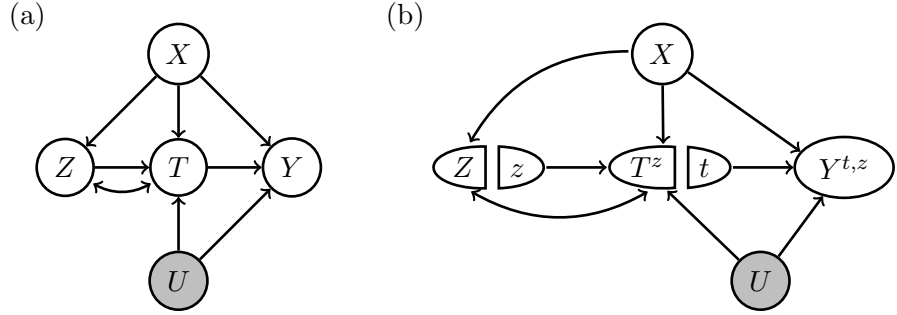
\begin{figure}\centering
\begin{tikzpicture}[node distance=5mm]
\node[name=indA]  at (-8.5,2) {(a)};
\node[name=indB]  at (-3.5,2) {(b)};
\node[name=T0, circle, draw, line width=1pt]  at (-6.5,0) {$T$};
\node[name=Y0, circle, draw, line width=1pt]  at (-5,0) {$Y$};

\node[name=U0, circle, draw, fill=lightgray, line width=1pt] at (-6.5,-1.5) {$U$};
\node[circle, draw, line width=1pt] (X0) at (-6.5,1.5) {$X$};
\node[circle, draw, line width=1pt] (Z0) at (-8,0) {$Z$};

\draw[->, black, line width=1pt] (U0) -- (T0);
\draw[->, black, line width=1pt] (U0) -- (Y0);
\draw[->, black, line width=1pt] (Z0) -- (T0);
\draw[<->, line width=1pt, shorten >=1pt, shorten <=1pt]
    (Z0) to[bend right=30] (T0);
\draw[->, black, line width=1pt] (X0) -- (Z0);
\draw[->, black, line width=1pt] (X0) -- (Y0);
\draw[->, black, line width=1pt] (X0) -- (T0);

\draw[->,black, line width=1pt] (T0) -- (Y0); 

\node[name=T,shape=swig vsplit, swig vsplit={gap=5pt}, line width=1pt]{
  \nodepart{left}{$T^z$}
  \nodepart{right}{$t$}
};

\node[name=Z,shape=swig vsplit, swig vsplit={gap=5pt}, line width=1pt, left=8mm of T]{
  \nodepart{left}{$Z$}
  \nodepart{right}{$z$}
};

\node[name=Y, ellipse, draw, right=8mm of T, line width=1pt] {$Y^{t,z}$};

\node[name=U, circle, draw, fill=lightgray, line width=1pt] at (1.2,-1.5) {$U$};
\node[circle, draw, line width=1pt] (X) at (-0.1,1.5) {$X$};

\draw[->, black, line width=1pt] (U) -- (T.280);
\draw[->, black, line width=1pt] (U) -- (Y);
\draw[->, line width=1pt] (Z) -- (T.180);
\draw[->, line width=1pt] (X) -- (T.90);
\draw[->, line width=1pt] (X) -- (Y.160);
\draw[->, line width=1pt] (T) -- (Y.180);
\draw[->, line width=1pt, shorten >=1pt, shorten <=1pt] (X) to[bend right=30] (Z.120);

\draw[<->, line width=1pt, shorten >=1pt, shorten <=1pt](Z.240) to[bend right=30] (T.240);
\end{tikzpicture}
\caption{Causal graphs representing the instrumental-variable model, adapted
from~\citet{wang2018bounded} (the bidirected arrow between $Z$
and $T$ represents possible unmeasured common causes of $Z$ and $T$; variables
$X$, $Z$, $T$, and $Y$ are observed; $U$ is unobserved): (a) a causal directed
acyclic graph with a bidirected arrow \citep{pearl2009causality}; (b) a
single-world intervention graph under $(Z,T)=(z,t)$ with a bidirected arrow
\citep{richardson2013single}, where $T^z$ denotes treatment under $Z=z$.}
\label{fig:dag}
\end{figure}

Let $O_1,\dots,O_n$ be independent copies of $O=(Y,T,X,Z)$ drawn from an unknown observed-data distribution $P_0\in\mc{M}$, where the model $\mc{M}$ is defined below.
Here, $Y$ is the observed outcome, $T\in\{0,1\}$ is a binary treatment, $X$ is a vector of observed baseline covariates, and $Z\in\{0,1\}$ is a binary instrumental variable.
Let $Y^{t,z}$ denote the potential outcome under a joint intervention setting treatment to $t\in\{0,1\}$ and the instrument to $z\in\{0,1\}$.
We further define $Y^t$ as the potential outcome under an intervention setting treatment to $t$ while leaving the instrument at its naturally observed value. 
Let $\mc Y\subseteq\mathbb R$ denote the common outcome domain in which $Y$, $Y^t$, and $Y^{t,z}$ take values.
Throughout the paper, $P$ denotes a generic observed-data distribution for $O$, and $P_X$ its marginal distribution of $X$.
We use $Q$ to denote a full-data distribution containing the observed variables and the potential outcomes $\{Y^0,Y^1,Y^{t,z}:t,z\in\{0,1\}\}$.
We say that $Q$ is an extension of $P$ if the observed-data distribution induced by $Q$ is equal to $P$.
For any such extension, define
\[
\mu^1(Q):=\E_Q(Y^1),\qquad
\mu^0(Q):=\E_Q(Y^0),\qquad
\psi(Q):=\mu^1(Q)-\mu^0(Q),
\]
where $\psi(Q)$ is the average treatment effect. Figure~\ref{fig:dag} gives a schematic graphical representation of this IV setting and of the joint intervention defining \(Y^{t,z}\).

Because multiple full-data extensions of the same observed-data distribution may yield different values of these parameters, $\mu^1(Q)$, $\mu^0(Q)$, and $\psi(Q)$ are generally not point identified by $P$. We first characterize the identified set for $\mu^1(Q)$ under a marginal sensitivity model for $Y^1$. The corresponding results for $\mu^0(Q)$ follow by relabeling treatment, and the two sets of results are then combined to obtain the sharp identified set for the ATE.

To connect the observed data with the latent potential outcomes, we impose the following assumptions on the observed- and full-data distributions under consideration.

\begin{assumption}[Consistency and composition]\label{ass:consist}
$Q$-almost surely, (i) $Y=Y^{T,Z}$ and
(ii) $Y^t=Y^{t,Z}$ for each $t\in\{0,1\}$.
\end{assumption}
\begin{assumption}[Treatment and instrument positivity]\label{ass:positivity}
There exists $\epsilon\in(0,1/2)$ such that, $P$-almost surely, (i) $P(T=1\mid X,Z)\in[\epsilon,1-\epsilon]$ and (ii) $P(Z=1\mid X)\in[\epsilon,1-\epsilon]$.
\end{assumption}
\begin{assumption}[Distributional exclusion restriction]
\label{ass:iv_excl}
For every $t,z\in\{0,1\}$ and $y\in\mc{Y}$,
\[
Q(Y^{t,z}\le y\mid X)=Q(Y^t\le y\mid X)
\]
$Q$-almost surely.
\end{assumption}

\begin{assumption}[Instrument exchangeability]
\label{ass:iv_exchan}
For every $t,z\in\{0,1\}$, $Y^{t,z}\indep_Q Z\mid X$.
\end{assumption}

Consistency links the observed outcome to the individual’s realized treatment–instrument pair, while composition defines a treatment-only intervention as retaining the individual’s natural instrument value. Together they imply \(Y=(1-T)Y^0+TY^1\). Assumption~\ref{ass:positivity} requires treatment overlap within each \((X,Z)\) stratum and instrument overlap within each \(X\) stratum, respectively, thereby ensuring that every \((T,Z)\) combination has probability bounded away from zero given \(X\).

Assumptions~\ref{ass:iv_excl} and~\ref{ass:iv_exchan} formalize the IV restrictions. Assumption~\ref{ass:iv_excl} concerns interventions on the instrument: after fixing treatment at \(t\), the conditional distribution of \(Y^{t,z}\) given \(X\) is the same for \(z=0\) and \(z=1\), and equals that of \(Y^t\). This condition is stronger than the conditional mean exclusion $\E(Y^{t,0}\mid X)=\E(Y^{t,1}\mid X)$, but weaker than the individual-level exclusion restriction \(Y^{t,0}=Y^{t,1}\) $Q$-almost surely \citep{swanson2018partial,balke1997bounds,levis2025covariate}. Assumption~\ref{ass:iv_exchan} instead concerns instrument assignment: conditional on \(X\), the observed instrument is as-if randomized with respect to each \(Y^{t,z}\) separately, a weaker requirement than joint
exchangeability of the potential outcomes as used to derive the
Balke--Pearl bounds \citep{balke1997bounds,swanson2018partial,levis2025covariate}. Thus, exclusion compares potential outcomes under different interventions on \(Z\), whereas exchangeability compares the same potential outcome across observed instrument groups. For example, in randomized trials with noncompliance, instrument exchangeability holds by design when \(Z\) denotes randomized assignment and \(T\) denotes treatment received, whereas distributional exclusion must still be justified in context. Together with composition, these assumptions imply $Y^t\indep_Q Z\mid X$ for each $t\in\{0,1\}$, which is the cross-stratum restriction used by the IV-MSM.

Crucially, we do not assume $Y^t\indep_Q T\mid X$. Therefore, the causal means and the ATE are generally not point identified by the observed-data distribution. We instead characterize their identified sets under user-specified restrictions on the magnitude of unmeasured confounding.

Specifically, we impose the MSM of \citet{tan2006distributional} separately within each instrument stratum.
Through Bayes' rule, its odds-ratio formulation is equivalent to bounds on density ratios involving the potential outcome.
The following outcome-overlap assumption ensures that these ratios, and the cross-instrument observed density ratios introduced in Section~\ref{sec:compatibility_dual}, are well defined.

\begin{assumption}[Outcome overlap]\label{ass:abs_cont}
For $t,z\in\{0,1\}$ and $P_X$-almost every $x$,
\[
Q(Y^t\mid T=1-t,X=x,Z=z)
\ll
Q(Y^t\mid T=t,X=x,Z=z).
\]
In addition, for $t\in\{0,1\}$ and $P_X$-almost every $x$,
\[
Q(Y^t\mid T=t,X=x,Z=1)
\ll
Q(Y^t\mid T=t,X=x,Z=0).
\]
\end{assumption}

The first condition requires that the missing potential-outcome distribution assign no probability to sets having zero probability under its observed counterpart, within each covariate and instrument stratum.
The second imposes the analogous requirement across instrument strata within each treatment arm, using $Z=0$ as the reference.
By consistency and composition, the second condition concerns observed outcome distributions and can, in principle, be refuted by the observed-data law.
The first involves missing potential outcomes and cannot be refuted from observed data alone.

For $z\in\{0,1\}$, let
$\pi_{z,P}(X)=P(T=1\mid X,Z=z)$ denote the observed propensity score in instrument stratum $z$. For a full-data extension $Q$ of $P$, Assumptions~\ref{ass:positivity}(i) and~\ref{ass:abs_cont} imply the existence of the Radon--Nikodym derivative
\begin{equation}
\lambda^1_{z,Q}(Y^1,X)
:=
\frac{dQ(Y^1\mid T=0,X,Z=z)}
     {dQ(Y^1\mid T=1,X,Z=z)}
=
\frac{\pi_{z,P}(X)}{1-\pi_{z,P}(X)}
\frac{1-Q(T=1\mid Y^1,X,Z=z)}
     {Q(T=1\mid Y^1,X,Z=z)}.
\label{def_lambda1}
\end{equation}
The superscript $1$ indicates that the ratio concerns $Y^1$, whereas the subscript $z$ indexes the instrument stratum. Because $\lambda^1_{z,Q}$ is a Radon--Nikodym derivative, it integrates to one under its denominator measure. By consistency and composition, this normalization can be written as
\begin{equation*}
\E_P\left[
\left.
\frac{T}{\pi_{z,P}(X)}
\lambda^1_{z,Q}(Y,X)
\,\right|\,X,Z=z
\right]
=
\E_Q\left[
\lambda^1_{z,Q}(Y^1,X)
\mid T=1,X,Z=z
\right]
=1
\end{equation*}
$P$-almost surely.

We impose the following instrumental-variable restricted MSM on $Y^1$.

\begin{assumption}[Instrumental-variable restricted MSM for $Y^1$]
\label{ass:msm_iv_dual}
For fixed sensitivity parameters $\Lambda_0\ge1$ and $\Lambda_1\ge1$, the full-data law $Q$ satisfies
\begin{equation}
\Lambda_z^{-1}
\le
\lambda^1_{z,Q}(Y^1,X)
\le
\Lambda_z,
\qquad z\in\{0,1\},
\label{msm_tan_dual}
\end{equation}
$Q$-almost surely.
\end{assumption}

The sensitivity parameter $\Lambda_z$ bounds the magnitude of unmeasured confounding within instrument stratum $Z=z$. Allowing $\Lambda_0$ and $\Lambda_1$ to differ accommodates asymmetric substantive knowledge across the two instrument strata. Although the sensitivity parameters could in principle differ between $Y^0$ and $Y^1$, throughout the paper we impose the same pair $(\Lambda_0,\Lambda_1)$ on both potential outcomes in the absence of substantive information supporting outcome-specific sensitivity bounds, without requiring the corresponding latent density ratios to be equal.

For fixed $(\Lambda_0,\Lambda_1)$, let
$\mc{M}_1(\Lambda_0,\Lambda_1)$ denote the collection of observed-data
distributions $P$ admitting a full-data extension $Q$ satisfying
Assumptions~\ref{ass:consist}--\ref{ass:msm_iv_dual}, with the
potential-outcome restrictions imposed for $Y^1$.
Define $\mc{M}_0(\Lambda_0,\Lambda_1)$ by imposing the corresponding
assumptions for $Y^0$, with treatment relabeled as $1-T$.
The observed-data model for the ATE is
$\mc{M}(\Lambda_0,\Lambda_1)
=\mc{M}_0(\Lambda_0,\Lambda_1)\cap\mc{M}_1(\Lambda_0,\Lambda_1)$.
When no ambiguity is possible, we write this model simply as $\mc{M}$.

The reciprocal bounds $[\Lambda_z^{-1},\Lambda_z]$ impose symmetry on the multiplicative scale. If $\Lambda_z=1$, then
$\lambda^1_{z,Q}(Y^1,X)=1$ $Q$-almost surely, implying
$Y^1\indep_Q T\mid X,Z=z$. Thus $\Lambda_z=1$ corresponds to no unmeasured confounding for $Y^1$ within instrument stratum $z$. As $\Lambda_z$ increases, the sensitivity restriction is progressively relaxed. Setting $\Lambda_z=+\infty$ removes the sensitivity restriction in that stratum. Under the common-sensitivity convention, the same interpretation applies to the latent density ratios involving $Y^0$ in $\mc{M}_0(\Lambda_0,\Lambda_1)$.

\section{Compatibility Restrictions under the IV-MSM}\label{sec:compatibility_dual}

For fixed sensitivity parameters $(\Lambda_0,\Lambda_1)$, the IV-MSM and the structural IV assumptions jointly restrict the full-data extensions $Q$ of a given observed-data distribution $P$. For some observed-data distributions, no extension satisfying all these restrictions exists. In this section, we focus on $Y^1$ and derive observable compatibility restrictions determining whether $P$ can belong to $\mc{M}_1(\Lambda_0,\Lambda_1)$. The corresponding restrictions for $Y^0$ follow by relabeling treatment.

We take $Z=0$ as the baseline instrument stratum and define the observed density ratio
\[
\gamma_{1,P}(Y,X):=
\frac{dP(Y\mid T=1,X,Z=1)}
     {dP(Y\mid T=1,X,Z=0)},
\]
which is well-defined under Assumptions~\ref{ass:positivity} and \ref{ass:abs_cont}. This Radon--Nikodym derivative quantifies the relative change in the outcome distribution among treated individuals when moving from the baseline stratum $Z=0$ to the comparison stratum $Z=1$. Under Assumptions~\ref{ass:consist}--\ref{ass:iv_exchan}, this observed distributional shift induces an affine relationship between the latent density ratios in the two instrument strata. To express this relationship, define
\begin{equation}
A_P(X,Y) := \frac{1-\pi_{1,P}(X)}{1-\pi_{0,P}(X)}\gamma_{1,P}(Y,X), \quad \text{and} \quad B_P(X,Y) := \frac{\pi_{1,P}(X)\gamma_{1,P}(Y,X)-\pi_{0,P}(X)}{1-\pi_{0,P}(X)}. \label{def:AB}
\end{equation}
Under Assumption~\ref{ass:consist}, $Y=Y^1$ almost surely among treated
individuals, so the functions $\gamma_{1,P}(Y,X)$, $A_P(X,Y)$, and
$B_P(X,Y)$ may equivalently be evaluated at $Y^1$. Using $A_P$ and $B_P$, define pointwise compatibility envelopes
\[
L_P(x,y):=
\max\left\{
\Lambda_0^{-1},
A_P(x,y)\Lambda_1^{-1}+B_P(x,y)
\right\},
\qquad
U_P(x,y):=
\min\left\{
\Lambda_0,
A_P(x,y)\Lambda_1+B_P(x,y)
\right\}.
\]
The following proposition establishes the affine relationship between the latent density ratios, with $A_P$ and $B_P$ as its slope and intercept, and shows that compatible latent density ratios must lie between these envelopes.

\begin{proposition}[Compatibility under the IV-MSM]
\label{prop:comp_msm_iv}
Let $P\in\mc{M}_1(\Lambda_0,\Lambda_1)$, and let $Q$ be a
full-data extension of $P$ satisfying the restrictions defining this model.
Then the latent density ratios satisfy
\[
\lambda^1_{0,Q}(Y^1,X)
=
A_P(X,Y^1)\lambda^1_{1,Q}(Y^1,X)
+
B_P(X,Y^1)
\qquad Q\text{-almost surely}.
\]
Consequently,
\begin{equation}
L_P(X,Y^1)
\le
\lambda^1_{0,Q}(Y^1,X)
\le
U_P(X,Y^1)
\qquad Q\text{-almost surely}.
\label{msm_iv_bounds}
\end{equation}
\end{proposition}

Proposition~\ref{prop:comp_msm_iv} establishes that the latent density
ratio in the baseline stratum must lie between the pointwise compatibility
envelopes $L_P(X,Y)$ and $U_P(X,Y)$. Compatibility therefore requires
these envelopes not to cross and
the latent density ratio to integrate to one. Define the conditional
envelope averages
$\widetilde L_P(X)
:=\E_P\{L_P(X,Y)\mid T=1,X,Z=0\}$ and
$\widetilde U_P(X)
:=\E_P\{U_P(X,Y)\mid T=1,X,Z=0\}$.
The following corollary expresses these requirements as observable
restrictions.

\begin{corollary}[Observable compatibility restrictions]
\label{coro:falsification_dual}
Suppose $P\in\mc{M}_1(\Lambda_0,\Lambda_1)$ with $\Lambda_0,\Lambda_1<+\infty$. Then
\[
\frac{\pi_{0,P}(x)+\{1-\pi_{0,P}(x)\}/\Lambda_0}
     {\pi_{1,P}(x)+\{1-\pi_{1,P}(x)\}\Lambda_1}
\le \gamma_{1,P}(y,x)\le
\frac{\pi_{0,P}(x)+\{1-\pi_{0,P}(x)\}\Lambda_0}
     {\pi_{1,P}(x)+\{1-\pi_{1,P}(x)\}/\Lambda_1}
\]
for $P_X$-almost every $x$ and $P(Y\mid T=1,X=x,Z=0)$-almost every $y$. Moreover, $\widetilde L_P(X)\le1\le\widetilde U_P(X)$ $P$-almost surely.
\end{corollary}

Corollary~\ref{coro:falsification_dual} provides necessary population-level conditions for $P\in\mc{M}_1(\Lambda_0,\Lambda_1)$. The permitted range of $\gamma_{1,P}$ depends on both the sensitivity parameters and the treatment propensities. At $(\Lambda_0,\Lambda_1)=(1,1)$, it collapses to $\gamma_{1,P}(Y,X)=1$ almost surely under $P(X,Y\mid T=1,Z=0)$; increasing either sensitivity parameter widens this range. A violation of the density-ratio inequalities or the normalization restrictions implies incompatibility with the joint model. It does not identify which component fails; under the IV model, it rules out the proposed sensitivity specification.
Among specifications that remain compatible, the choice of $(\Lambda_0,\Lambda_1)$ remains a substantive modeling decision.

\begin{remark}[Single-arm limit and non-binding IV constraints]
If $\Lambda_1=+\infty$, no finite sensitivity restriction is imposed in
the $Z=1$ stratum. Compatibility still requires $\gamma_{1,P}(Y,X)>0$ almost surely under $P(X,Y\mid T=1,Z=0)$: otherwise, the affine relationship in Proposition~\ref{prop:comp_msm_iv} would force the nonnegative latent density ratio $\lambda^1_{0,Q}$ to be negative.
The envelopes then reduce to
$L_P(X,Y)=\max\{\Lambda_0^{-1},B_P(X,Y)\}$ and $U_P(X,Y)=\Lambda_0$.
If $B_P(X,Y)\le\Lambda_0^{-1}$ almost surely under $P(X,Y\mid T=1,Z=0)$, then $L_P(X,Y)=\Lambda_0^{-1}$, so the cross-stratum restriction supplied by the instrument is non-binding. The resulting sharp bounds therefore reduce to those obtained from the marginal sensitivity restriction in the baseline stratum alone.
\end{remark}

\section{Sharp Bounds}\label{sec:sharp_bounds}

Having characterized the pointwise compatibility restrictions, we now translate them into bounds on the causal parameter. We restrict attention to observed-data distributions satisfying \(\E_P(|Y|)<\infty\) and take \(Z=0\) as the baseline stratum, with \(\Lambda_0<+\infty\), while the comparison-stratum parameter \(\Lambda_1\) may equal \(+\infty\). Let $\mc{E}_P(\Lambda_0,\Lambda_1)$ be the collection of measurable functions $\lambda$ satisfying
\[
\E_P\left\{
\frac{T\lambda(Y,X)}{\pi_{0,P}(X)}
\,\middle|\,X,Z=0
\right\}=1,
\qquad
L_P(X,Y)\le\lambda(Y,X)\le U_P(X,Y)
\]
where the normalization holds $P$-almost surely and the envelope
inequalities hold almost surely under $P(X,Y\mid T=1,Z=0)$.

For notational convenience, throughout the remainder of the paper we define $e_P(X):=P(Z=0\mid X)$, so that inverse weighting in the baseline stratum is written as $I\{Z=0\}/e_P(X)$. For any function $\lambda\in\mc{E}_P(\Lambda_0,\Lambda_1)$, we define the functional
\[
R_P(\lambda) := \E_{P}\left[ \frac{I(Z=0)}{e_P(X)} \left\{ TY + T\frac{1-\pi_{0,P}(X)}{\pi_{0,P}(X)}\lambda(Y,X)Y \right\} \right].
\]

The following proposition adapts the sharpness constructions of
\citet{dorn2023sharp} and \citet{tan2024model} to the IV-MSM.

\begin{proposition}[Sharpness mapping]
\label{prop:sharpness_msm_iv}
Let $P\in\mc{M}_1(\Lambda_0,\Lambda_1)$.
\begin{enumerate}[(i)]
\item For every $\lambda\in\mc{E}_P(\Lambda_0,\Lambda_1)$, there exists a
full-data extension $Q$ of $P$ satisfying the restrictions defining
$\mc{M}_1(\Lambda_0,\Lambda_1)$ such that
$\lambda^1_{0,Q}(Y^1,X)=\lambda(Y^1,X)$ $Q$-almost surely and
$\mu^1(Q)=R_P(\lambda)$.
\item Conversely, for every such full-data extension $Q$ of $P$,
$\lambda^1_{0,Q}\in\mc{E}_P(\Lambda_0,\Lambda_1)$ and
$\mu^1(Q)=R_P(\lambda^1_{0,Q})$.
\end{enumerate}
\end{proposition}

Proposition~\ref{prop:sharpness_msm_iv} implies that the sharp identified set for $\mu^1(Q)$ under
$\mc{M}_1(\Lambda_0,\Lambda_1)$ is $[\mu^{1-}(P),\mu^{1+}(P)]$, where
\[
\mu^{1-}(P):=\inf_{\lambda\in\mc{E}_P(\Lambda_0,\Lambda_1)}R_P(\lambda)\qquad\text{and}\qquad\mu^{1+}(P):=\sup_{\lambda\in\mc{E}_P(\Lambda_0,\Lambda_1)}R_P(\lambda).
\]
Characterizing these bounds reduces to finding density ratios $\lambda_P^\pm\in\mc{E}_P(\Lambda_0,\Lambda_1)$ that achieve these extremes.

By linearity of $R_P(\lambda)$, optimization has a threshold
characterization analogous to the Neyman--Pearson lemma:
minimization places as much weight as permitted on smaller
outcomes, whereas maximization favors larger outcomes. When $Y\mid(T=1,X,Z=0)$ is atomless, finite threshold functions
$q^-_P$ and $q^+_P$ satisfying conditional normalization exist
on the set where $\widetilde L_P(X)<1<\widetilde U_P(X)$.
These thresholds need not be unique. Writing
$\Delta_P(X,Y):=U_P(X,Y)-L_P(X,Y)$, a pair of optimal density
ratios on this set is given by
\[
\begin{aligned}
\lambda^-_P(Y,X)
&=L_P(X,Y)+\Delta_P(X,Y)I\{Y\le q^-_P(X)\},\\
\lambda^+_P(Y,X)
&=U_P(X,Y)-\Delta_P(X,Y)I\{Y\le q^+_P(X)\}.
\end{aligned}
\]

To determine the thresholds required by the normalization constraint, we use the position of one between the conditional envelope averages \(\widetilde L_P(X)\) and \(\widetilde U_P(X)\). Corollary~\ref{coro:falsification_dual} shows that model compatibility requires $\widetilde{L}_P(X) \le 1 \le \widetilde{U}_P(X)$ almost surely. Define the structural weight parameter
\[
\tau_P(X):=
\begin{cases}
\dfrac{\widetilde{U}_P(X)-1}
{\widetilde{U}_P(X)-\widetilde{L}_P(X)},
&\text{if }\widetilde{L}_P(X)<\widetilde{U}_P(X),\\
0,&\text{if }\widetilde{L}_P(X)=\widetilde{U}_P(X).
\end{cases}
\]
Compatibility guarantees that $\tau_P(X)\in[0,1]$ $P$-almost surely. For $P_X$-almost every $x$ with
$\widetilde{L}_P(x)=\widetilde{U}_P(x)$, compatibility requires both quantities to equal one. The nonnegativity of
$\Delta_P(X,Y)$ then implies $L_P(x,Y)=U_P(x,Y)$ almost surely under
$P(Y\mid T=1,X=x,Z=0)$, so the value assigned to $\tau_P(x)$ is immaterial.

For $P_X$-almost every $x$ at which either normalization inequality
binds, the feasible density ratio is uniquely determined:
$\widetilde L_P(x)=1$ implies $\lambda(Y,x)=L_P(x,Y)$, whereas
$\widetilde U_P(x)=1$ implies $\lambda(Y,x)=U_P(x,Y)$, almost
surely under $P(Y\mid T=1,X=x,Z=0)$. Thus, the conditional contributions to the lower and upper sharp bounds
coincide at such covariate values. The threshold construction is needed only on the nondegenerate set where
$\widetilde{L}_P(X)<1<\widetilde{U}_P(X)$.

On the nondegenerate set, the total mass available to shift between the
envelopes is $\widetilde U_P(X)-\widetilde L_P(X)$. For $\lambda_P^+$, the excess mass that must be subtracted from the upper envelope is $\widetilde U_P(X)-1$, corresponding to a fraction $\tau_P(X)$ of the available envelope gap. For $\lambda_P^-$, the deficit that must be added
to the lower envelope is $1-\widetilde L_P(X)$, corresponding to the remaining fraction $1-\tau_P(X)$. Accordingly, \(q_P^-(X)\) and \(q_P^+(X)\) are the \(\{1-\tau_P(X)\}\)-th and \(\tau_P(X)\)-th conditional quantiles of \(Y\), respectively, under the conditional distribution \(Y\mid(T=1,X,Z=0)\) weighted by \(\Delta_P(X,Y)\).

Equivalently, these thresholds minimize the corresponding weighted
check-loss objectives. Let
$\rho_\nu(y,q):=(y-q)\{\nu-I(y\le q)\}$ denote the check-loss function. When finite measurable threshold functions $q_P^-$ and $q_P^+$ exist, they satisfy
$$
\begin{aligned}
        q^-_P &\in \arg\min_{q(\cdot)} \E_P \left[ \frac{I(Z=0)}{e_P(X)} \frac{T\{1-\pi_{0,P}(X)\}}{\pi_{0,P}(X)} \Delta_P(X,Y) \rho_{1-\tau_P(X)}\{Y, q(X)\} \right], \\
        q^+_P &\in \arg\min_{q(\cdot)} \E_P \left[ \frac{I(Z=0)}{e_P(X)} \frac{T\{1-\pi_{0,P}(X)\}}{\pi_{0,P}(X)}\Delta_P(X,Y) \rho_{\tau_P(X)}\{Y, q(X)\} \right].
\end{aligned}
$$

Substituting the optimal density ratios into $R_P(\lambda)$ yields
check-loss representations of the sharp bounds. Let $\mc{Q}$ be the
class of measurable functions from $\mc{X}$ to $\mathbb{R}$, and write
$V_P(X,Y)=1+\frac{1-\pi_{0,P}(X)}{\pi_{0,P}(X)}
[\tau_P(X)L_P(X,Y)+\{1-\tau_P(X)\}U_P(X,Y)]$.
For a fixed threshold $q\in\mc{Q}$, define the lower- and upper-bound
objective functionals
\begin{equation}
\begin{aligned}
\nu_P^{1-}(q)
&:=\E_P\left[\frac{I(Z=0)}{e_P(X)}T
\left\{YV_P(X,Y)
-\frac{1-\pi_{0,P}(X)}{\pi_{0,P}(X)}
\Delta_P(X,Y)\rho_{1-\tau_P(X)}(Y,q(X))\right\}\right],\\
\nu_P^{1+}(q)
&:=\E_P\left[\frac{I(Z=0)}{e_P(X)}T
\left\{YV_P(X,Y)
+\frac{1-\pi_{0,P}(X)}{\pi_{0,P}(X)}
\Delta_P(X,Y)\rho_{\tau_P(X)}(Y,q(X))\right\}\right].
\end{aligned}
\label{eq:fixed_threshold_objectives}
\end{equation}

The following theorem expresses the sharp bounds as optimized values of these fixed-threshold functionals. Its atomlessness assumption avoids the need for fractional allocation at outcome mass points. The theorem separately gives sufficient conditions for finite optimal thresholds to exist, without requiring uniqueness. Results for binary outcomes are presented in Section~\ref{sec:binary_bounds}, with extensions to finite ordered outcomes in Appendix~\ref{sec:supp_finite_extensions}.

\begin{theorem}[Representations of the Sharp Bounds]
\label{thm:sharp_bounds_reps}
Suppose $P\in\mc{M}_1(\Lambda_0,\Lambda_1)$, $\E_P(|Y|)<\infty$, and the conditional distribution of $Y\mid(T=1,X,Z=0)$ is atomless for $P_X$-almost every $x$. Then the sharp bounds are
\[
\mu^{1-}(P)=\sup_{q\in\mc{Q}}\nu_P^{1-}(q)\qquad\text{and}\qquad\mu^{1+}(P)=\inf_{q\in\mc{Q}}\nu_P^{1+}(q).
\]

If, for $P_X$-almost every $x$, either
$\widetilde L_P(x)=\widetilde U_P(x)=1$ or
$\widetilde L_P(x)<1<\widetilde U_P(x)$, the supremum and infimum
are attained by finite measurable threshold functions $q_P^-$ and
$q_P^+$, respectively. In that case,
$\mu^{1-}(P)=\nu_P^{1-}(q_P^-)$ and
$\mu^{1+}(P)=\nu_P^{1+}(q_P^+)$.
\end{theorem}

The corresponding sharp bounds for $\mu^0(Q)$ follow by relabeling treatment. Combining the two sets of bounds yields the sharp identified set for the ATE.

\begin{corollary}[Sharp Bounds for the ATE]
\label{cor:sharp_ate}
Suppose $P\in\mc{M}(\Lambda_0,\Lambda_1)$, and let
$\mu^{0-}(P)$ and $\mu^{0+}(P)$ denote the sharp bounds for $\mu^0(Q)$
obtained by relabeling treatment. Then the sharp identified set for $\psi(Q)$ is
$[\psi^-(P),\psi^+(P)]$, where
\[
\psi^-(P):=\mu^{1-}(P)-\mu^{0+}(P),
\qquad
\psi^+(P):=\mu^{1+}(P)-\mu^{0-}(P).
\]
\end{corollary}

The bounds in Corollary~\ref{cor:sharp_ate} are jointly sharp because the full-data constructions for $Y^1$ and $Y^0$ can be coupled conditionally on $(Y,T,X,Z)$ without changing their observed-data margins or the restrictions imposed on either potential outcome.

The optimization representations in Theorem~\ref{thm:sharp_bounds_reps} characterize the sharp bounds whether or not finite threshold functions attain the extrema. For estimation, however, optimization over the unrestricted class of measurable functions is generally infeasible. Restricting the optimization to a tractable function class produces conservative outer bounds, as formalized below.

\begin{corollary}[Valid Bounds under Restricted Function Classes]\label{cor:restricted_classes}
    Let $\widetilde{\mc{Q}} \subseteq \mc{Q}$ be a nonempty, prespecified function class, such as finite-dimensional linear or spline class. Define the restricted-class bounds:
\[
\widetilde\mu^{1-}(P)=\sup_{q\in\widetilde{\mc{Q}}}\nu_P^{1-}(q)\qquad\text{and}\qquad\widetilde\mu^{1+}(P)=\inf_{q\in\widetilde{\mc{Q}}}\nu_P^{1+}(q).
\]
    Then the restricted-class bounds satisfy
    $$ \widetilde{\mu}^{1-}(P) \le \mu^{1-}(P) \le \mu^{1+}(P) \le \widetilde{\mu}^{1+}(P). $$
\end{corollary}

Thus restricting the threshold-function class may sacrifice sharpness but cannot invalidate the resulting population bounds.

\begin{remark}[Reduction to conventional MSM sharp bounds]
Let $\widetilde X$ denote the adjustment variables in a conventional MSM analysis. Its sharp-bound expressions are recovered by replacing $\pi_{0,P}(X)$ with $\pi_P(\widetilde X):=P(T=1\mid\widetilde X)$, removing the baseline-stratum weight $I\{Z=0\}/e_P(X)$, and setting $L_P=\Lambda^{-1}$ and $U_P=\Lambda$. These substitutions give $\Delta_P=\Lambda-\Lambda^{-1}$, $\tau_P=\Lambda/(\Lambda+1)$ for $\Lambda>1$, and $V_P=1/\pi_P(\widetilde X)$, recovering the conventional MSM sharp bounds of \citet{dorn2023sharp} and \citet{tan2024model}. Taking $\widetilde X=X$ ignores $Z$, whereas taking $\widetilde X=(X,Z)$ additionally adjusts for $Z$.
\end{remark}

\begin{remark}[Combining IV and sensitivity restrictions]
\label{rem:combining_iv_msm}
The IV-MSM combines the stratum-specific marginal sensitivity restrictions through the affine relationship induced by the IV assumptions, constraining $\lambda^1_{0,Q}(Y^1,X)$ to $[L_P(X,Y^1),U_P(X,Y^1)]$. It therefore reduces the feasible set of full-data extensions relative to an IV analysis without sensitivity restrictions. Consequently, its identified set for $\mu^1(Q)$ is contained in the corresponding IV-only identified set, with strict inclusion whenever the additional sensitivity restrictions change either extremal value.

When $\Lambda_0=\Lambda_1=\Lambda$, the sharp IV-MSM
interval for $\mu^1(Q)$ is also contained in the corresponding conventional-MSM interval at the same $\Lambda$, whether adjusting for $(X,Z)$ or $X$ alone.
The former follows because the IV-MSM is a submodel of
the conventional MSM adjusting for $(X,Z)$. The latter follows from $Y^t\indep_Q Z\mid X$, implied by
Assumptions~\ref{ass:consist}, \ref{ass:iv_excl},
and~\ref{ass:iv_exchan}, together with Jensen's inequality applied to the lower and upper treatment-probability limits corresponding to the odds restrictions; the argument is provided in Appendix~\ref{sec:proofs_identification}.
\end{remark}

\section{Estimation and Inference}\label{sec:estimation}

We develop estimation and inference for the restricted-class bounds $\widetilde\mu^{1-}(P)$ and $\widetilde\mu^{1+}(P)$ defined in Corollary~\ref{cor:restricted_classes}. The corresponding bounds for $\mu^0$ follow by relabeling treatment, and the ATE bounds are obtained using the contrasts in Section~\ref{sec:sharp_bounds}. Throughout this section, $(\Lambda_0,\Lambda_1)$ is fixed and finite, with the same pair applied to both potential outcomes. For estimation of the restricted-class bounds, we focus on $(\Lambda_0,\Lambda_1)\neq(1,1)$. At $(1,1)$, the model implies $Y^t\indep T\mid X,Z$, so the ATE is point identified and usual point-identified estimators apply.

Our primary inferential result provides one-sided lower confidence limits for the lower bounds and one-sided upper confidence limits for the upper bounds, for both the potential-outcome means and the ATE.
Combining these limits with an appropriate allocation of the error probability yields outward confidence sets covering the restricted-class intervals and hence the sharp identified intervals. Under the conditions of
Theorem~\ref{thm:restricted_crossfit_inference}, these guarantees do not require pathwise differentiability of the optimized bounds or that the learned thresholds attain the population optimum, but coverage may be conservative.
Under stronger conditions, we establish regular and efficient estimation of the restricted-class bounds and asymptotically valid two-sided confidence intervals.
We first derive the influence functions underlying these results and then develop cross-fitted one-step estimation and inference. We conclude with a level-controlled compatibility test for candidate sensitivity-parameter pairs.

\subsection{Pathwise differentiability of the restricted-class bounds}
\label{subsec:sharp_pathwise}

We examine pathwise differentiability first for fixed-threshold functionals and then for their optimized values over $\widetilde{\mc Q}$. Throughout, this class is nonempty,
deterministic, and held fixed as $P$ varies. We use the same class for the lower and upper bounds on each potential-outcome mean. The class may differ between potential outcomes, but we suppress this distinction in the notation.

The corresponding restricted ATE bounds are
$\widetilde\psi^-=\widetilde\mu^{1-}-\widetilde\mu^{0+}$ and
$\widetilde\psi^+=\widetilde\mu^{1+}-\widetilde\mu^{0-}$.
By Corollary~\ref{cor:restricted_classes},
$\widetilde\psi^-(P)\le\psi^-(P)\le\psi^+(P)\le\widetilde\psi^+(P)$.
These targets depend on the population law and the specified
classes, not on the particular thresholds learned from a sample.
Selecting the target-defining class from the estimation data
would require additional analysis.

For $c>0$, write $\Gamma_{c,P}(x,y)=\{\gamma_{1,P}(y,x)[\pi_{1,P}(x)+
\{1-\pi_{1,P}(x)\}c]-\pi_{0,P}(x)\}/\{1-\pi_{0,P}(x)\}$. With this notation, the envelopes defined in Section~\ref{sec:compatibility_dual} can be written as $L_P(x,y)=\max\{\Lambda_0^{-1},\Gamma_{\Lambda_1^{-1},P}(x,y)\}$ and
$U_P(x,y)=\min\{\Lambda_0,\Gamma_{\Lambda_1,P}(x,y)\}$.

Following the threshold representation in
Section~\ref{sec:sharp_bounds}, for $b\in\{-,+\}$ let $q^b$ be a fixed threshold function, not necessarily optimal. Write
$a_{U,q}^-(y,x)=I\{y\le q^-(x)\}$,
$a_{U,q}^+(y,x)=I\{y>q^+(x)\}$, and
$a_{L,q}^b(y,x)=1-a_{U,q}^b(y,x)$. The associated envelope allocation is
$\lambda_{P,q}^b(y,x)
=a_{U,q}^b(y,x)U_P(x,y)+a_{L,q}^b(y,x)L_P(x,y)$, which need not satisfy the conditional normalization constraint. Using this allocation, define
\begin{equation}
\begin{aligned}
\theta_{P,q}^b(x)
&=q^b(x)+\E_P\!\left[
\lambda_{P,q}^b(Y,x)\{Y-q^b(x)\}
\mid T=1,X=x,Z=0\right],\\
M_{P,q}^b(x)
&=\pi_{0,P}(x)m_{0,P}(x)
+\{1-\pi_{0,P}(x)\}\theta_{P,q}^b(x),
\end{aligned}
\label{eq:sharp_theta_M}
\end{equation}
where $m_{0,P}(x)=\E_P(Y\mid T=1,X=x,Z=0)$. A direct rearrangement of the representations in~\eqref{eq:fixed_threshold_objectives} gives $\nu_P^{1b}(q^b)=\E_P\{M_{P,q}^b(X)\}$. Because $\lambda_{P,q}^b$ need not be conditionally normalized for an
arbitrary $q^b$, $\theta_{P,q}^b(x)$ does not generally reduce to
$\E_P\{\lambda_{P,q}^b(Y,x)Y\mid T=1,X=x,Z=0\}$.

To state the efficient influence function of the fixed-threshold functional compactly, we first describe how perturbations of the nuisance functions entering $\Gamma_{c,P}$ propagate through the envelopes. Because $L_P$ and $U_P$ are defined by a maximum and a minimum, respectively, their first-order changes depend on which term determines each envelope. Define the corresponding active-term indicators $\omega_{L,P}=I\{\Gamma_{\Lambda_1^{-1},P}>\Lambda_0^{-1}\}$ and $\omega_{U,P}=I\{\Gamma_{\Lambda_1,P}<\Lambda_0\}$. For $\zeta\in\{\gamma_1,\pi_0,\pi_1\}$, let
$\partial_\zeta\Gamma_{c,P}$ denote the ordinary derivative of
$\Gamma_{c,P}$ with respect to the indicated nuisance function, and define $\dot L_{P,\zeta}=\omega_{L,P}\partial_\zeta
\Gamma_{\Lambda_1^{-1},P}$ and
$\dot U_{P,\zeta}=\omega_{U,P}\partial_\zeta
\Gamma_{\Lambda_1,P}$.

Let $D_{P,q,\zeta}^b=\{Y-q^b(X)\}
(a_{U,q}^b\dot U_{P,\zeta}+a_{L,q}^b\dot L_{P,\zeta})$ and define
\begin{equation}
\begin{aligned}
H_{0,P,q}^b(Y,X)
={}&\pi_{0,P}(X)Y+\{1-\pi_{0,P}(X)\}
\left[\lambda_{P,q}^b(Y,X)\{Y-q^b(X)\}
-\gamma_{1,P}(Y,X)D_{P,q,\gamma_1}^b(Y,X)\right],\\
H_{1,P,q}^b(Y,X)
={}&\{1-\pi_{0,P}(X)\}D_{P,q,\gamma_1}^b(Y,X).
\end{aligned}
\label{eq:sharp_H}
\end{equation}
Write $\overline H_{z,P,q}^b(x)=\E_P\{H_{z,P,q}^b(Y,x)\mid
T=1,X=x,Z=z\}$ and $\overline D_{P,q,\zeta}^b(x)=
\E_P\{D_{P,q,\zeta}^b(Y,x)\mid T=1,X=x,Z=0\}$. Finally, let
$C_{0,P,q}^b=m_{0,P}-\theta_{P,q}^b+(1-\pi_{0,P})
\overline D_{P,q,\pi_0}^b$ and
$C_{1,P,q}^b=(1-\pi_{0,P})\overline D_{P,q,\pi_1}^b$.

\begin{theorem}[Pathwise differentiability of restricted-class bounds]
\label{thm:eif_fixed_q} Fix $b\in\{-,+\}$. Suppose that $P\in\mc M_1(\Lambda_0,\Lambda_1)$, the outcome domain $\mc Y$ is bounded, $\widetilde{\mc Q}$ consists of bounded measurable threshold functions, and the two envelope-tie sets have probability zero among observations with $T=1,Z=0$; that is, $P\{\Gamma_{\Lambda_1^{-1},P}=\Lambda_0^{-1}\mid T=1,Z=0\}=0$ and $P\{\Gamma_{\Lambda_1,P}=\Lambda_0\mid T=1,Z=0\}=0$.
Then:
\begin{enumerate}[(i)]
    \item For every fixed $q^b\in\widetilde{\mc Q}$, the fixed-threshold functional $\nu_P^{1b}(q^b)$ is pathwise differentiable in the nonparametric observed-data model, with efficient influence function
\begin{equation}
\begin{aligned}
\phi_{P,q}^{1b}(O)
={}&M_{P,q}^b(X)-\nu_P^{1b}(q^b)\\
&+\frac{I(Z=0)}{e_P(X)}
\left[
C_{0,P,q}^b(X)\{T-\pi_{0,P}(X)\}
+\frac{T}{\pi_{0,P}(X)}
\{H_{0,P,q}^b(Y,X)-\overline H_{0,P,q}^b(X)\}
\right]\\
&+\frac{I(Z=1)}{1-e_P(X)}
\left[
C_{1,P,q}^b(X)\{T-\pi_{1,P}(X)\}
+\frac{T}{\pi_{1,P}(X)}
\{H_{1,P,q}^b(Y,X)-\overline H_{1,P,q}^b(X)\}
\right].
\end{aligned}
\label{eq:eif_fixed_q}
\end{equation}
\item Let
$\widetilde{\mc Q}_P^{b,\star}=\{q\in\widetilde{\mc Q}:\nu_P^{1b}(q)=\widetilde\mu^{1b}(P)\}$ be nonempty, and fix $\widetilde q_P^b\in\widetilde{\mc Q}_P^{b,\star}$.
Suppose that, along every two-sided regular parametric submodel $\{P_s:s\in(-\delta,\delta)\}$ through $P$, the change due to reoptimizing is negligible to first order, in the sense that
\begin{equation}
\widetilde\mu^{1b}(P_s)
-
\nu_{P_s}^{1b}(\widetilde q_P^b)
=
o(|s|).
\label{eq:first_order_optimizer}
\end{equation}
Then $\widetilde\mu^{1b}$ is pathwise differentiable at $P$, with efficient influence function $\phi_{P,\widetilde q_P^b}^{1b}$.
\end{enumerate}
\end{theorem}

The no-tie conditions ensure that one argument strictly determines each envelope almost surely among observations with $T=1,Z=0$, removing the directional ambiguity in differentiating the maximum and minimum operations.
The fixed-threshold influence function in part~\textup{(i)} underlies our cross-fitted estimator, one-sided confidence limits, and outward confidence sets in Section~\ref{subsec:sharp_crossfit}; these do not require condition~\eqref{eq:first_order_optimizer}.
The boundedness requirements on the outcome domain and threshold functions are sufficient but not necessary.
More general moment conditions allowing unbounded outcomes and threshold functions are given in Appendix~\ref{sec:proofs_identification}.

Part~\textup{(ii)} provides the additional differentiability result used to establish efficiency for the optimized restricted-class bounds.
In condition~\eqref{eq:first_order_optimizer}, $\widetilde\mu^{1b}(P_s)$ reoptimizes the objective under $P_s$, whereas $\nu_{P_s}^{1b}(\widetilde q_P^b)$ retains a threshold optimal at $P$.
The condition requires reoptimization to contribute no first-order change, so the fixed-threshold influence function is also the efficient influence function of the optimized bound. Sufficient conditions for \eqref{eq:first_order_optimizer} to hold for the threshold classes used in the numerical experiments are given in Appendix~\ref{subsec:supp_optimizer_regularity}.

\subsection{One-step cross-fitted estimation}
\label{subsec:sharp_crossfit}

To accommodate flexible estimation of the nuisance and threshold functions,
we use cross-fitting. Let
\[
\eta_P=
\left(
e_P,\pi_{0,P},\pi_{1,P},\gamma_{1,P},m_{0,P},
\theta_{P,q}^b,
\overline H_{0,P,q}^b,\overline H_{1,P,q}^b,
\overline D_{P,q,\pi_0}^b,\overline D_{P,q,\pi_1}^b
\right)
\]
collect the nuisance functions required for a supplied threshold $q^b$. Define the uncentered influence function $\Phi^{1b}(O;\eta_P,q^b)
=
\nu_P^{1b}(q^b)+\phi_{P,q}^{1b}(O)$. Although $M_{P,q}^b$ appears in the influence function, it need not be
estimated through an additional regression, since the identity
\begin{equation}
M_{P,q}^b(x)
=
\overline H_{0,P,q}^b(x)
+\overline H_{1,P,q}^b(x)
+\{1-\pi_{0,P}(x)\}q^b(x)
\label{eq:linked_M}
\end{equation}
holds for every fixed threshold. We therefore construct $\widehat M^b$ through equation~\eqref{eq:linked_M}, using the same fitted conditional means and propensity score that appear in the estimated influence function.

Partition the observations into a fixed number $K$ of approximately equal
folds, with fold sizes $n_k$ satisfying $n_k/n\to 1/K$. Using only the training sample $-k$, estimate the nuisances and learn $\widehat q_{-k}^b\in\widetilde{\mc Q}$.
For $i$ in fold $k(i)$, let
$\widehat\Phi_i^{1b}=\Phi^{1b}(O_i;\widehat\eta_{-k(i)},
\widehat q_{-k(i)}^b)$, and define the cross-fitted one-step estimator
$\widehat{\widetilde\mu}^{1b}=n^{-1}\sum_i\widehat\Phi_i^{1b}$. Analogously, obtain $\widehat{\widetilde\mu}^{0b}$, and form
$\widehat{\widetilde\psi}^-=\widehat{\widetilde\mu}^{1-}
-\widehat{\widetilde\mu}^{0+}$ and
$\widehat{\widetilde\psi}^+=\widehat{\widetilde\mu}^{1+}
-\widehat{\widetilde\mu}^{0-}$.

To describe the error from learning the thresholds, write
$\overline\nu_n^{1b}=\sum_k(n_k/n)\nu_P^{1b}(\widehat q_{-k}^b)$.
This fold-average population value is random and differs from the fixed target
$\widetilde\mu^{1b}(P)$. Define the nonnegative within-class regrets
\begin{equation}
r_{\widetilde{\mc Q},n}^{1-}
=\widetilde\mu^{1-}(P)-\overline\nu_n^{1-},
\qquad
r_{\widetilde{\mc Q},n}^{1+}
=\overline\nu_n^{1+}-\widetilde\mu^{1+}(P).
\label{eq:within_class_regret}
\end{equation}
They measure the loss from using the learned fold-specific thresholds rather than population-optimal thresholds within $\widetilde{\mc Q}$. Likewise, the fold-average ATE values, defined by the same contrasts, satisfy
$\overline\psi_n^-\le\widetilde\psi^-(P)\le
\widetilde\psi^+(P)\le\overline\psi_n^+$ for every training sample.

We next give a condition controlling estimation of the exact envelopes. Adapting the margin condition used by
\citet{levis2025covariate} for inference on nonsmooth bound functionals, define the distances to the two envelope ties by $d_{L,P}(x,y)
=
\big|
\Gamma_{\Lambda_1^{-1},P}(x,y)-\Lambda_0^{-1}
\big|$ and $d_{U,P}(x,y)
=
\left|
\Gamma_{\Lambda_1,P}(x,y)-\Lambda_0
\right|$.

\begin{assumption}[Envelope margin condition]
\label{ass:exact_envelope_margin}
There exist constants $C<\infty$, $\kappa>0$, and $t_0>0$ such that
\[
P\{d_{L,P}(X,Y)\le t\mid T=1,Z=0\}
+
P\{d_{U,P}(X,Y)\le t\mid T=1,Z=0\}
\le Ct^\kappa
\]
for every $t\in[0,t_0]$.
\end{assumption}

This condition controls how often the two arguments defining either envelope are nearly equal. Such near-ties matter because a small estimation error in $\Gamma_{c,P}$ can change which argument determines the maximum or minimum. The distances depend on \((\Lambda_0,\Lambda_1)\), so the condition is evaluated separately at each candidate sensitivity-parameter pair. It permits arbitrarily small distances to a tie, but requires the probability of lying within distance $t$ of either tie to be of order at most $t^\kappa$. For example, it holds with $\kappa=1$ if, conditional on $(T=1,Z=0)$, the signed differences between the competing envelope arguments have densities bounded near zero. We therefore expect the condition to be plausible in many settings where these differences vary continuously and no structural restriction forces them to equal zero. The condition excludes exact ties with positive probability, as required
by Theorem~\ref{thm:eif_fixed_q}, and additionally controls concentration
near them.

A separate issue arises from learning the threshold functions. With cross-fitting, each $\widehat q_{-k}^b$ is estimated from a different training sample and may therefore induce a different fold-specific uncentered influence function. For a measurable function $f(O)$, write $\|f\|_{P,2}=\{\E_P[f(O)^2]\}^{1/2}$. To obtain a common first-order term across folds, we impose the following stabilization condition.

\begin{assumption}[Influence-function stabilization]
\label{ass:sharp_if_stabilization}
There exists a fixed $q_\infty^b\in\widetilde{\mc Q}$ such that
\[
\max_{1\le k\le K}
\left\|
\Phi^{1b}(\,\cdot\,;\eta_P,\widehat q_{-k}^b)
-
\Phi^{1b}(\,\cdot\,;\eta_P,q_\infty^b)
\right\|_{P,2}
=o_p(1).
\]
\end{assumption}

This assumption holds, for example, if the learned thresholds converge to a unique within-class optimizer and the uncentered influence function is $L_2(P)$-continuous at that optimizer. It also permits nonunique optimizers if the learning procedure consistently selects one of them or if all optimizers that may be selected induce the same uncentered influence function. It may fail when different folds continue to select optimizers that induce different influence functions. The assumption concerns stabilization of the influence function and does not itself require \(q_\infty^b\) to be a population optimizer. The latter condition
is needed for the efficiency conclusion below. By comparison, the cross-fitted analysis of \citet{dorn2025doubly} under the conventional MSM assumes that the fold-specific conditional quantile estimators themselves converge to a common fixed function, which, together with suitable continuity, implies the stabilization condition above. 

We now state the resulting cross-fitted expansion and inferential conclusions.

\begin{theorem}[Cross-fitted inference for restricted-class bounds]
\label{thm:restricted_crossfit_inference}
Suppose the observations are i.i.d. and the true and fitted instrument and treatment probabilities are uniformly bounded away from zero and one.
Assume the boundedness conditions of Theorem~\ref{thm:eif_fixed_q}\textup{(i)}, Assumption~\ref{ass:exact_envelope_margin}, and Assumption~\ref{ass:sharp_if_stabilization}, and
$\|\Phi^{1b}(\cdot;\widehat\eta_{-k},\widehat q_{-k}^b)
-\Phi^{1b}(\cdot;\eta_P,\widehat q_{-k}^b)\|_{P,2}=o_p(1)$
uniformly over folds. 
Suppose the fold-averaged nuisance drift satisfies
\[
R_n^{1b}
:=
\sum_{k=1}^K\frac{n_k}{n}
\E_P\left[
\Phi^{1b}(O;\widehat\eta_{-k},\widehat q_{-k}^b)
-
\Phi^{1b}(O;\eta_P,\widehat q_{-k}^b)
\right]=o_p(n^{-1/2}).
\]
Then
\[
\sqrt n\big(
\widehat{\widetilde\mu}^{1b}
-
\overline\nu_n^{1b}
\big)
\rightsquigarrow
N(0,\sigma_{1b}^2),\quad\text{with}\quad \sigma_{1b}^2
=
\Var_P\big\{
\phi_{P,q_\infty^b}^{1b}(O)
\big\},
\]
where $q_\infty^b$ is the fixed threshold specified in Assumption~\ref{ass:sharp_if_stabilization}.

If, in addition, $r_{\widetilde{\mc Q},n}^{1b}$ defined in~\eqref{eq:within_class_regret} is 
\(o_p(n^{-1/2})\), then
\[
\sqrt n\left\{
\widehat{\widetilde\mu}^{1b}
-
\widetilde\mu^{1b}(P)
\right\}
\rightsquigarrow
N(0,\sigma_{1b}^2).
\]

If also \(q_\infty^b\in\widetilde{\mc Q}_P^{b,\star}\) and condition~\eqref{eq:first_order_optimizer} holds with \(\widetilde q_P^b=q_\infty^b\), then the estimator is regular and efficient for \(\widetilde\mu^{1b}(P)\).

The conclusions extend jointly to the four potential-outcome bounds and
their ATE contrasts under the corresponding conditions.
\end{theorem}

The first conclusion supports one-sided inference because
$\overline\nu_n^{1-}\le\widetilde\mu^{1-}(P)$ and
$\overline\nu_n^{1+}\ge\widetilde\mu^{1+}(P)$, with analogous
ordering for the ATE bounds. This guarantee does not require
root-$n$-negligible within-class regret. Under the additional regret
condition, the learned-threshold value and the restricted-class bound
differ by $o_p(n^{-1/2})$, giving asymptotic normality about the restricted-class bound with
the same variance. Sufficient conditions for
$R_n^{1b}=o_p(n^{-1/2})$ are provided in Appendix~\ref{subsubsec:supp_estimation_assumptions}.

\paragraph{Variance estimation.}
For each potential-outcome bound, define
\[
\widehat\sigma_{tb}^{\,2}
=\frac1n\sum_{i=1}^n
\left(
\widehat\Phi_i^{tb}-\widehat{\widetilde\mu}^{tb}
\right)^2,
\qquad
\widehat{\mathrm{se}}_{tb}
=\frac{\widehat\sigma_{tb}}{\sqrt n},
\qquad t\in\{0,1\},\ b\in\{-,+\}.
\]
For the ATE bounds, form the row-level contrasts
$\widehat\Phi_i^{\psi,-}
=\widehat\Phi_i^{1-}-\widehat\Phi_i^{0+}$
and
$\widehat\Phi_i^{\psi,+}
=\widehat\Phi_i^{1+}-\widehat\Phi_i^{0-}$.
Since $n^{-1}\sum_i\widehat\Phi_i^{\psi,b}
=\widehat{\widetilde\psi}^b$, define
\begin{equation}
\widehat\sigma_{\psi,b}^{\,2}
=\frac1n\sum_{i=1}^n
\left(
\widehat\Phi_i^{\psi,b}-\widehat{\widetilde\psi}^b
\right)^2,
\qquad
\widehat{\mathrm{se}}_b
=\frac{\widehat\sigma_{\psi,b}}{\sqrt n}.
\label{eq:ate_bound_variance}
\end{equation}
These contrast-based variances retain the covariance between the two
potential-outcome estimators.

\begin{corollary}[One-sided confidence limits and outward confidence sets]
\label{cor:outward_confidence_set}
Suppose the conditions for the first conclusion of
Theorem~\ref{thm:restricted_crossfit_inference} hold for the four
potential-outcome bounds, the ATE contrasts have positive limiting
variances, and their standard-error estimators are consistent.
Define
\[
L_{n,\alpha}
=\widehat{\widetilde\psi}^-
-z_{1-\alpha}\widehat{\mathrm{se}}_-,
\qquad
U_{n,\alpha}
=\widehat{\widetilde\psi}^+
+z_{1-\alpha}\widehat{\mathrm{se}}_+.
\]
Then
\[
\liminf_{n\to\infty}
P^{\otimes n}\{L_{n,\alpha}\le\widetilde\psi^-(P)\}
\ge1-\alpha,
\qquad
\liminf_{n\to\infty}
P^{\otimes n}\{\widetilde\psi^+(P)\le U_{n,\alpha}\}
\ge1-\alpha.
\]
Moreover, the outward confidence set $\mc C_{1-\alpha}^{\mathrm{out}}
:=
[L_{n,\alpha/2},U_{n,\alpha/2}]$
satisfies
\[
\liminf_{n\to\infty}
P^{\otimes n}\!\left\{
[\widetilde\psi^-(P),\widetilde\psi^+(P)]
\subseteq\mc C_{1-\alpha}^{\mathrm{out}}
\right\}
\ge1-\alpha.
\]
\end{corollary}

Analogous one-sided limits and outward confidence sets apply to each potential-outcome interval, using
$\widehat{\widetilde\mu}^{tb}$ and
$\widehat{\mathrm{se}}_{tb}$, with positive limiting variances and consistent standard errors.

These coverage guarantees also hold under regular $n^{-1/2}$-local alternatives without requiring condition~\eqref{eq:first_order_optimizer}. This follows by comparing the estimator with a fixed-threshold functional and using the outward ordering, as established in the supplement. Coverage may therefore remain valid, but conservative, when the optimized bound is not pathwise differentiable.
Because the restricted-class interval contains the sharp identified interval, the outward confidence set also provides coverage for the latter.

\paragraph{Two-sided confidence intervals and efficiency.}
Under the additional negligible-regret conditions in Theorem~\ref{thm:restricted_crossfit_inference}, the two-sided Wald intervals $\widehat{\widetilde\psi}^b\pm z_{1-\alpha/2}\widehat{\mathrm{se}}_b$ have asymptotic coverage $1-\alpha$ for the individual restricted-class bounds $\widetilde\psi^b(P)$ at the fixed law $P$.
Analogous intervals apply to the potential-outcome bounds using $\widehat{\widetilde\mu}^{tb}$ and $\widehat{\mathrm{se}}_{tb}$. 
Under the theorem's additional conditions for regularity and efficiency, these two-sided coverage guarantees also hold under regular local alternatives.
For these intervals to cover the unrestricted sharp bounds,
one additionally needs control of the approximation error from restricting the threshold class, as discussed in the supplement.

An analogous distinction arises under the conventional MSM: consistent estimation of the relevant conditional quantiles supports inference for individual sharp bounds, whereas inconsistent quantile estimation may still yield valid but conservative outward inference under the corresponding nuisance and regularity conditions
\citep{dorn2023sharp,tan2024model,dorn2025doubly}.

\subsection{Compatibility testing}
\label{sec:compatibility_test}

For a prespecified finite candidate pair, we test the null hypothesis $H_0(\Lambda_0,\Lambda_1): P\in\mc M(\Lambda_0,\Lambda_1)$. Corollary~\ref{coro:falsification_dual} gives eight observable inequalities across the two treatment arms: lower and upper bounds on $\gamma_{t,P}$ and the normalization restrictions $\widetilde L_{t,P}(X)\le1\le\widetilde U_{t,P}(X)$.

Following conditional moment-inequality methods \citep{andrews2013inference}, we use nonnegative weighting functions to detect localized violations. For each inequality, a function is selected from a prespecified finite collection using training observations and evaluated with the fitted nuisances on held-out observations. The test uses the maximum of eight standardized cross-fitted moments, calibrated by a Gaussian multiplier procedure. Under the conditions given in the supplement, it has asymptotic size at most $\alpha$ and is consistent against fixed alternatives for which the witness-selection rule selects a moment whose population violation remains bounded away from zero with probability tending to one.
The supplement provides the construction and theoretical details.

Rejection refutes the joint candidate model. Under the IV model, it rules out the proposed sensitivity pair; at $(1,1)$, it provides evidence against the absence of unmeasured confounding. Nonrejection does not establish compatibility.
\section{Sharp Bounds for Binary Outcomes}
\label{sec:binary_bounds}

When the observed and potential outcomes are binary, identification and
estimation simplify because each conditional outcome distribution is
determined by a single probability. Suppose $\mc{Y}=\{0,1\}$. For $z\in\{0,1\}$, define the observed probability $a_{z,P}(x):=P(Y=1\mid T=1,X=x,Z=z)$. Accordingly, at outcome values with positive probability under the baseline
treated-outcome law, $ \gamma_{1,P}(1,x)=a_{1,P}(x)/a_{0,P}(x)$ and $\gamma_{1,P}(0,x)=\{1-a_{1,P}(x)\}/\{1-a_{0,P}(x)\}$, where each ratio is needed only when its denominator is positive.

For a full-data extension $Q$ of $P$, let $\vartheta_Q^1(x):=Q(Y^1=1\mid T=0,X=x,Z=0)$. At outcome values with positive probability under the baseline treated
distribution, the density ratio is
\[
  \lambda^1_{0,Q}(1,x)
  =\frac{\vartheta_Q^1(x)}{a_{0,P}(x)},
  \qquad
  \lambda^1_{0,Q}(0,x)
  =\frac{1-\vartheta_Q^1(x)}{1-a_{0,P}(x)}.
\]
Thus the pointwise envelope restrictions in Proposition~\ref{prop:comp_msm_iv}, together with normalization, reduce to an
interval restriction on $\vartheta_Q^1(x)$. On the event
$0<a_{0,P}(x)<1$, define
\[
\begin{aligned}
  \vartheta_P^{1-}(x)
  &:=
  \max\left\{
    a_{0,P}(x)L_P(x,1),
    1-\{1-a_{0,P}(x)\}U_P(x,0)
  \right\},\\
  \vartheta_P^{1+}(x)
  &:=
  \min\left\{
    a_{0,P}(x)U_P(x,1),
    1-\{1-a_{0,P}(x)\}L_P(x,0)
  \right\}.
\end{aligned}
\]
If $a_{0,P}(x)=0$, set
$\vartheta_P^{1-}(x)=\vartheta_P^{1+}(x)=0$; if $a_{0,P}(x)=1$, set
$\vartheta_P^{1-}(x)=\vartheta_P^{1+}(x)=1$. These endpoint values are forced
by Assumption~\ref{ass:abs_cont}. With these conventions, the pointwise envelope restrictions are equivalent
to $\vartheta_P^{1-}(x)
\le
\vartheta_Q^1(x)
\le
\vartheta_P^{1+}(x)$. Consequently, a feasible value of $\vartheta_Q^1(x)$ exists if and only if
$\vartheta_P^{1-}(x)\le\vartheta_P^{1+}(x)$.

\begin{proposition}[Sharp bounds for binary outcomes]
\label{prop:binary_bounds}
Suppose $\mc{Y}=\{0,1\}$ and
$P\in\mc{M}_1(\Lambda_0,\Lambda_1)$.  Then the sharp identified set for
$\mu^1(Q)$ is $[\mu^{1-}(P),\mu^{1+}(P)]$, where
\[
  \mu^{1b}(P)
  =\E_P\left[
    \pi_{0,P}(X)a_{0,P}(X)
    +\{1-\pi_{0,P}(X)\}\vartheta_P^{1b}(X)
  \right],
  \qquad b\in\{-,+\}.
\]
\end{proposition}

To see the result, observe that every feasible binary density ratio is in
one-to-one correspondence with a function $\vartheta_Q^1$ lying between
$\vartheta_P^{1-}$ and $\vartheta_P^{1+}$.  Conditional on $X$, the causal
mean is affine and increasing in this function. Its extrema are therefore attained by setting
$\vartheta_Q^1=\vartheta_P^{1-}$ and
$\vartheta_Q^1=\vartheta_P^{1+}$, respectively. These endpoint functions
induce feasible density ratios in $\mc E_P(\Lambda_0,\Lambda_1)$, and
Proposition~\ref{prop:sharpness_msm_iv} ensures that each such density ratio
is realized by a full-data extension of $P$.

The sharp bounds for $\mu^0(Q)$ follow by relabeling treatment, and the sharp
ATE bounds are obtained as in Corollary~\ref{cor:sharp_ate}. If $0<a_{z,P}(X)<1$ almost surely for both instrument strata,
the bounds converge to the natural IV bounds as
$\Lambda_0,\Lambda_1\to\infty$:
\[
\E_P\!\left[\max_{z\in\{0,1\}}\pi_{z,P}(X)a_{z,P}(X)\right]
\le \mu^1(Q) \le
\E_P\!\left[\min_{z\in\{0,1\}}\left\{\pi_{z,P}(X)a_{z,P}(X)
+1-\pi_{z,P}(X)\right\}\right]
\]
\citep{robins1989analysis,manski1990nonparametric}.  For finite sensitivity
parameters, the IV-MSM can further narrow this interval.

\paragraph{Estimation and inference.} For binary outcomes, the explicit bound formulas yield a simplified version of the cross-fitted one-step construction in Section~\ref{sec:estimation}, with no threshold estimation. The corresponding inference results hold without the threshold-related conditions, but require an additional margin condition for the maxima and minima defining $\vartheta_P^{1-}$ and $\vartheta_P^{1+}$. The specialized construction and conditions are given in the supplement.

\paragraph{Simultaneous sensitivity bands.}
Along $\Lambda_0=\Lambda_1=\Lambda$, the sharp lower ATE bound is nonincreasing, the upper bound is nondecreasing, and both are $2$-Lipschitz on any compatible interval. These properties allow simultaneous outward confidence limits on a fixed sensitivity grid to be extended between grid points while preserving coverage over the entire interval, without requiring a uniform asymptotic expansion. The joint grid calibration and its extension to an entire interval are provided in Appendix~\ref{sec:supp_binary_bands}.

\section{Numerical Experiments}
\label{sec:numerical_experiments}

We conduct numerical experiments to evaluate the finite-sample performance of the one-step IV-MSM estimators, outward confidence sets, and compatibility test, varying the instrument--treatment association
and the instrument--outcome association conditional on treatment
and covariates.
For continuous outcomes, the targets are restricted-class ATE bounds; for binary outcomes, they are sharp ATE bounds.
We compare our procedure with the conventional-MSM estimators of \citet{tan2024model} and \citet{dorn2025doubly}.
Each experiment uses 500 replications, with $n=6000$ for bound estimation and $n\in\{6000,12000\}$ for compatibility testing.

For each $i=1,\dots,n$, we generate a covariate vector $X_i=(X_{i1},\ldots,X_{i5})^\top$ with independent components drawn from a standard-normal distribution truncated to $[-2,2]$, and independently generate $Z_i\sim\operatorname{Bernoulli}(0.5)$. Define
\[
\begin{split}
g_T(x)&=0.14x_1+0.12x_2+0.10x_3+0.08x_4+0.06x_5,\\
g_Y(x)&=0.20x_1-0.25x_2+0.30x_3+0.15x_4-0.20x_5,\\
h(x)&=1+0.6\sin(x_1x_2)+0.4\sin(x_3x_4).
\end{split}
\]
We generated treatment as $T_i\sim\text{Bernoulli}(p_i)$ where
\[
\log\frac{p_i}{1-p_i}=g_T(X_i)+a(Z_i-1/2).
\]
For continuous outcomes, we set $Y_i=g_Y(X_i)+0.75T_i+R_i$ where,  conditional on $(X_i,Z_i,T_i)$, $R_i$ is drawn from $N\{Z_i(2T_i-1)b h(X_i),1\}$ truncated to $[-2,2]$.
For binary outcomes, we retain the same distribution of $(X_i,Z_i,T_i)$ and generate $Y_i\sim\text{Bernoulli}(\widetilde p_i)$ where
\[
\log\frac{\widetilde p_i}{1-\widetilde p_i}
=-0.40+1.50T_i+g_Y(X_i)+Z_i(2T_i-1)b h(X_i).
\]
These specifications define the observed-data law rather than a full-data causal model; the coefficients $0.75$ and $1.50$ are not specified as population ATEs.

The parameter $a$ controls IV relevance, with conditional treatment odds ratio $e^a$, while $b$ controls the instrument--outcome association given treatment and covariates, which can arise through collider stratification under the IV assumptions.
When $b=0$, $\gamma_t(y,x)=1$ for both treatment arms, and the observed-data law is compatible with every $\Lambda\ge1$.
We consider $(a,b)\in\{(0.55,0.075),(1.10,0.075),(1.65,0.075),(1.10,0.15)\}$. 
The first three settings vary instrument relevance at fixed $b$; the fourth increases the conditional instrument--outcome association at $a=1.10$.
We use $\Lambda_0=\Lambda_1=\Lambda$ and evaluate estimation of the lower and upper ATE bounds and inference using outward confidence sets at $\Lambda\in\{1.8,2,2.5,3\}$. For the continuous-outcome setting $(a,b)=(1.10,0.15)$, the first three sensitivity values are
incompatible; only $\Lambda=3$ is included in identified-interval coverage comparisons.
All four values are compatible with the IV assumptions for the other settings.

Our one-step estimators use five-fold cross-fitting and a kernel-based learner for $\gamma_t$.
For continuous outcomes, thresholds are estimated by weighted check-loss optimization over a fixed additive natural-spline class, which also defines the restricted-class population target.
For binary outcomes, we use the sharp bound formulas without threshold estimation.
The main conventional-MSM comparisons adjust for $X$.
Appendix~\ref{subsec:supp_main_experiments} provides nuisance-estimation and tuning details, comparisons additionally adjusting for $Z$, and results using an alternative spline-based learner for $\gamma_t$ and, for continuous outcomes, a radial-basis function threshold class with its corresponding population target.
Coverage is evaluated against each method's population interval; comparisons of confidence-set width therefore reflect differences in population bounds as well as estimation uncertainty.

\begin{figure}[!htb]
\centering
\includegraphics[width=\linewidth]{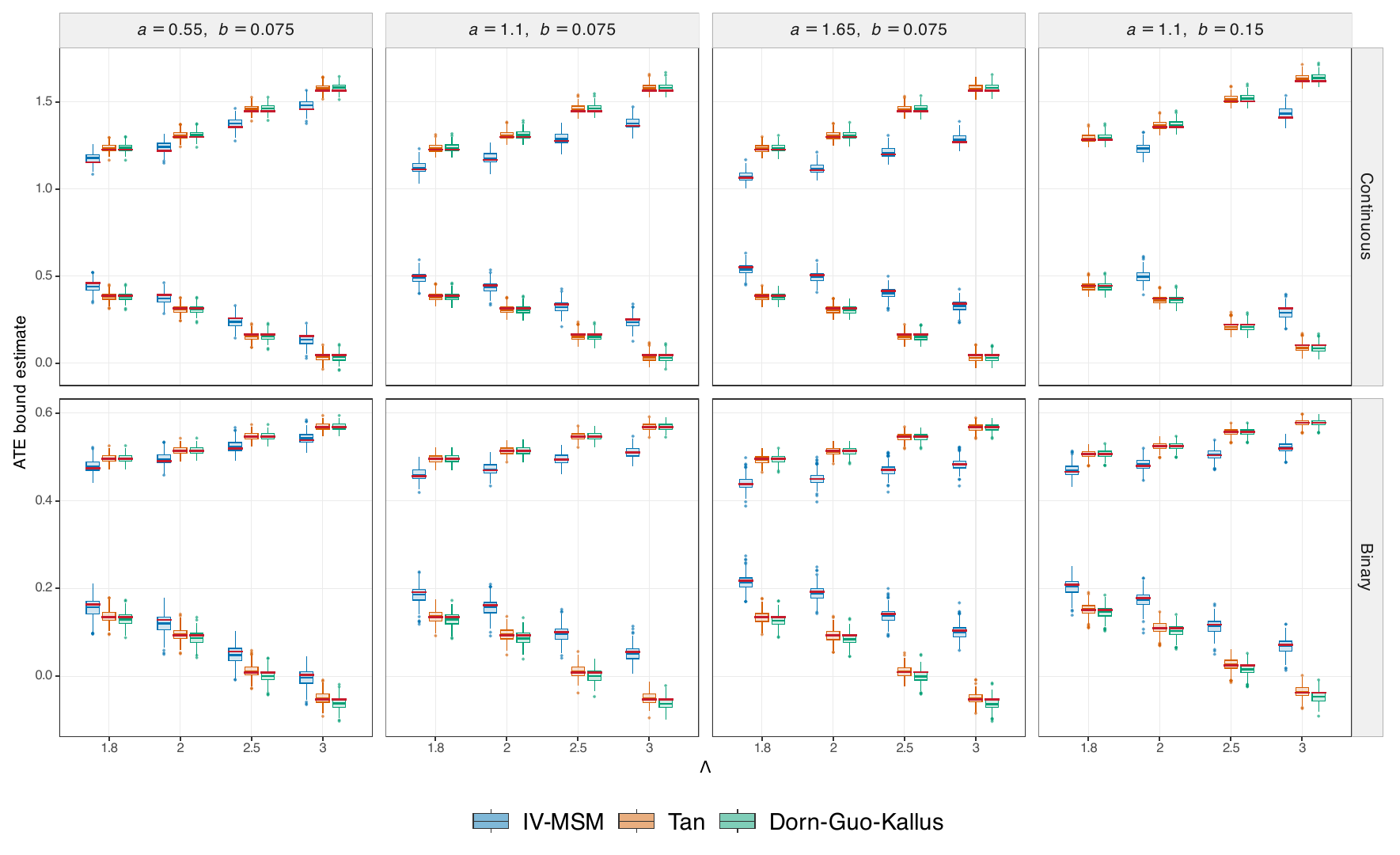}
\caption{ATE bound estimates for continuous (top) and binary (bottom)
outcomes, with $n=6000$ and 500 replications for each outcome
type and $(a,b)$ combination. Red segments mark method-specific
population bounds. Lower- and upper-bound estimates appear within each method's column. Continuous
$(a,b)=(1.10,0.15)$ includes diagnostic IV-MSM estimates at $\Lambda=2$
without a population reference.}
\label{fig:simulation_endpoints}
\end{figure}

\begin{figure}[!htb]
\centering
\includegraphics[width=\linewidth]{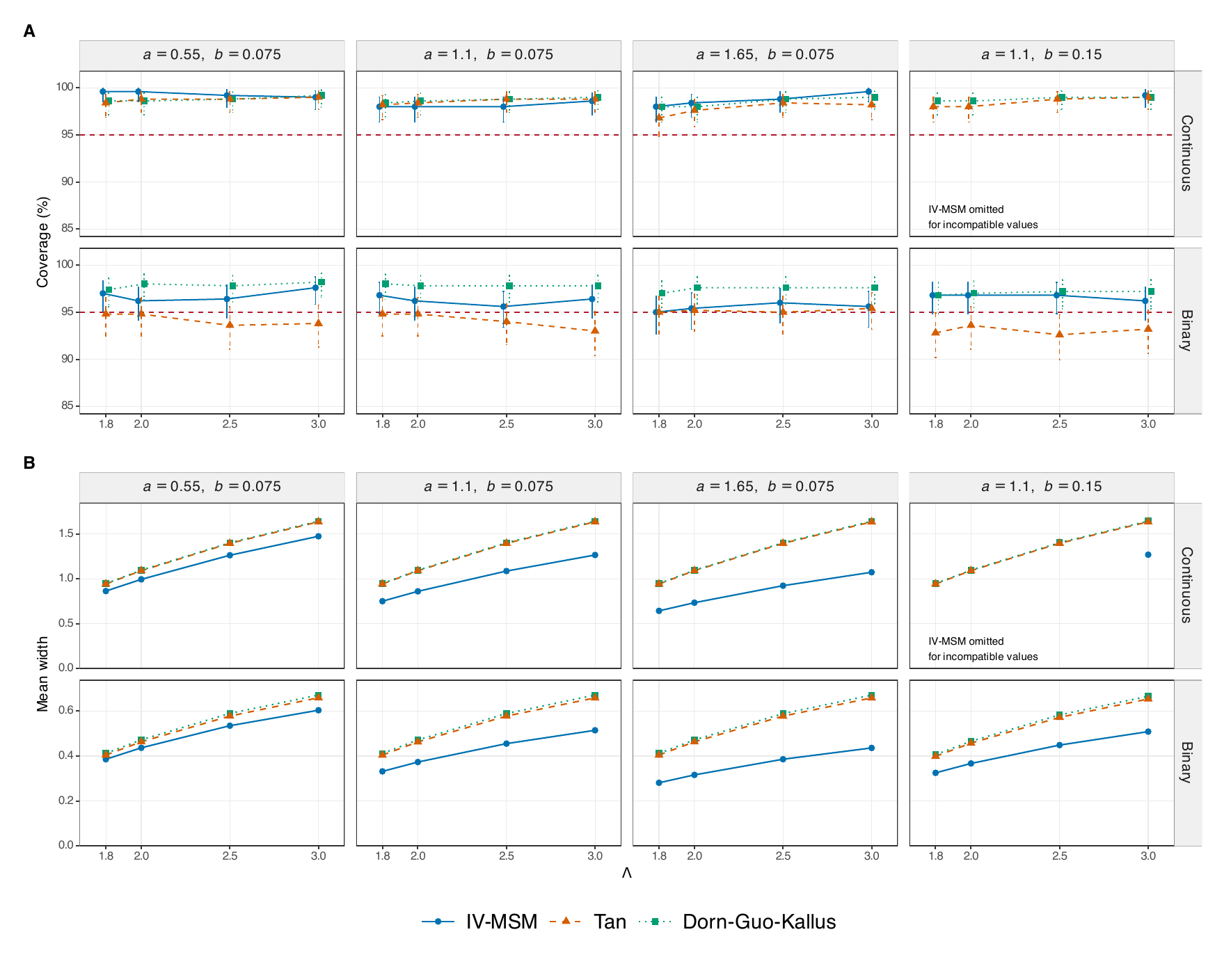}
\caption{Outward coverage (A) and mean width (B) of 95\% ATE confidence
sets, with $n=6000$ and 500 replications for each outcome
type and $(a,b)$ combination. Rows distinguish outcome
types and columns distinguish settings. Bars in A are exact 95\% binomial
Monte Carlo intervals; the dashed line marks 95\% coverage. Incompatible
IV-MSM cells are omitted from both panels.}
\label{fig:simulation_x5_outward}
\end{figure}

Figures~\ref{fig:simulation_endpoints} and~\ref{fig:simulation_x5_outward} summarize bound estimation, outward coverage, and mean confidence-set width.
At fixed $b$, increasing instrument relevance narrows the IV-MSM population intervals and confidence sets, while conventional-MSM intervals change comparatively little.
Larger $\Lambda$ yields wider population intervals.
Over compatible settings, IV-MSM outward coverage ranges from 98.0\% to 99.6\% for continuous outcomes and from 95.0\% to 97.6\% for binary outcomes. Tan's procedure exhibits below-nominal outward coverage
in some binary-outcome settings.
The IV-MSM produces shorter confidence sets in every matching compatible setting, with reductions relative to the conventional-MSM procedures of 8.4--34.6\% for continuous outcomes and 4.6--35.0\% for binary outcomes.
Comparisons additionally adjusting for $Z$ retain this width advantage (see Appendix~\ref{subsec:supp_main_experiments}).
In the fourth continuous-outcome setting, the IV-MSM estimates at the incompatible value $\Lambda=2$ form a narrower interval
than at larger $\Lambda$. However, we omit it from coverage and mean confidence-set width comparisons because no compatible population interval exists at this value.

Two-sided 95\% confidence intervals for the individual bounds exhibit undercoverage.
For the IV-MSM, lower- and upper-bound coverage ranges are 85.6--93.0\% and 84.0--93.6\%, respectively, for continuous outcomes, and 89.6--95.0\% and 89.8--94.8\% for binary outcomes.
Undercoverage also occurs for the conventional-MSM procedures.
Detailed results regarding coverage of two-sided confidence intervals are provided in Appendix~\ref{subsec:supp_main_experiments}.

\begin{figure}[!htb]
\centering
\includegraphics[width=\linewidth]
{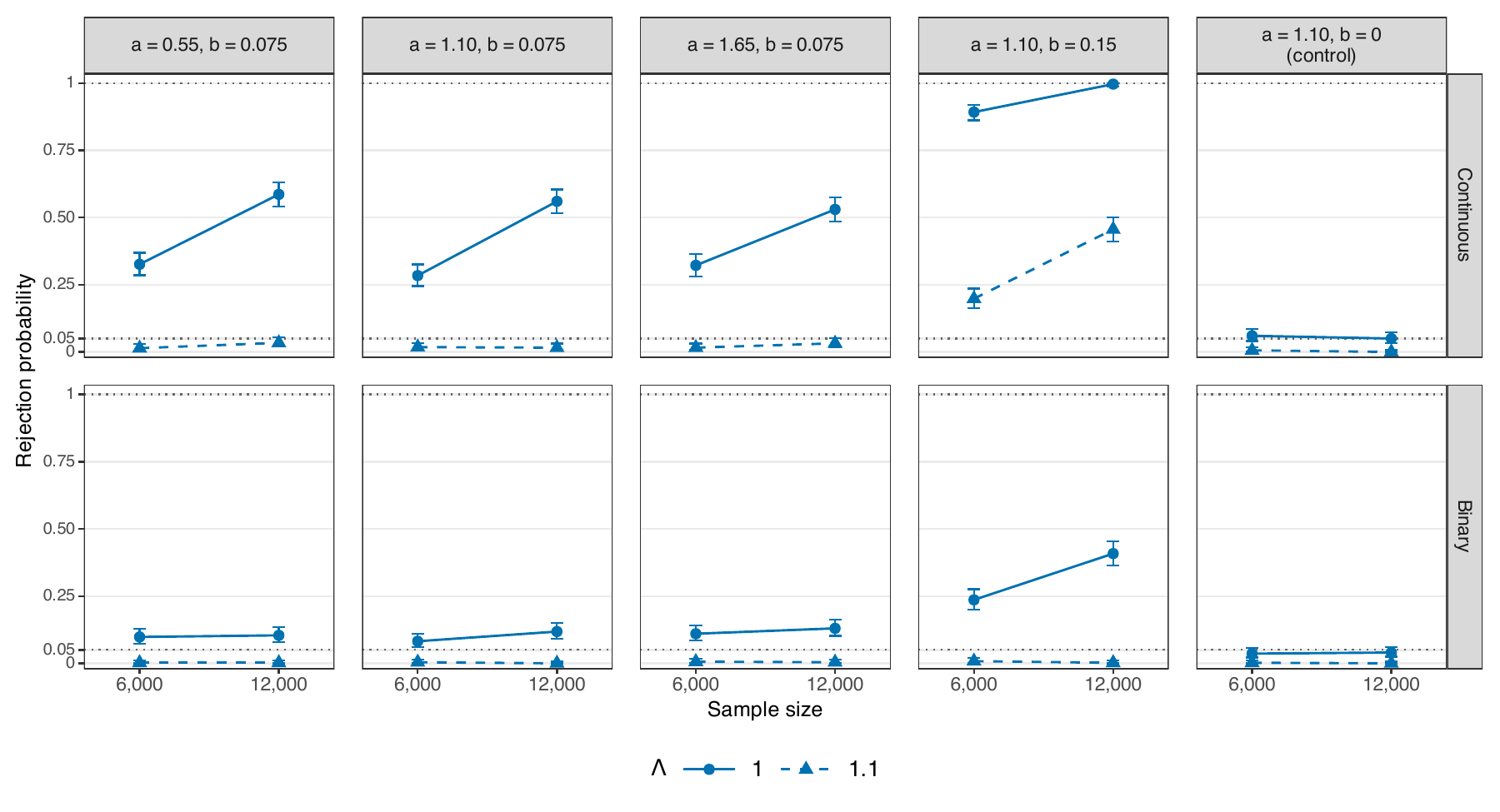}
\caption{Estimated compatibility-test rejection probabilities based on 500 replications for each outcome
type and $(a,b)$ combination. Columns identify $(a,b)$ settings and rows distinguish outcome types. Circles and solid lines indicate $\Lambda=1$; triangles and dashed lines indicate $\Lambda=1.1$.
Bars are exact 95\% binomial Monte Carlo intervals. Horizontal dotted lines mark 0.05 and 1. The $b=0$ setting is a compatible control.}
\label{fig:compatibility_test_results}
\end{figure}

For compatibility testing, we evaluate the level-$0.05$ pointwise maximum test of Section~\ref{sec:compatibility_test}, using five outer folds and 399 Gaussian multiplier draws.
Figure~\ref{fig:compatibility_test_results} compares rejection probabilities at $n=6000$ and $n=12000$ for $\Lambda\in\{1,1.1\}$.
Additional sensitivity values and witness-function details are provided in the supplement.
A separate $(a,b)=(1.10,0)$ control for each outcome type assesses null rejection; at $\Lambda=1$, its rejection probabilities
remain close to the nominal level.

Increasing the sample size improves power at $\Lambda=1$, particularly for continuous outcomes.
Increasing $a$ at fixed $b$ does not yield a monotone increase in rejection, whereas doubling $b$ substantially increases rejection for both outcome types.
At $\Lambda=1.1$, rejection remains low except in the fourth continuous-outcome setting, where the larger sample produces a substantial improvement.
Thus, instrument relevance alone does not determine the test's ability to detect violations, and power against some incompatible specifications remains limited even at $n=12000$.
In a targeted follow-up for the fourth binary-outcome setting at $\Lambda=1$, increasing the sample size to $n=24000$ raised rejection to $71.4\%$, compared with $23.6\%$ and $40.8\%$ at $n=6000$ and $n=12000$, respectively.
Nonrejection should not be interpreted as confirmation of compatibility.

Appendix~\ref{sec:supp_experiments} presents an expanded scalar-covariate experiment covering more combinations of instrument relevance and instrument--outcome association, together with experiments for smoothed bounds, finite ordered outcomes, and finite discrete instruments.
Code for the population calculations, simulations, and figures is available at \url{https://github.com/aolivasm/iv-msm-sa}.

\section{Application to the Oregon Health Insurance Experiment}
\label{sec:oregon_application}

In 2008, the state of Oregon drew names by lottery from a waiting list of uninsured, low-income adults, giving those selected an opportunity to apply for Oregon Health Plan Standard.
Selection did not guarantee enrollment, and individuals in either lottery group could obtain Medicaid through other programs \citep{finkelstein2012oregon,taubman2014medicaid}.
We use lottery selection, randomized conditional on household size, as the
instrument $Z$; coverage by any Medicaid program as treatment $T$; and any emergency-department visit as the binary outcome $Y$.
Our goal is to estimate the average effect of Medicaid coverage on the probability of an emergency-department visit during the study period (March 10, 2008---September 30, 2009). Because the lottery changed but did not determine Medicaid coverage, comparisons by coverage may remain confounded.

Randomization conditional on household size supports instrument
exchangeability. The exclusion restriction is plausible if lottery
selection affects emergency-department use only through Medicaid
coverage, as assumed in the original IV analysis
\citep{taubman2014medicaid}; it would fail if selection changed
care-seeking through other pathways.

Using the public-use administrative data for the Portland-area emergency-department study sample \citep{finkelstein2018replication}, we retain one adult per household and exclude 18 adults from three-person households lacking treated lottery-control observations, leaving 22,170 adults. We adjust for age, sex, prior emergency-department use, household
size, and lottery-draw indicators.

We assess compatibility for $\Lambda_0=\Lambda_1=\Lambda$ at $\Lambda\in\{1,1.1,1.2\}$ using the test in Section~\ref{sec:compatibility_test}.
The corresponding $p$-values are $0.030$, $0.236$, and $0.980$. At the 5\% level, the test rejects only $\Lambda=1$, corresponding to no unmeasured confounding, under the IV model.
For reference, under no unmeasured confounding, a separate augmented inverse-probability-weighted analysis adjusting for $X$ estimates an increase of $0.158$ in the probability of emergency-department use, with a 95\% confidence interval of $[0.141,0.175]$.

We construct 95\% simultaneous outward confidence bands over $\Lambda\in[1.2,2]$, using the IV-MSM estimator and the conventional-MSM estimators of Tan and Dorn--Guo--Kallus, adjusting for $X$ \citep{tan2024model,dorn2025doubly}. We calibrate the bands on a grid with spacing $0.01$ and extend them between grid points using monotonicity and the Lipschitz property described in Section~\ref{sec:binary_bounds}. Application-specific implementation details, smoothed IV-MSM results, and comparisons additionally adjusting for $Z$ are provided in Appendix~\ref{sec:supp_oregon_application}.

Figure~\ref{fig:oregon_ed_results} shows the outward bands and their widths for the IV-MSM and its competitors. At $\Lambda=1.60$, the IV-MSM band is $[0.00046,0.26976]$, whereas the Tan and Dorn--Guo--Kallus bands are $[-0.04852,0.32242]$ and $[-0.01979,0.32818]$, respectively. The IV-MSM band first includes zero just above $\Lambda=1.60$, compared with approximately $1.38$ for Tan and $1.49$ for Dorn--Guo--Kallus. Across $\Lambda\in[1.2,2]$, the IV-MSM band is approximately 24--41\% narrower than Tan's and 20--38\% narrower than Dorn--Guo--Kallus's. For comparison, the IV-only bound estimates are $[-0.2354,0.5128]$, with a 95\% outward confidence set of $[-0.2468,0.5242]$.

At $\Lambda=1.60$, the IV-MSM permits Medicaid-coverage odds
conditional on a potential outcome to range from $1/1.6$ to
$1.6$ times the reference odds given the observed covariates
and lottery status.
To interpret this sensitivity value on the outcome scale,
we compare the estimated potential-outcome mean bounds under
the joint IV and sensitivity restrictions with the standardized means under no unmeasured confounding after adjustment for the observed covariates and lottery status.
Under the latter assumption, the estimated proportion of adults who would have an emergency-department visit is 47.66\% if everyone had Medicaid coverage and 31.98\% if no one had coverage.
The estimated IV-MSM bounds permit departures from these
benchmarks of up to 10.64 percentage points downward or
7.28 points upward under Medicaid coverage, and 1.46 points downward or 2.63 points upward without coverage.
These are estimated identification ranges, not confidence
intervals or estimates of the confounding actually present;
their scientific plausibility requires substantive judgment.

These findings complement the original IV analysis, which
estimated the local average treatment effect of Medicaid
coverage among individuals whose enrollment was induced by
the lottery \citep{taubman2014medicaid}.
This interpretation relies on the IV assumptions and treatment monotonicity.
Our analysis instead bounds the ATE in the population
represented by the study sample, rather than only among individuals whose Medicaid enrollment was induced by the lottery.
It does not require treatment monotonicity, but retains our IV assumptions and imposes explicit sensitivity restrictions on the degree of unmeasured confounding.
Under these assumptions, the simultaneous IV-MSM band excludes zero until slightly above $\Lambda=1.60$, whereas the IV-only confidence set includes zero.
The comparisons with the conventional MSM adjusting for
$(X,Z)$ provided in Appendix~\ref{sec:supp_oregon_application} isolate the additional information supplied by the IV assumptions.
The tipping point marks where the confidence band ceases
to exclude zero, not where a null effect is established.

\begin{figure}[!htb]
\centering
\includegraphics[width=\linewidth]{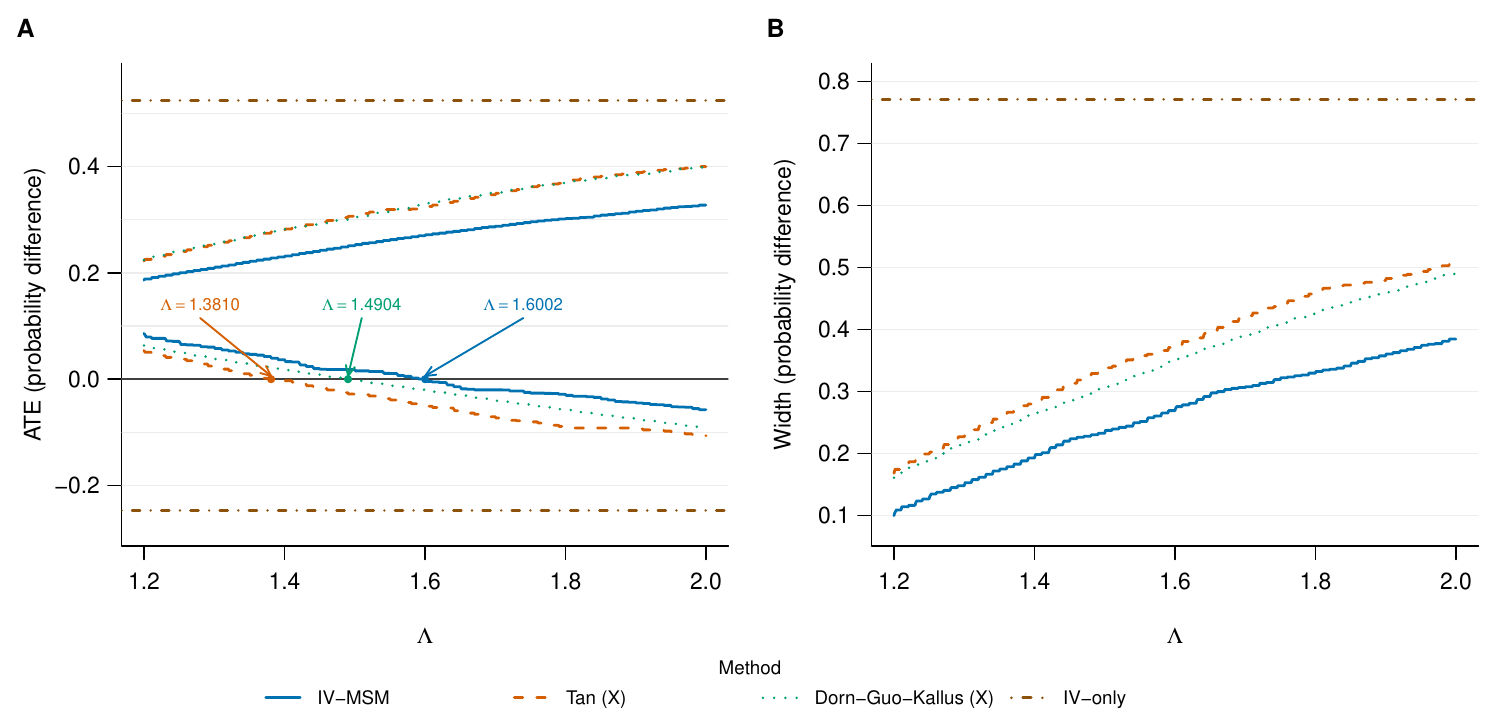}
\caption{Sensitivity analysis for the ATE of Medicaid coverage on
emergency-department use in the Oregon Health Insurance Experiment.
Panel~(a) shows 95\% simultaneous outward confidence bands over
$\Lambda\in[1.2,2]$, and panel~(b) shows their widths.
The conventional-MSM estimators adjust for $X$.
The horizontal reference lines show the IV-only 95\% outward
confidence set and its width.}
\label{fig:oregon_ed_results}
\end{figure}

\section{Discussion}\label{sec:discussion}

In this work, we developed the IV-MSM, a framework that combines instrumental-variable assumptions with interpretable restrictions on unmeasured confounding to study population treatment effects.
It allows an instrument to strengthen sensitivity analysis without changing the target population or imposing treatment monotonicity or treatment-effect homogeneity.
We characterized the sharp identified sets for the potential-outcome means and the ATE and developed cross-fitted one-step estimators that accommodate flexible, nonparametric estimation of the nuisance functions under suitable convergence-rate conditions.
We established asymptotically valid, potentially conservative outward confidence sets and, under stronger conditions, regular, efficient estimation and inference for the restricted-class bounds.

A closely related sensitivity model was proposed by \cite{rosenbaum1987sensitivity}.
At fixed observed covariates, it bounds pairwise treatment-odds ratios across latent-confounder values, whereas the MSM bounds treatment odds conditional on a potential outcome relative to the observed treatment odds.
For each potential outcome, the MSM bounds pairwise treatment-odds ratios across its values by $\Lambda^2$ at fixed observed covariates.
The converse need not hold, so this implication does not make the two models equivalent \citep{tan2006distributional,zhao2019sensitivity}. Appendix~\ref{sec:rosenbaum_model} provides a more detailed comparison.
\citet{yadlowsky2022bounds} develop population bounds and semiparametric estimation under a Rosenbaum-type restriction.
Extending the IV-MSM framework to such restrictions is a natural direction for future work.

Our estimation results build on methods for the conventional MSM \citep{dorn2023sharp,tan2024model,dorn2025doubly}.
For continuous outcomes, prespecified threshold classes provide tractable outer bounds when the weighted conditional quantiles defining the sharp bounds are difficult to estimate.
Under the stated regularity conditions, outward inference does not require the learned thresholds to attain the population optimum or the optimized bounds to be pathwise differentiable.
The numerical experiments illustrate this distinction: valid outward coverage can coexist with substantially lower coverage of two-sided confidence intervals for the individual bounds.

\citet{zhang2026average} point out that the MSM can yield conservative bounds because its sensitivity parameter must accommodate the largest treatment-selection departure, even if that departure occurs only in a low-probability region.
They propose an average-case alternative that constrains an integrated second moment of a treatment-selection weight and illustrate its potential to yield narrower effect bounds. Appendix~\ref{sec:supp_sensitivity_model_comparison} provides a contrasting example in which the MSM yields narrower population bounds for the same observed distribution under the stated calibration.
Together, these examples show that relative informativeness depends on the distribution and parameter calibration.
The choice between these models should reflect which features of treatment selection investigators are prepared to constrain, rather than which model produces narrower bounds in a particular example.

Other approaches parameterize unmeasured confounding through the treatment and outcome associations of a hypothetical confounder \citep{imbens2003sensitivity}, bias factors or omitted-variable associations \citep{ding2016sensitivity,cinelli2020making}, departures from conditional independence \citep{masten2018identification}, or selection functions relating observed and missing potential-outcome distributions \citep{scharfstein1999adjusting,robins2000sensitivity,franks2020flexible}. Appendix~\ref{sec:other_sa_models} discusses these approaches and their relationship to the IV-MSM.

Our compatibility test connects to tests of observable IV restrictions and instrument-based assessments of unconfoundedness \citep{bonet2001instrumentality,kitagawa2015test,guo2014using,
black2022simple}.
Unlike sensitivity analyses for instrument invalidity, we maintain exclusion and instrument exchangeability \citep{small2007sensitivity,conley2012plausibly}.
Stronger joint IV assumptions may further tighten identification, as in the Balke--Pearl framework and its covariate-adjusted extensions \citep{balke1997bounds,levis2025covariate}; Appendix~\ref{sec:lit_iv_assumptions_test} discusses how
their IV assumptions differ from ours.
For binary outcomes, Appendix~\ref{subsec:supp_binary_joint_exchangeability}
characterizes the sharp bounds under conditional joint
exchangeability of $(Y^0,Y^1)$ with $Z$ given $X$.
This strengthening leaves the sharp bounds unchanged whenever $1\le\Lambda_0,\Lambda_1\le3$, but can tighten them outside this range.

We implement the test by adapting existing methods to our compatibility restrictions \citep{andrews2013inference}. 
Although asymptotically valid under the stated conditions, it had limited finite-sample power, particularly against incompatible sensitivity values above one.
In the Oregon Health Insurance Experiment application, only $\Lambda=1$ was rejected; failure to reject larger values does not establish compatibility.

Limited overlap and nuisance estimation remain practical challenges, while restricted threshold classes and smoothing can widen the bounds.
Compatibility also does not determine which sensitivity values are scientifically plausible.
Reporting results across a range of values, with simultaneous bands where available, shows how the conclusions depend on the assumed degree of unmeasured confounding.
When varying a common sensitivity parameter, the tipping point at which the outward confidence band first includes the null effect summarizes the robustness of the inferential conclusion.
In the Oregon Health Insurance Experiment application, the IV-MSM supported a positive effect over a wider range of sensitivity values than the conventional-MSM analyses adjusting for the same covariates and lottery status.

Appendix~\ref{sec:supp_finite_extensions} extends the framework to finite ordered outcomes and finite discrete instruments.
The potential-outcome mean bounds also yield sharp bounds for the average treatment effects among the treated and untreated and, for binary outcomes, marginal causal risk ratios and odds ratios, provided the relevant denominators are nonzero.
Future work could extend the framework to continuous instruments, longitudinal treatments, and censored time-to-event outcomes, and develop more powerful compatibility tests.

\appendix
\section*{Appendix}

\input{supp_discussion}

\input{supp_inference}

\input{supp_extensions}

\input{supp_experiments}

\input{supp_application}

\input{supp_main_proofs}

\input{supp_additional_proofs}

\bibliographystyle{apalike}
\bibliography{references}

\end{document}

%% file: supp_discussion.tex
\section{Comparison with related sensitivity frameworks}
\label{sec:supp_related_frameworks}

Sensitivity analyses differ in the features of treatment selection and outcome variation they constrain.
We compare the IV-MSM with related approaches, distinguishing restrictions on the full-data law from methods used to estimate the resulting bounds.

\subsection{Rosenbaum-type sensitivity models}\label{sec:rosenbaum_model}

We formalize the comparison between Rosenbaum-type restrictions and the conventional MSM through their normalized density-ratio representations \citep{tan2006distributional,zhao2019sensitivity}.
Neither model requires an instrument.

Consider the potential outcome $Y^1$ and let $\mathcal C_P$ be the class of positive measurable functions $\lambda$ satisfying $\E_P\{\lambda(Y,X)\mid T=1,X\}=1$.
Such functions represent candidate density ratios of $Y^1\mid T=0,X$ relative to $Y^1\mid T=1,X$.
For this potential outcome, the MSM and Rosenbaum-type classes are
\[
\begin{aligned}
\mathcal C_{\mathrm{MSM}}(\Lambda)=\{\lambda\in\mathcal C_P:
       \Lambda^{-1}\le\lambda\le\Lambda\}\quad\text{and}\quad
\mathcal C_{\mathrm R}(\Gamma)=\left\{\lambda\in\mathcal C_P:
  \frac{\operatorname*{ess\,sup}_y\lambda(y,X)}
       {\operatorname*{ess\,inf}_y\lambda(y,X)}
  \le\Gamma\right\}.
\end{aligned}
\]
The essential extrema and pointwise inequalities are understood under $P(Y\mid T=1,X=x)$ for $P_X$-almost every $x$.
By Bayes' rule, pairwise treatment-odds ratios conditional on $Y^1$ are ratios of the corresponding density-ratio values in reverse order.
Thus, $\mathcal C_{\mathrm R}(\Gamma)$ bounds their pairwise variation, whereas $\mathcal C_{\mathrm{MSM}}(\Lambda)$ fixes their permitted range relative to one.

The following nesting relationship is established by \citet{zhao2019sensitivity}; we express it here in density-ratio notation:
\[
\mathcal C_{\mathrm R}(\Lambda)
\subseteq
\mathcal C_{\mathrm{MSM}}(\Lambda)
\subseteq
\mathcal C_{\mathrm R}(\Lambda^2).
\]
Normalization implies that the essential infimum of $\lambda$ cannot exceed one and its essential supremum cannot be below one.
Together with the pairwise restriction, this yields the first inclusion.
The second follows because the largest ratio between two values in $[\Lambda^{-1},\Lambda]$ is $\Lambda^2$.
Thus, $\Lambda^2$ is an implied pairwise bound, not an exact conversion between the sensitivity models.

An IV-restricted Rosenbaum-type model could impose the pairwise density-ratio restriction within each instrument stratum and require compatibility with the IV assumptions. The affine identity in Proposition~\ref{prop:comp_msm_iv} would still apply, since it follows from the IV restrictions rather than the MSM.
However, the optimization would need to enforce the pairwise restriction itself, rather than only its implied pointwise bounds.
Deriving sharp bounds and developing inference for this extension remain topics for future work.

\subsection{Average-case restrictions and a contrasting example} \label{sec:supp_sensitivity_model_comparison}

The average-case model of \citet{zhang2026average} and the MSM both restrict how treatment assignment may depend on potential outcomes given observed covariates.
Whereas the MSM imposes a pointwise bound on the treatment-odds ratio, the average-case model bounds an integrated second moment of a treatment-selection weight.
To describe it, consider $Y^1$ without an instrument and write $\pi_P(x)=P(T=1\mid X=x)$.
Define
\[
h_Q(y,x)
=\frac{\pi_P(x)}{Q(T=1\mid Y^1=y,X=x)}
=\pi_P(x)+\{1-\pi_P(x)\}\lambda_Q^1(y,x),
\]
where $\lambda_Q^1$ is the density ratio of $Y^1\mid T=0,X$ relative to $Y^1\mid T=1,X$.
By Bayes' rule, $h_Q$ reweights the observed treated-outcome distribution to the population distribution of $Y^1$ given $X$.
Consequently,
\[
\E_Q(Y^1)
=\E_P\!\left[
  \E_P\{Yh_Q(Y,X)\mid T=1,X\}
\right].
\]
In addition to $h_Q\ge\pi_P$ and the conditional normalization $\E_P\{h_Q(Y,X)\mid T=1,X\}=1$, the average-case model imposes
\[
\E_P\!\left[
  \E_P\{h_Q(Y,X)^2\mid T=1,X\}
\right]\le\Sigma,
\qquad \Sigma\ge1.
\]
The outer expectation averages over the population distribution of $X$.
By normalization, this restriction equivalently bounds the corresponding average of $(h_Q-1)^2$ by $\Sigma-1$.
Thus, $\Sigma=1$ forces $h_Q=1$, whereas larger values permit departures whose squared magnitude is controlled on average.
Unlike the MSM, this restriction does not impose a fixed pointwise upper bound on the selection weight.

The authors illustrate the potential advantage of this restriction when large selection departures occur in low-probability regions.
We give a contrasting example, applying the model separately to each potential outcome by relabeling treatment.
Both models are calibrated to the smallest sensitivity parameter containing the same specified full-data law.

There are no covariates or instrument.
For $c\in[0,3/4]$, put $\Lambda_c=1+c$, $q_c=\Lambda_c/(1+\Lambda_c)$, and define the density on $[0,1]$ by
\[
r_c(y)=
\begin{cases}
\Lambda_c^{-1},&0\le y<q_c,\\
\Lambda_c,&q_c\le y\le1.
\end{cases}
\]
Its integral is one.
Let $T\sim\operatorname{Bernoulli}(1/2)$, set both observed laws $Y^t\mid T=t$ to $\operatorname{Unif}[0,1]$, and give the missing laws $Y^1\mid T=0$ and $Y^0\mid T=1$ densities $r_c(y)$ and $r_c(1-y)$, respectively.
One admissible joint distribution is obtained by taking $Y^0$ and $Y^1$ to be independent conditional on $T$; other dependence structures preserving the specified conditional distributions are also possible.
With consistency, the observed distribution is the same for every $c$:
\[
P(T=1)=\tfrac12,\qquad
Y\mid T=t\sim\operatorname{Unif}[0,1].
\]

In this example, the selection weights reduce to $h_t(y)=\{1+\lambda^t(y)\}/2$, where $\lambda^t$ is the missing-to-observed density ratio for $Y^t$.
The average-case restrictions become $h_t\ge1/2$, $\int_0^1h_t(y)\,dy=1$, and $\int_0^1h_t(y)^2\,dy\le\Sigma$, whereas the MSM imposes $\Lambda^{-1}\le\lambda^t\le\Lambda$.
Under each model, we use the same sensitivity parameter separately for both potential outcomes; the second-moment budget is not shared between treatment arms.

For the specified full-data law, the smallest admissible parameters are
\[
\Lambda_c=1+c,\qquad
\Sigma_c=1+\frac{c^2}{4(1+c)}.
\]
The sharp population ATE intervals are
\[
\begin{aligned}
\mathcal I_{\mathrm{MSM}}(c)
=\left[-\frac{c}{2(2+c)},\,
          \frac{c}{2(2+c)}\right]\qquad\text{and}\qquad
\mathcal I_{\mathrm{avg}}(c)
=\left[-\frac{c}{\sqrt{12(1+c)}},\,
          \frac{c}{\sqrt{12(1+c)}}\right].
\end{aligned}
\]
For the MSM, placing the larger density ratio on the largest outcomes maximizes each mean; reflection minimizes it.
For the average-case model, Cauchy--Schwarz gives
\[
\left|\int_0^1 yh(y)\,dy-\tfrac12\right|
\le \sqrt{\frac{\Sigma_c-1}{12}}.
\]
Equality is attained by $h_\pm(y)=1\pm c\sqrt{3/(1+c)}(y-1/2)$.
For the stated range of $c$, these weights satisfy $h_\pm\ge1/2$, so they define valid missing-outcome densities $2h_\pm-1$.
The opposite mean extremes are jointly attainable by choosing the two missing distributions separately and coupling them with the observed outcomes within treatment groups.

For every $c>0$, the ratio of interval widths is
\[
\frac{\operatorname{width}\{\mathcal I_{\mathrm{avg}}(c)\}}
     {\operatorname{width}\{\mathcal I_{\mathrm{MSM}}(c)\}}
=\frac{2+c}{\sqrt{3(1+c)}}>1.
\]
For example, $c=1/2$ gives $\Lambda_c=1.5$ and $\Sigma_c=25/24$, with intervals $[-0.1000,0.1000]$ and $[-0.1179,0.1179]$, respectively.

This ordering has a direct explanation. The MSM restriction and normalization imply
\[
\int_0^1h(y)^2\,dy
\le 1+\frac{(\Lambda-1)^2}{4\Lambda}.
\]
Indeed, this follows by integrating $\{h-(1+\Lambda^{-1})/2\}\{(1+\Lambda)/2-h\}\ge0$.
Our constructed density ratios attain this second-moment ceiling.
At the resulting calibration, the average-case model contains all MSM-admissible weights and additionally permits weights outside their pointwise range.

The example does not establish a general advantage of the MSM.
It shows that average-case calibration need not produce narrower bounds even when both parameters are calibrated to the same
full-data law.
These are oracle calibrations: the common observed distribution also admits the $c=0$ completion, so its smallest compatible parameters are $\Lambda=\Sigma=1$.
Relative informativeness depends on the full-data configuration and calibration, whereas the scientific choice concerns which features of selection investigators are prepared to constrain.

\subsection{Selection functions and other confounding parameterizations}\label{sec:other_sa_models}

Selection-function approaches specify how treatment or observation depends on an unobserved outcome \citep{scharfstein1999adjusting,robins2000sensitivity,franks2020flexible}.
Without an instrument, consider
\[
\operatorname{logit}Q(T=1\mid Y^1=y,X=x)
=\alpha(x)+s(y,x),
\]
where $s$ is a prespecified sensitivity function.
Bayes' rule and conditional normalization give
\[
\lambda_Q^1(y,x)
=
\frac{\exp\{-s(y,x)\}}
{\E_P[\exp\{-s(Y,x)\}\mid T=1,X=x]},
\]
provided the denominator is finite and positive.
For a fixed $s$, normalization determines $\alpha(x)$, and the density ratio determines the missing distribution of $Y^1\mid T=0,X$ from the observed distribution of $Y\mid T=1,X$.

For example, \citet{franks2020flexible} consider logistic selection models linear in specified functions of the potential outcome, including linear and quadratic terms, together with flexible observed-outcome models based on mixtures of exponential families.
A choice such as $s(y,x)=\eta y$ produces an exponential tilt of the observed outcome distribution, while richer specifications can allow nonlinear selection and outcome--covariate interactions.

The MSM instead bounds the missing-to-observed density ratio without specifying its shape.
This distinction is useful when investigators can constrain the magnitude of selection but cannot credibly specify its functional form.
For any selection function whose implied density ratio satisfies the MSM restriction, the resulting missing distribution belongs to the MSM's admissible class.
The approaches are not generally nested, however, because a selection-function specification need not satisfy a given MSM bound.
The IV-MSM additionally requires agreement of the completed potential-outcome distributions across instrument strata, conditional on observed covariates.

Other sensitivity analyses introduce a hypothetical confounder and vary its associations with treatment and outcome, with calibration informed by observed covariates \citep{imbens2003sensitivity}.
Bias-factor approaches constrain associations on risk-ratio scales \citep{ding2016sensitivity}, while omitted-variable approaches express sensitivity through partial associations with treatment and outcome \citep{cinelli2020making}.
\citet{masten2018identification} instead relax conditional independence by bounding departures of treatment probabilities conditional on potential outcomes from their observed values.
These parameters constrain different quantities and cannot generally be translated into a common numerical sensitivity scale.

The enhanced MSM of \citet{zhang2025enhanced} supplements the MSM with restrictions on conditional outcome-mean variation across latent-confounder values, yielding sharp bounds that can be narrower than the conventional MSM bounds.
The additional parameters allow investigators to explore assumptions about outcome-mean variation, potentially informed by substantive knowledge, but require justification alongside the treatment-selection restriction.
The IV-MSM framework could also be extended to incorporate these restrictions, combining outcome-mean constraints with the information supplied by an instrument.

\subsection{Conventional-MSM estimation and restricted threshold classes}

Our estimation approach builds on sharp identification and inference under the conventional MSM \citep{dorn2023sharp,tan2024model,dorn2025doubly}.
A useful connection is Tan's finite-moment relaxation of conditional density-ratio normalization. Retaining a finite collection of balance conditions in place of the full conditional restriction leads to a corresponding finite-dimensional dual threshold class \citep{tan2024model}.
Our restricted-class construction similarly makes the optimization tractable, but its objective retains the IV-implied envelopes.

Restricting the threshold class changes the population target: the resulting interval contains the unrestricted sharp interval.
Outward inference and two-sided inference for the individual bounds have different requirements.
Under the conditions established in the paper, outward confidence limits remain valid without optimal threshold learning, whereas two-sided inference for the restricted-class bounds requires the within-class optimization error to be sufficiently small.
These distinctions are related to conservative validity under imperfect quantile estimation in conventional-MSM procedures.

With the same stratum-specific sensitivity restrictions, the IV-MSM is contained in the conventional MSM conditioning on $(X,Z)$ because it additionally requires $Y^t\indep Z\mid X$.
No analogous nesting follows for the conventional MSM conditioning only on $X$, which uses a different observed propensity score and selection restriction.
Nevertheless, conventional-MSM analyses adjusting for $X$ and $(X,Z)$ often gave similar results in our numerical experiments and data application.

\subsection{IV assumptions and compatibility testing}\label{sec:lit_iv_assumptions_test}

Compatibility testing asks whether the observed distribution admits a full-data law satisfying the proposed restrictions.
This question underlies instrumentality inequalities and tests of IV models \citep{bonet2001instrumentality,kitagawa2015test}.
Our test assesses the joint restrictions imposed by the IV assumptions and a candidate sensitivity-parameter pair.
Under the IV model, rejection rules out the proposed
sensitivity-parameter pair.

Related procedures use instruments to assess unconfoundedness or treatment selection \citep{guo2014using,donald2014testing,deluna2014testing,black2022simple}.
Their null hypotheses and maintained assumptions differ, including restrictions involving compliance groups or one-sided noncompliance.
In our framework, $(\Lambda_0,\Lambda_1)=(1,1)$ imposes $Y^t\indep T\mid X,Z$ for both potential outcomes, while larger values permit specified departures.
Rejection refutes the joint candidate model; it identifies a failure of the sensitivity specification only if the IV assumptions are maintained.
Nonrejection does not establish compatibility or determine which sensitivity values are scientifically plausible.

The distinction between marginal and joint IV assumptions is also important.
Our sharpness construction sets $Y^{t,0}=Y^{t,1}=Y^t$, so strengthening distributional exclusion to individual-level exclusion alone does not tighten the bounds.
The construction preserves instrument exchangeability separately for each potential outcome but does not establish joint exchangeability of $(Y^0,Y^1)$ with $Z$ given $X$.
Such joint restrictions connect to the assumptions underlying
Balke--Pearl bounds and their covariate-adjusted extensions
\citep{balke1997bounds,levis2025covariate}.
Appendix~\ref{subsec:supp_binary_joint_exchangeability}
provides identification results under this strengthening for
binary outcomes.
Finally, the IV-MSM maintains exclusion and instrument exchangeability rather than assessing sensitivity to their failure \citep{small2007sensitivity,conley2012plausibly}.

%% file: supp_inference.tex
\section{Additional Regularity Results and Inference}
\label{sec:supp_extensions_regularity}

This section provides additional inferential procedures and regularity results supporting the estimation and inference results in the main text.
Sections~\ref{sec:supp_smoothed_bounds} and~\ref{sec:supp_compatibility_test} develop smoothed-bound estimation and the compatibility-test construction, respectively; both also apply to binary outcomes.
Section~\ref{sec:supp_binary_bands} develops simultaneous sensitivity bands for binary outcomes.
Finally, Section~\ref{subsec:supp_optimizer_regularity} provides sufficient conditions for the first-order optimizer condition in Theorem~\ref{thm:eif_fixed_q}, and Section~\ref{subsubsec:supp_estimation_assumptions} provides sufficient nuisance-rate conditions for root-$n$-negligible drift.

\subsection{Smoothed restricted-class bounds and estimation}
\label{sec:supp_smoothed_bounds}

A fixed smooth outer approximation removes the discontinuous selection of the terms attaining the envelope maxima and minima.
For a prespecified $\varepsilon\in(0,\Lambda_0^{-1})$, put $h_\varepsilon=\varepsilon/\log2$ and define
\begin{equation}
\begin{aligned}
L_{P,\varepsilon}
&=h_\varepsilon\log\!\left[
\{\exp(\Lambda_0^{-1}/h_\varepsilon)
+\exp(\Gamma_{\Lambda_1^{-1},P}/h_\varepsilon)\}/2\right],\\
U_{P,\varepsilon}
&=-h_\varepsilon\log\!\left[
\{\exp(-\Lambda_0/h_\varepsilon)
+\exp(-\Gamma_{\Lambda_1,P}/h_\varepsilon)\}/2\right].
\end{aligned}
\label{eq:smoothed_envelopes}
\end{equation}
By the log-sum-exp inequalities and compatibility, $L_P-\varepsilon\le L_{P,\varepsilon}\le L_P\le U_P\le U_{P,\varepsilon}\le U_P+\varepsilon$.

Replace $(L_P,U_P)$ by these envelopes in the fixed-threshold functional~\eqref{eq:fixed_threshold_objectives} to obtain $\nu_{P,\varepsilon}^{1b}(q)$.
Optimize over the same specified class to define $\widetilde\mu_\varepsilon^{1-}(P)=\sup_{q\in\widetilde{\mc Q}}\nu_{P,\varepsilon}^{1-}(q)$ and $\widetilde\mu_\varepsilon^{1+}(P)=\inf_{q\in\widetilde{\mc Q}}\nu_{P,\varepsilon}^{1+}(q)$.
For $q^-,q^+\in\widetilde{\mc Q}$,
\begin{equation}
\begin{aligned}
\nu_{P,\varepsilon}^{1-}(q^-)
\le\widetilde\mu_\varepsilon^{1-}(P)
\le\widetilde\mu^{1-}(P),\qquad
\widetilde\mu^{1+}(P)
\le\widetilde\mu_\varepsilon^{1+}(P)
\le\nu_{P,\varepsilon}^{1+}(q^+).
\end{aligned}
\label{eq:fixed_q_outer_bounds}
\end{equation}
Smoothing and threshold restriction therefore define distinct outward relaxations.

With $\operatorname{expit}(u)=\{1+\exp(-u)\}^{-1}$, replace the exact envelope derivative indicators by
\[
\begin{aligned}
\omega_{L,P,\varepsilon}
=\operatorname{expit}\!\left\{
\frac{\Gamma_{\Lambda_1^{-1},P}-\Lambda_0^{-1}}{h_\varepsilon}
\right\},\qquad
\omega_{U,P,\varepsilon}
=\operatorname{expit}\!\left\{
\frac{\Lambda_0-\Gamma_{\Lambda_1,P}}{h_\varepsilon}
\right\}.
\end{aligned}
\]
Making these substitutions in $\lambda,\theta,M,D,H,C$ and equation~\eqref{eq:eif_fixed_q} gives $\phi_{P,q,\varepsilon}^{1b}$.
Under the positivity and boundedness conditions of Theorem~\ref{thm:eif_fixed_q}\textup{(i)}, this is the efficient influence function of the smoothed fixed-threshold functional even at envelope ties.
For the restricted optimum, Theorem~\ref{thm:eif_fixed_q}\textup{(ii)} applies to its smoothed within-class optimizer set. The sufficient conditions in Section~\ref{subsec:supp_optimizer_regularity} likewise apply after
replacing the fixed-threshold functionals, gradients, optimized values, and optimizer sets by their $\varepsilon$ counterparts.
Smoothing does not eliminate distinct optimizer gradients.

Estimation and inference follow Section~\ref{sec:estimation}, replacing the envelopes, objectives, and influence functions by their smoothed counterparts.
For fixed $\varepsilon>0$, the envelope no-tie and margin conditions are unnecessary; the remaining regularity conditions still apply.
Sufficient conditions for negligible nuisance drift, including their dependence on $\varepsilon$, are discussed in Section~\ref{subsubsec:supp_estimation_assumptions}.
Allowing $\varepsilon=\varepsilon_n\to0$ requires a separate asymptotic analysis and is not considered here.
Under the corresponding inference conditions, the resulting outward confidence set covers the smoothed restricted-class interval and, by its outward ordering, the exact restricted-class and sharp intervals.

\subsection{Compatibility-test details}
\label{sec:supp_compatibility_test}

The test uses observed conditional moment inequalities and does not involve a threshold function $q$ or optimization over a threshold class.
It therefore applies to continuous, binary, and other finite-valued outcomes, under the test-specific regularity conditions below.
For a prespecified candidate pair, consider $H_0(\Lambda_0,\Lambda_1):P\in\mc M(\Lambda_0,\Lambda_1)$.
The test uses the exact envelopes because this null concerns the original IV-MSM, rather than its smoothed outer approximation.
For $t\in\{0,1\}$, write $\pi_{z,P}^t(x)=P(T=t\mid X=x,Z=z)$ and let $\underline\gamma_{t,P}(x)$ and $\overline\gamma_{t,P}(x)$ denote the lower and upper bounds on $\gamma_{t,P}$ in Corollary~\ref{coro:falsification_dual}, applied to treatment arm $t$.

For a measurable allocation $r:\mc Y\times\mc X\to[0,1]$, let $m_{z,t,P}^r(x)=\E_P\{r(Y,X)\mid T=t,X=x,Z=z\}$.
Define
\[
c_{t,P}^{\ell}
=\frac{\pi_{1,P}^t+(1-\pi_{1,P}^t)/\Lambda_1}{1-\pi_{0,P}^t},
\qquad
c_{t,P}^{u}
=\frac{\pi_{1,P}^t+(1-\pi_{1,P}^t)\Lambda_1}{1-\pi_{0,P}^t},
\qquad
b_{t,P}=-\frac{\pi_{0,P}^t}{1-\pi_{0,P}^t}.
\]
These density-ratio bounds and the two conditional-normalization restrictions imply the four conditional moments
\begin{align}
\theta_{t,P}^{\mathrm{pw},-}(X;r)
&=\underline\gamma_{t,P}m_{0,t,P}^r-m_{1,t,P}^r,\notag\\
\theta_{t,P}^{\mathrm{pw},+}(X;r)
&=m_{1,t,P}^r-\overline\gamma_{t,P}m_{0,t,P}^r,\notag\\
\theta_{t,P}^{\mathrm{norm},-}(X;r)
&=c_{t,P}^{\ell}m_{1,t,P}^r
+(b_{t,P}-\Lambda_0^{-1})m_{0,t,P}^r
-(1-\Lambda_0^{-1}),\notag\\
\theta_{t,P}^{\mathrm{norm},+}(X;r)
&=-c_{t,P}^{u}m_{1,t,P}^r
+(\Lambda_0-b_{t,P})m_{0,t,P}^r
-(\Lambda_0-1).
\label{eq:supp_compatibility_moment_bases}
\end{align}
Under the null, all four are nonpositive almost surely for every $r$.

The allocation $r$ selects an outcome--covariate region.
A nonnegative localization $h(X)$ then prevents a violation confined to a covariate region from being averaged with slack elsewhere.
We call $w=(r,h)$ a witness and, for $j\in\mathcal J:=\{(\mathrm{pw},-),(\mathrm{pw},+), (\mathrm{norm},-),(\mathrm{norm},+)\}$, define
\begin{equation*}
\Theta_{t,j}(P;w)
=\E_P[h(X)\theta_{t,j,P}(X;r)]\le0
\quad\text{under }H_0.
\end{equation*}

For completeness, we give the uncentered influence function used to evaluate a fixed witness.
Put $e_{0,P}=e_P=P(Z=0\mid X)$, $e_{1,P}=1-e_P$, and $D_t=I(T=t)$.
Regard $\theta_{t,j,P}$ as a function of $(m_0,m_1,p_0,p_1)$, where $p_z$ abbreviates $\pi_{z,P}^t$, and let $\partial_v\theta_{t,j,P}$ denote its ordinary partial derivative.

\begin{lemma}[Fixed-witness influence function]
\label{lem:supp_compatibility_fixed_witness_if}
For fixed bounded $w=(r,h)$ and under overlap and square integrability, an uncentered influence function for $\Theta_{t,j}(P;w)$ is
\begin{align}
\Phi_{t,j}^{\mathrm{comp}}(O;P,w)
={}&h(X)\left[\theta_{t,j,P}(X;r)
+\sum_{z=0}^1
\partial_{m_z}\theta_{t,j,P}(X;r)
\frac{I(Z=z,T=t)}{e_{z,P}(X)\pi_{z,P}^t(X)}
\{r(Y,X)-m_{z,t,P}^r(X)\}
\right.\notag\\
&\left.\hspace{8mm}
+\sum_{z=0}^1
\partial_{p_z}\theta_{t,j,P}(X;r)
\frac{I(Z=z)}{e_{z,P}(X)}
\{D_t-\pi_{z,P}^t(X)\}
\right].
\label{eq:supp_compatibility_uncentered_if}
\end{align}
Its expectation is $\Theta_{t,j}(P;w)$. The required derivatives follow directly from equation~\eqref{eq:supp_compatibility_moment_bases}.
For the two pointwise moments, respectively, they are
\[
\begin{aligned}
(\partial_{m_0},\partial_{m_1},\partial_{p_z})
&=(\underline\gamma_{t,P},-1,
m_{0,t,P}^r\partial_{p_z}\underline\gamma_{t,P}),\\
(\partial_{m_0},\partial_{m_1},\partial_{p_z})
&=(-\overline\gamma_{t,P},1,
-m_{0,t,P}^r\partial_{p_z}\overline\gamma_{t,P}).
\end{aligned}
\]
For the two normalization moments, differentiate $c_{t,P}^{\ell}$, $c_{t,P}^{u}$, and $b_{t,P}$.
\end{lemma}

Split the sample into a fixed number of folds. On the observations outside fold $k$, estimate the nuisance functions and select one witness $\widehat w_{-k,t,j}$ from a finite candidate collection for each of the eight pairs $(t,j)$.
Freeze that choice before evaluating equation~\eqref{eq:supp_compatibility_uncentered_if} on fold $k$.
For $i$ in fold $k(i)$, write
\[
\widehat\Phi_{i,t,j}^{\mathrm{comp}}
=\Phi_{t,j}^{\mathrm{comp}}
\{O_i;\widehat P_{-k(i)},\widehat w_{-k(i),t,j}\},
\quad
\overline\Phi_{t,j}^{\mathrm{comp}}
=\frac1n\sum_{i=1}^n\widehat\Phi_{i,t,j}^{\mathrm{comp}},
\]
and let $\widehat\sigma_{t,j}$ be their empirical standard deviation.
The recommended statistic and its Gaussian multiplier analogue are
\begin{equation}
T_n=\max_{t,j}
\frac{\sqrt n\,\overline\Phi_{t,j}^{\mathrm{comp}}}
{\widehat\sigma_{t,j}},
\qquad
T_n^*=\max_{t,j}\frac1{\sqrt n}\sum_{i=1}^n\xi_i
\frac{\widehat\Phi_{i,t,j}^{\mathrm{comp}}
-\overline\Phi_{t,j}^{\mathrm{comp}}}
{\widehat\sigma_{t,j}},
\label{eq:supp_learned_witness_test}
\end{equation}
where the $\xi_i$ are independent standard normal multipliers. Reject when $T_n$ exceeds the conditional $(1-\alpha)$ quantile $c_{1-\alpha}^*$ of $T_n^*$.
This construction adapts the logic of conditional moment-inequality inference \citep{andrews2013inference}, with witness selection separated from its held-out evaluation.

\begin{theorem}[Validity of the learned-witness compatibility test]
\label{thm:learned_witness_compatibility}
Fix $(\Lambda_0,\Lambda_1)$, the number of folds, and a finite collection of witness-selection rules.
Suppose the true and fitted instrument and treatment probabilities are uniformly bounded away from zero and one; every selected witness is measurable with respect to its training observations; the eight held-out uncentered influence functions have nondegenerate variances and uniformly bounded $(2+\delta)$th moments for some $\delta>0$; their nuisance remainders are $o_p(n^{-1/2})$ uniformly over the eight moments; and the conditional Gaussian multiplier approximation is consistent for their joint centered limit.
Then, for every $P\in\mc M(\Lambda_0,\Lambda_1)$ satisfying these conditions,
\[
\limsup_{n\to\infty}
P\{T_n>c_{1-\alpha}^*\}\le\alpha.
\]
\end{theorem}

Multiplier consistency is imposed directly here; deriving primitive sufficient conditions for the learned-witness procedure is left for future work.

The test in equation~\eqref{eq:supp_learned_witness_test} accounts jointly for both treatment arms and all four inequality families. Rejection provides evidence against an observable necessary implication of the joint sensitivity model and IV assumptions.
Under the IV model, rejection rules out the proposed
sensitivity-parameter pair. Nonrejection does not establish compatibility, and power may be limited for violations concentrated on regions of small probability.

\subsection{Simultaneous sensitivity bands for binary outcomes}
\label{sec:supp_binary_bands}

Consider a compact interval $\mathcal I=[\Lambda_{\min},\Lambda_{\max}]
\subset[1,\infty)$ of compatible common sensitivity parameters
$\Lambda_0=\Lambda_1=\Lambda$ for both potential outcomes.
Let $\psi^-(\Lambda)$ and $\psi^+(\Lambda)$ denote the sharp ATE
bounds for outcomes coded zero and one.

\begin{proposition}[Shape of the binary-outcome bounds]
\label{prop:supp_binary_bound_shape}
The lower bound $\psi^-$ is nonincreasing and the upper bound $\psi^+$ is nondecreasing on $\mathcal I$.
Both are Lipschitz with constant at most $2$.
\end{proposition}

Fix a finite grid $\lambda_1<\cdots<\lambda_J$ in $\mathcal I$, including its endpoints.
Suppose the binary bound estimators admit the joint expansions
\[
\widehat\psi_j^b-\psi^b(\lambda_j)
=\frac1n\sum_{i=1}^n\phi_j^b(O_i)+o_p(n^{-1/2}),
\qquad b\in\{-,+\},\quad j=1,\ldots,J,
\]
where the fixed score vector has mean zero, finite second moments, and positive coordinate variances.
Assume consistent estimation of its covariance matrix and the bound standard errors.
These conditions follow from the binary one-step inference results when their assumptions hold at every grid point.

Let $(G_j^-,G_j^+)_{j=1}^J$ be the standardized Gaussian limit and let $c_{1-\alpha}$ be the $(1-\alpha)$ quantile of $\max_j\{G_j^-,-G_j^+\}$.
Estimate this quantile consistently using the estimated joint covariance or a centered Gaussian multiplier procedure with common multipliers across all bounds and grid points.
Define
\[
\ell_j=\widehat\psi_j^-
-\widehat c_{1-\alpha}\widehat{\mathrm{se}}_j^-,
\qquad
u_j=\widehat\psi_j^+
+\widehat c_{1-\alpha}\widehat{\mathrm{se}}_j^+.
\]

\begin{proposition}[Finite-grid and continuous outward bands]
\label{prop:supp_binary_continuous_bands}
Under the preceding joint expansion and covariance conditions, and continuity of the limiting maximum distribution at its critical quantile, the grid limits cover both population bounds simultaneously with asymptotic probability at least $1-\alpha$.
Writing $(v)_+=\max(v,0)$, define
\[
L_n(\Lambda)=\max_{1\le j\le J}
\{\ell_j-2(\Lambda-\lambda_j)_+\},
\qquad
U_n(\Lambda)=\min_{1\le j\le J}
\{u_j+2(\Lambda-\lambda_j)_+\}.
\]
Then
\[
\liminf_{n\to\infty}P\!\left\{
L_n(\Lambda)\le\psi^-(\Lambda)\le\psi^+(\Lambda)\le U_n(\Lambda)
\text{ for all }\Lambda\in\mathcal I
\right\}\ge1-\alpha.
\]
\end{proposition}

The extension uses monotonicity and the known Lipschitz constant, rather than interpolation of pointwise intervals.
It requires no uniform asymptotic expansion or no-tie condition between grid points.
The grid is fixed as $n$ increases; a growing grid requires additional theory.
For a lower-only band, calibrate $\max_jG_j^-$ and use the same lower extension.

\subsection{Sufficient conditions for the first-order optimizer condition}
\label{subsec:supp_optimizer_regularity}

We provide sufficient conditions under which condition~\eqref{eq:first_order_optimizer} in Theorem~\ref{thm:eif_fixed_q} holds for threshold classes of the form $\widetilde{\mc Q}=\{q_\beta(x)=B(x)^\top\beta:\beta\in\mathbb R^d\}$, where $B$ is a prespecified finite-dimensional basis, including an intercept. This family includes linear, additive natural-spline, and finite-dimensional RBF classes, encompassing the threshold classes used in the numerical experiments.
The basis is held fixed as $P$ varies; in particular, spline knots and, for RBF classes, the centers, bandwidth, and number of basis functions are prespecified.

In addition to the conditions of Theorem~\ref{thm:eif_fixed_q}\textup{(i)}, assume that $Y$ is bounded and that the following conditions hold:
\begin{enumerate}
\item $B(X)$ is bounded and $\E_P\{B(X)B(X)^\top\}$ is positive definite.
\item For some $\delta>0$, $\widetilde L_P(X)\le1-\delta$ and $\widetilde U_P(X)\ge1+\delta$ almost surely.
\item At a population-optimal coefficient vector $\beta_P^b$, the conditional density $f_{0,P}(y\mid x)$ of $Y\mid(T=1,X=x,Z=0)$ and the envelope width $\Delta_P(x,y)=U_P(x,y)-L_P(x,y)$ are continuous in $y$ near $q_{\beta_P^b}(x)$ for $P_X$-almost every $x$.
The density is uniformly bounded in a neighborhood of these thresholds, and
\[
f_{0,P}\{q_{\beta_P^b}(x)\mid x\}>0,
\qquad
\Delta_P\{x,q_{\beta_P^b}(x)\}>0
\]
for $P_X$-almost every $x$.
\end{enumerate}

The first condition excludes redundant basis directions.
The second places the conditional normalization restrictions strictly inside their feasible range and ensures that an optimal coefficient vector exists without imposing an artificial bound on the coefficients.
The third requires positive outcome density and envelope width at the optimal threshold.
Together with the first condition, it gives strict local curvature in every coefficient direction. Since the lower and upper objectives are globally concave and convex, respectively, the optimizer is unique. We show below that these conditions also imply that reoptimization contributes no first-order change, as required by condition~\eqref{eq:first_order_optimizer}.

\paragraph{Existence of an optimizer.}
Write $F_P^b(\beta)=\nu_P^{1b}(q_\beta)$ and suppose $|Y|\le M$.
For a fixed covariate value $x$ and scalar threshold $q$, the conditional lower-bound objective is
\[
\begin{aligned}
G_P^-(x,q)
={}&\pi_{0,P}(x)m_{0,P}(x)
+\{1-\pi_{0,P}(x)\}
\Big[
q+\E_P\big\{
U_P(x,Y)(Y-q)I(Y\le q)\\
&\hspace{43mm}
+L_P(x,Y)(Y-q)I(Y>q)
\mid T=1,X=x,Z=0
\big\}
\Big].
\end{aligned}
\]
Thus $F_P^-(\beta)=\E_P\{G_P^-(X,B(X)^\top\beta)\}$.
This objective is defined for every threshold value; evaluating it does not require the threshold to produce a normalized density ratio.

For $q>M$, every outcome satisfies $Y\le q$, giving
\[
G_P^-(x,q)-G_P^-(x,M)
=
\{1-\pi_{0,P}(x)\}
\{1-\widetilde U_P(x)\}(q-M)
\le-\epsilon\delta(q-M).
\]
Here treatment positivity gives $1-\pi_{0,P}(x)\ge\epsilon$, and the second condition gives $\widetilde U_P(x)\ge1+\delta$.
Similarly, for $q<-M$,
\[
G_P^-(x,q)-G_P^-(x,-M)
=
\{1-\pi_{0,P}(x)\}
\{1-\widetilde L_P(x)\}(q+M)
\le-\epsilon\delta(-M-q).
\]
Thus the objective decreases at least linearly as the threshold moves beyond either end of the outcome support.

For $|q|\le M$, bounded outcomes and envelopes imply $|G_P^-(x,q)|\le C_0$ for a constant $C_0$ independent of $x$.
Combining this bound with the preceding inequalities gives, for every $q$,
\[
G_P^-(x,q)
\le C_0-\epsilon\delta(|q|-M)_+
\le C_0+\epsilon\delta M-\epsilon\delta|q|.
\]
Consequently, for $C=C_0+\epsilon\delta M$,
\[
F_P^-(\beta)
\le C-\epsilon\delta\,\E_P|B(X)^\top\beta|.
\]

To relate this bound to the coefficient vector, let
$\|B(X)\|\le K_B$ almost surely and let $\lambda_B>0$ be the smallest eigenvalue of $\E_P\{B(X)B(X)^\top\}$.
For $\beta\ne0$,
\[
\lambda_B\|\beta\|^2
\le \E_P\{(B(X)^\top\beta)^2\}
\le \E_P\{\|B(X)\|\,\|\beta\|\,|B(X)^\top\beta|\}
\le K_B\|\beta\|\,\E_P|B(X)^\top\beta|.
\]
where the second inequality follows from $|B(X)^\top\beta|\le \Vert B(X)\Vert\,\|\beta\|$ by Cauchy--Schwarz.
Therefore $\E_P|B(X)^\top\beta|\ge(\lambda_B/K_B)\|\beta\|$, and hence $F_P^-(\beta)\to-\infty$ as $\|\beta\|\to\infty$.

For the upper-bound objective, the same argument gives $F_P^+(\beta)\ge-C+\epsilon\delta\,\E_P|B(X)^\top\beta|$, so $F_P^+(\beta)\to+\infty$ as $\|\beta\|\to\infty$.
Both objectives are continuous in $\beta$. Their respective maximum and minimum must therefore be attained at finite coefficient vectors.

\paragraph{Uniqueness of the optimizer.}
The curvature at an optimizer is determined by the $d\times d$ matrix
\[
J_P^b
=\E_P\!\left[
\{1-\pi_{0,P}(X)\}
\Delta_P\{X,q_{\beta_P^b}(X)\}
f_{0,P}\{q_{\beta_P^b}(X)\mid X\}
B(X)B(X)^\top
\right].
\]
The scalar multiplying $B(X)B(X)^\top$ is strictly positive almost surely.
Hence, for every nonzero vector $v$, $v^\top J_P^b v>0$, because the first condition ensures
$\E_P\{(v^\top B(X))^2\}>0$.
Differentiating the objectives at their optimizers gives
\[
\nabla_\beta^2F_P^-(\beta_P^-)=-J_P^-,
\qquad
\nabla_\beta^2F_P^+(\beta_P^+)=J_P^+.
\]
The bounded density near the thresholds, bounded basis, and bounded envelopes justify differentiation under the expectation.
Thus, the lower objective has strictly negative curvature and the upper objective strictly positive curvature in every coefficient direction at the optimizer.
A second optimizer is impossible: global concavity or convexity would make the objective constant along the segment joining the two optimizers, contradicting this strict curvature.

\paragraph{First-order effect of reoptimization.}
Consider a regular submodel $P_s$ through $P$, on the same bounded outcome support and covariate domain on which $B$ is bounded.
Write $\|P_s-P\|_{\mathrm{TV}}=\int|p_s-p|\,d\nu$ for the total-variation norm, where $p_s$ and $p$ are densities relative to a common dominating measure $\nu$.
Regularity implies $\|P_s-P\|_{\mathrm{TV}}=O(|s|)$.

We first control how perturbing $P$ changes differences between threshold objectives.
Since
\[
|q_\beta(x)-q_{\beta'}(x)|
\le K_B\|\beta-\beta'\|,
\]
the same bound holds for the differences between $(y-q_\beta(x))_+$ and $(y-q_{\beta'}(x))_+$, and between $(q_\beta(x)-y)_+$ and $(q_{\beta'}(x)-y)_+$.
Moreover, positivity and the triangle inequality give
\[
\begin{aligned}
&\left\|
P_{s,X}(dx)f_{z,P_s}(y\mid x)\,dy
-P_X(dx)f_{z,P}(y\mid x)\,dy
\right\|_{\mathrm{TV}}
\le (1+2/\epsilon^2)\|P_s-P\|_{\mathrm{TV}},\\
&\E_P|\pi_{z,P_s}(X)-\pi_{z,P}(X)|
\le (2/\epsilon)\|P_s-P\|_{\mathrm{TV}}.
\end{aligned}
\]
These inequalities follow by adding and subtracting the corresponding joint distributions and using $P(T=1,Z=z\mid X)\ge\epsilon^2$ and $P(Z=z\mid X)\ge\epsilon$.

Substituting the definitions of $L_P$ and $U_P$ and using $\gamma_{1,P}f_{0,P}=f_{1,P}$ expresses $(1-\pi_{0,P})L_Pf_{0,P}$ and $(1-\pi_{0,P})U_Pf_{0,P}$ as maxima and minima of linear combinations of $f_{0,P}$ and $f_{1,P}$ weighted by treatment probabilities.
Because maxima and minima are Lipschitz in their arguments, the preceding inequalities also control changes in these weighted envelope measures.
Applying these bounds to the threshold representation gives
\[
\begin{aligned}
\big|
\{F_{P_s}^b(\beta)-F_{P_s}^b(\beta')\}
-\{F_P^b(\beta)-F_P^b(\beta')\}
\big|
&\le C\|P_s-P\|_{\mathrm{TV}}\|\beta-\beta'\|,\\
|F_{P_s}^b(\beta)-F_P^b(\beta)|
&\le C\|P_s-P\|_{\mathrm{TV}}(1+\|\beta\|),
\end{aligned}
\]
for a constant $C$ independent of the coefficient vectors and sufficiently small $s$.

For the lower endpoint, the previously established bound $F_P^-(\beta)\le C_0-c\|\beta\|$, with $c>0$, and the second inequality above ensure that coefficients approaching the perturbed supremum remain in a common bounded set.
Choose $\beta_s$ such that
\[
0\le
\widetilde\mu^{1-}(P_s)-F_{P_s}^-(\beta_s)
\le s^2.
\]
The objectives converge uniformly on that bounded set.
Since $F_P^-$ has a unique optimizer, it follows that
$\beta_s\to\beta_P^-$.

Using the first inequality above and optimality of $\beta_P^-$ at $P$,
\[
\begin{aligned}
0
&\le
\widetilde\mu^{1-}(P_s)-F_{P_s}^-(\beta_P^-)\\
&\le
s^2+\{F_P^-(\beta_s)-F_P^-(\beta_P^-)\}
+C\|P_s-P\|_{\mathrm{TV}}\|\beta_s-\beta_P^-\|\\
&\le
s^2+C\|P_s-P\|_{\mathrm{TV}}\|\beta_s-\beta_P^-\|
=o(|s|).
\end{aligned}
\]
Applying the same argument to the negative of the upper objective establishes condition~\eqref{eq:first_order_optimizer} for both endpoints.

\subsection{Sufficient conditions for negligible nuisance drift}
\label{subsubsec:supp_estimation_assumptions}

We decompose $R_n^{1b}$ into regression and propensity estimation error and the error from estimating the envelope inputs. Throughout this subsection, expectations are taken under $P$ with the training-sample fits held fixed. For each fold, abbreviate $q=\widehat q_{-k}^b$ and $\widehat\xi=(\widehat\pi_0,\widehat\pi_1,\widehat\gamma_1)$, suppressing the fold index.
Put $e_{0,P}=e_P$, $e_{1,P}=1-e_P$, and define $\widehat e_z$ analogously.

Form $\widehat H_z^b$ and $\widehat D_\zeta^b$ using $(\widehat\xi,q)$ in the main-text definitions, and let
\[
\overline H_z^{b,\dagger}(x)
=\E_P\{\widehat H_z^b(Y,x)\mid T=1,X=x,Z=z\}.
\]
Here $\widehat H_z^b$ is constructed using the fitted nuisance functions, but its conditional expectation is taken under the true distribution $P$, holding the training sample fixed.
Thus, $\overline H_z^{b,\dagger}$ is the population target of the fitted regression $\widehat{\overline H}_z^b$.
If $\widehat\xi=\xi_P$, then $\overline H_z^{b,\dagger}=\overline H_{z,P,q}^b$ for the same supplied threshold $q$, because $\widehat H_z^b=H_{z,P,q}^b$.
Define $C_z^{b,\dagger}$ similarly: in the definition of $C_{z,P,q}^b$, replace the envelope inputs by $\widehat\xi$ and evaluate every conditional expectation under $P$.
In particular, use $m_{0,P}$, $\theta^\dagger=q+\E_P\{\widehat\lambda_q^b(Y,X)(Y-q)\mid T=1,X,Z=0\}$, and the conditional means of $\widehat D_\zeta^b$.
Write $\widehat{\overline H}_z^b$ and $\widehat C_z^b$ for their fitted counterparts, and set $M^{b,\dagger}=\overline H_0^{b,\dagger}+\overline H_1^{b,\dagger}+(1-\widehat\pi_0)q$.
As in the main text, $\widehat M^b$ must use this same linked construction with the fitted conditional means.

\paragraph{Two-component decomposition.}
Direct conditional expectation of the uncentered influence function gives
\begin{equation}
R_n^{1b}=\sum_{k=1}^K\frac{n_k}{n}
\{R_{\mathrm{reg},-k}^{1b}+R_{\mathrm{env},-k}^{1b}\},
\label{eq:supp_drift_decomposition}
\end{equation}
where, with fold indices suppressed,
\begin{align}
R_{\mathrm{reg}}^{1b}
={}&\sum_{z=0}^1\E_P\left[
\left\{\frac{e_{z,P}\pi_{z,P}}{\widehat e_z\widehat\pi_z}-1\right\}
\{\overline H_z^{b,\dagger}-\widehat{\overline H}_z^b\}
\right]+\sum_{z=0}^1\E_P\left[
\left\{\frac{e_{z,P}}{\widehat e_z}\widehat C_z^b
-C_z^{b,\dagger}\right\}
(\pi_{z,P}-\widehat\pi_z)
\right],
\label{eq:supp_regression_drift}\\
R_{\mathrm{env}}^{1b}
={}&\E_P\{M^{b,\dagger}-M_{P,q}^b\}
+\sum_{z=0}^1\E_P\{C_z^{b,\dagger}
(\pi_{z,P}-\widehat\pi_z)\}.
\label{eq:supp_envelope_drift}
\end{align}
To verify the decomposition, take the conditional expectation of the estimated uncentered influence function given $X$, holding the training-sample fits fixed. Its outcome-residual term satisfies
\[
\begin{aligned}
\E_P\!\left[
\frac{I(Z=z)T}{\widehat e_z(X)\widehat\pi_z(X)}
\{\widehat H_z^b(Y,X)-\widehat{\overline H}_z^b(X)\}
\,\middle|\,X
\right]=
\frac{e_{z,P}(X)\pi_{z,P}(X)}
{\widehat e_z(X)\widehat\pi_z(X)}
\{\overline H_z^{b,\dagger}(X)
-\widehat{\overline H}_z^b(X)\}.
\end{aligned}
\]
Indeed, the indicator restricts the expectation to observations with $T=1,Z=z$, whose conditional probability given $X$ is $e_{z,P}(X)\pi_{z,P}(X)$.
Within this group, the conditional mean of $\widehat H_z^b(Y,X)$ is $\overline H_z^{b,\dagger}(X)$.

Likewise, the treatment-residual term satisfies
\[
\begin{aligned}
\E_P\!\left[
\frac{I(Z=z)}{\widehat e_z(X)}
\widehat C_z^b(X)\{T-\widehat\pi_z(X)\}
\,\middle|\,X
\right]=
\frac{e_{z,P}(X)}{\widehat e_z(X)}
\widehat C_z^b(X)
\{\pi_{z,P}(X)-\widehat\pi_z(X)\},
\end{aligned}
\]
because $P(Z=z\mid X)=e_{z,P}(X)$ and $\E_P(T\mid X,Z=z)=\pi_{z,P}(X)$.

Combining these expressions with $\widehat M^b=\widehat{\overline H}_0^b+\widehat{\overline H}_1^b+(1-\widehat\pi_0)q$, subtracting $M_{P,q}^b$, and taking expectation over $X$ gives the fold-specific drift.
Adding and subtracting $\sum_{z=0}^1\E_P\{C_z^{b,\dagger}(\pi_{z,P}-\widehat\pi_z)\}$ then gives the two-component decomposition above.

\paragraph{Regression and propensity component.}
If $\widehat\xi=\xi_P$, then $R_{\mathrm{env}}^{1b}=0$ and the conditional bias reduces to
\[
\E_P\{\Phi^{1b}(O;\widehat\eta,q)\mid X\}-M_{P,q}^b(X)
=\sum_{z=0}^1
\left(\frac{e_{z,P}}{\widehat e_z}-1\right)
\{\overline H_{z,P,q}^b-\widehat{\overline H}_z^b\}.
\]
Thus, conditional on correct treatment propensities and density ratio, the fixed-threshold estimator is unbiased if either the instrument propensity or both conditional mean regressions are correct.
This holds for every supplied $q$, not only a population optimizer.

More generally, suppose true and fitted probabilities are uniformly
bounded away from zero and one and $\widehat C_z^b$ is uniformly bounded in probability
Define the maximal fold errors
\[
\begin{aligned}
u_n&=\max_k\|\widehat e_{-k}-e_P\|_{P,2},&
v_n&=\max_{k,z}\|\widehat\pi_{z,-k}-\pi_{z,P}\|_{P,2},\\
h_n&=\max_{k,z}\|\widehat{\overline H}_{z,-k}^b
-\overline H_{z,-k}^{b,\dagger}\|_{P,2},&
c_n&=\max_{k,z}\|\widehat C_{z,-k}^b
-C_{z,-k}^{b,\dagger}\|_{P,2}.
\end{aligned}
\]
Cauchy--Schwarz applied to \eqref{eq:supp_regression_drift}, using $(e_{z,P}/\widehat e_z)\widehat C_z^b-C_z^{b,\dagger}=(e_{z,P}/\widehat e_z-1)\widehat C_z^b+(\widehat C_z^b-C_z^{b,\dagger})$, yields
\[
\left|\sum_k\frac{n_k}{n}R_{\mathrm{reg},-k}^{1b}\right|
=O_p\{(u_n+v_n)h_n+(u_n+c_n)v_n\}.
\]

\paragraph{Envelope-input component.}
Assume bounded outcomes and fitted thresholds, uniformly bounded true and fitted density ratios, and the overlap conditions above.
For density-ratio errors use the baseline norm $\|g\|_{0,2}^2=\E_P[\E_P\{g(Y,X)^2\mid T=1,X,Z=0\}]$, and put
\[
d_n=v_n+\max_k\|\widehat\gamma_{1,-k}-\gamma_{1,P}\|_{0,2},
\qquad
a_n=\max_{k,\,c\in\{\Lambda_1^{-1},\Lambda_1\}}
\|\widehat\Gamma_{c,-k}-\Gamma_{c,P}\|_\infty.
\]
On regions where the fitted and true envelopes select the same arguments, the maps are smooth with bounded second derivatives.
The augmentation in~\eqref{eq:supp_envelope_drift} cancels their first-order error, leaving $O_p(d_n^2)$.
To see the density-ratio cancellation, use
\[
M^{b,\dagger}
=M_{\widehat\xi,q}^b
+(1-\widehat\pi_0)
\E_P\{(\gamma_{1,P}-\widehat\gamma_1)
\widehat D_{\gamma_1}^b\mid T=1,X,Z=0\},
\]
where $M_{\widehat\xi,q}^b=\widehat\pi_0m_{0,P}+(1-\widehat\pi_0)\theta^\dagger$.
The remaining $C_z^{b,\dagger}$ terms cancel the first-order treatment propensity errors.

For changes in the selected envelope argument, the elementary bound
\[
|\max(a,\widehat g)-\max(a,g)-I(\widehat g>a)(\widehat g-g)|
\le|\widehat g-g|I\{|g-a|\le|\widehat g-g|\}
\]
and its minimum analogue localize the remainder to within $a_n$ of a tie.
Assumption~\ref{ass:exact_envelope_margin} bounds the probability of this region by $O_p(a_n^\kappa)$, while its contribution has magnitude $O_p(a_n)$.
Consequently,
\[
\left|\sum_k\frac{n_k}{n}R_{\mathrm{env},-k}^{1b}\right|
=O_p(d_n^2+a_n^{1+\kappa}).
\]

Combining the two components, sufficient conditions for $R_n^{1b}=o_p(n^{-1/2})$ are
\[
(u_n+v_n)h_n+(u_n+c_n)v_n=o_p(n^{-1/2}),\qquad
d_n=o_p(n^{-1/4}),\qquad
a_n=o_p\{n^{-1/[2(1+\kappa)]}\}.
\]
For example, $o_p(n^{-1/4})$ rates for $u_n,v_n,h_n,c_n,d_n$, together with the displayed uniform envelope rate, suffice.
All regression targets here use the learned threshold and fitted pseudo-outcomes for that fold; no convergence rate for the threshold itself enters this drift calculation.

\paragraph{Smoothed envelopes.}
For the smoothed estimator, all pseudo-outcomes, regression targets, and derivative weights are formed using the smoothed envelopes.
The regression and propensity errors are controlled by the corresponding product-rate conditions above.
The envelope-input calculation instead uses the smooth derivatives of $L_{P,\varepsilon}$ and $U_{P,\varepsilon}$.

The second derivatives of the smoothed maximum and minimum with respect to their arguments are bounded by $1/(4h_\varepsilon)=(\log 2)/(4\varepsilon)$.
Under the boundedness and positivity conditions above, Taylor expansion and cancellation of the first-order nuisance errors therefore give
\[
\left|\sum_k\frac{n_k}{n}
R_{\mathrm{env},\varepsilon,-k}^{1b}\right|
=O_p\{(1+\varepsilon^{-1})d_n^2\}.
\]
Unlike the exact-envelope calculation, this bound does not require the margin condition or the uniform error bound $a_n$.
For fixed $\varepsilon>0$, the condition $d_n=o_p(n^{-1/4})$, together with the corresponding regression and propensity product-rate conditions, is sufficient for root-$n$-negligible nuisance drift.

The factor $\varepsilon^{-1}$ shows that smaller smoothing parameters require tighter control of the envelope-input errors.
If $\varepsilon=\varepsilon_n\to0$, controlling this contribution would require $\varepsilon_n^{-1}d_n^2=o_p(n^{-1/2})$.
This drift condition alone does not establish inference with a vanishing smoothing parameter; that setting requires a separate asymptotic analysis.

%% file: supp_extensions.tex
\section{Extensions of the Identification Framework}
\label{sec:supp_finite_extensions}

Section~\ref{subsec:supp_finite_outcomes} extends identification and estimation to outcomes with finite ordered support.
Section~\ref{subsec:supp_finite_iv} extends the IV-MSM to finite discrete instruments and gives the combined finite-outcome specialization.
Section~\ref{subsec:supp_binary_joint_exchangeability} characterizes sharp bounds for binary outcomes under an additional joint exchangeability assumption.
Proofs are provided separately in Section~\ref{sec:supp_extension_proofs}.

\subsection{Finite-support outcomes}
\label{subsec:supp_finite_outcomes}

Suppose that $Y$ has ordered support $\mc Y=\{y_1<\cdots<y_K\}$.
We state the results for both $t\in\{0,1\}$.
Let $\pi_{z,P}^t(x)=P(T=t\mid X=x,Z=z)$ and $a_{z,k,P}^t(x)=P(Y=y_k\mid T=t,X=x,Z=z)$ for $z\in\{0,1\}$ and $k\in\{1,\ldots,K\}$.
Thus $\pi_{z,P}^1=\pi_{z,P}$ and $\pi_{z,P}^0=1-\pi_{z,P}$ in the notation of the main text.
For binary outcomes, the category probability for $y_k=1$ and $t=1$ is $a_{z,P}$.
Under the IV assumptions, the conditional distribution of $Y^t$ given $X$ does not depend on $Z$.
Write its category probabilities under a compatible full-data law $Q$ as $h_{k,Q}^t(x)=Q(Y^t=y_k\mid X=x)$.
For each instrument stratum,
\[
h_{k,Q}^t
=\pi_{z,P}^t a_{z,k,P}^t
+(1-\pi_{z,P}^t)\vartheta_{z,k,Q}^t,
\]
where $\vartheta_{z,k,Q}^t(x)=Q(Y^t=y_k\mid T=1-t,X=x,Z=z)$.
The IV-MSM restricts $\vartheta_{z,k,Q}^t/ a_{z,k,P}^t$ to $[\Lambda_z^{-1},\Lambda_z]$, with the usual convention that zero observed mass forces zero missing mass.

For compactness, suppress $(t,P,x)$ and define
\begin{align}
\ell_k
&=\max_{z\in\{0,1\}}
\left\{
\pi_z a_{z,k},
\left[\pi_z+\frac{1-\pi_z}{\Lambda_z}\right] a_{z,k}
\right\},\notag\\
u_k
&=\min_{z\in\{0,1\}}
\left\{
\pi_z a_{z,k}+1-\pi_z,
\left[\pi_z+(1-\pi_z)\Lambda_z\right] a_{z,k}
\right\}.
\label{eq:supp_finite_outcome_envelopes}
\end{align}
The first lower candidate is redundant for finite $\Lambda_z$, but displaying it makes the probability-simplex restriction explicit and accommodates the IV-only limit.

Write $\ell=(\ell_1,\ldots,\ell_K)^\top$ and $u=(u_1,\ldots,u_K)^\top$.
For feasible lower and upper probability bounds $\ell\le u$, construct $h^-(\ell,u)$ by first assigning probability $\ell_k$ to each category.
Distribute the remaining probability $1-\sum_k\ell_k$ in increasing outcome order, adding to each category until its probability reaches $u_k$ or no mass remains.
This allocation minimizes the mean over all probability vectors between $\ell$ and $u$.
Construct $h^+(\ell,u)$ analogously, distributing the remaining probability in decreasing outcome order.
This allocation maximizes the mean over the same feasible probability vectors.
Define the resulting lower and upper means by
\[
\mathcal W^b(\ell,u)=\sum_{k=1}^K y_k h_k^b(\ell,u),
\qquad b\in\{-,+\}.
\]

\begin{proposition}[Finite-support compatibility and sharp bounds]
\label{prop:supp_finite_outcome_bounds}
Suppose Assumption~\ref{ass:positivity} holds.
For $P_X$-almost every $x$, there exists a conditional full-data extension of $P(\,\cdot\mid X=x)$ satisfying the assumptions defining $\mc M_t(\Lambda_0,\Lambda_1)$ if and only if
\[
\ell_{k,P}^t(x)\le u_{k,P}^t(x)\quad\text{for every }k
\qquad\text{and}\qquad
\sum_{k=1}^K\ell_{k,P}^t(x)\le1\le
\sum_{k=1}^Ku_{k,P}^t(x).
\]
When these conditions hold almost surely, the sharp bounds are
\begin{equation}
\mu^{tb}(P)
=\E_P\!\left[
\mathcal W^b\{\ell_P^t(X),u_P^t(X)\}
\right],
\qquad b\in\{-,+\}.
\label{eq:supp_finite_outcome_bounds}
\end{equation}
The corresponding ATE bounds are $\psi^-(P)=\mu^{1-}(P)-\mu^{0+}(P)$ and $\psi^+(P)=\mu^{1+}(P)-\mu^{0-}(P)$.
\end{proposition}

When $K=2$ and $(y_1,y_2)=(0,1)$, Proposition~\ref{prop:supp_finite_outcome_bounds} reduces to the binary-outcome formulas.
More generally, the finite allocation procedure gives the bounds without estimating a conditional threshold function.

\begin{assumption}[Finite-support bound regularity]
\label{ass:supp_finite_outcome_regularity}
For $P_X$-almost every $x$, there are no ties in the finite maxima and minima defining $\ell_P^t(x)$ and $u_P^t(x)$.
For each $b\in\{-,+\}$, the category receiving the last portion of probability mass in the construction of $h^b\{\ell_P^t(x),u_P^t(x)\}$ remains unchanged under sufficiently small perturbations of the treatment and conditional outcome probabilities for which the allocation remains feasible.
\end{assumption}

This is a sufficient condition for differentiability.
It may exclude ties in intermediate expressions that do not affect differentiability of the final bound.
For inference, let $M_P^{tb}(x)=\mathcal W^b\{\ell_P^t(x),u_P^t(x)\}$.
Under Assumption~\ref{ass:supp_finite_outcome_regularity}, write $\dot M_{\pi_z,P}^{tb}$ and $\dot M_{a_{z,k},P}^{tb}$ for its partial derivatives with respect to $\pi_{z,P}^t$ and $a_{z,k,P}^t$, respectively.
Also let $e_{z,P}(x)=P(Z=z\mid X=x)$, so that $e_{0,P}=e_P$ and $e_{1,P}=1-e_P$.

\begin{proposition}[Influence function for a finite-support outcome]
\label{prop:supp_finite_outcome_eif}
Suppose Assumptions~\ref{ass:positivity} and~\ref{ass:supp_finite_outcome_regularity} hold and the compatibility conditions in
Proposition~\ref{prop:supp_finite_outcome_bounds} hold almost surely.
Then $\mu^{tb}(P)$ is pathwise differentiable, with efficient influence function
\begin{align}
\phi_P^{tb}(O)
={}&M_P^{tb}(X)-\mu^{tb}(P)\notag\\
&+\sum_{z=0}^1\frac{I(Z=z)}{e_{z,P}(X)}
\left[
\dot M_{\pi_z,P}^{tb}(X)\{I(T=t)-\pi_{z,P}^t(X)\}
\right.\notag\\[-2mm]
&\hspace{39mm}\left.
+\frac{I(T=t)}{\pi_{z,P}^t(X)}
\sum_{k=1}^K\dot M_{a_{z,k},P}^{tb}(X)
\{I(Y=y_k)- a_{z,k,P}^t(X)\}
\right].
\label{eq:supp_finite_outcome_eif}
\end{align}
\end{proposition}

In equation~\eqref{eq:supp_finite_outcome_eif}, the coefficients $\dot M_{a_{z,k},P}^{tb}$ describe changes in the conditional outcome probabilities.
Since these probabilities sum to one, their perturbations sum to zero, so the derivative coefficients are determined only up to a common additive constant for each $z$.
This ambiguity does not affect the influence function, because $\sum_{k=1}^K\{I(Y=y_k)-a_{z,k,P}^t(X)\}=0$.

The cross-fitted one-step estimator averages the associated uncentered influence functions, following Section~\ref{sec:estimation}.
Its implementation requires estimation only of the instrument, treatment, and conditional outcome probabilities and their derived maps.

Assumption~\ref{ass:supp_finite_outcome_regularity} can be avoided by smoothing the finite maximum and minimum representations of the probability bounds and the resulting lower and upper means.
Replace each maximum by a log-mean-exp approximation lying below it and each minimum by its analogue lying above it, as in Section~\ref{sec:supp_smoothed_bounds}.
For every fixed smoothing parameter $\varepsilon>0$, this produces differentiable functionals $\mu_\varepsilon^{t-}(P)\le\mu^{t-}(P)$ and $\mu_\varepsilon^{t+}(P)\ge\mu^{t+}(P)$.
Their efficient influence functions are given by equation~\eqref{eq:supp_finite_outcome_eif}, with $M_P^{tb}$ and its partial derivatives replaced by their smoothed counterparts.
The cross-fitted one-step estimator averages the associated uncentered influence functions.
Thus, finite-support outcomes require neither density-ratio estimation nor conditional-threshold estimation; only the primitive conditional probabilities and their derived finite-dimensional maps are estimated.

\subsection{Finite discrete instruments}
\label{subsec:supp_finite_iv}

Now let $Z$ have finite support $\mc Z=\{z_1,\ldots,z_J\}$, where $J\ge2$, and assign a sensitivity parameter $1\le\Lambda_j<\infty$ to each instrument category.
We first extend the density-ratio formulation to finite instruments and then give the corresponding formulas for finite-support outcomes.
Fix $t\in\{0,1\}$ and define
\[
e_{j,P}(x)=P(Z=z_j\mid X=x),\qquad
\pi_{j,P}^t(x)=P(T=t\mid X=x,Z=z_j).
\]
Let $f_{j,P}^t(y\mid x)$ denote the conditional density of
$Y\mid(T=t,X=x,Z=z_j)$. Choose any reference category $j_0$ and put
$\gamma_{j,P}^t=f_{j,P}^t/f_{j_0,P}^t$, so that
$\gamma_{j_0,P}^t=1$.

Under a compatible full-data law, let $h_Q^t(y\mid x)$ denote the common
conditional density of $Y^t$ given $(X=x,Z=z_j)$, and define
$\rho_Q^t=h_Q^t/f_{j_0,P}^t$. The observed--missing mixture identity gives
\[
\rho_Q^t(y,x)
=\gamma_{j,P}^t(y,x)
\left[\pi_{j,P}^t(x)+\{1-\pi_{j,P}^t(x)\}\lambda_{j,Q}^t(y,x)\right],
\]
where the IV-MSM imposes
$\lambda_{j,Q}^t\in[\Lambda_j^{-1},\Lambda_j]$. Consequently, define
\begin{align}
A_{j,P}^t(x)
&=\pi_{j,P}^t(x)+\{1-\pi_{j,P}^t(x)\}\Lambda_j^{-1},
&
B_{j,P}^t(x)
&=\pi_{j,P}^t(x)+\{1-\pi_{j,P}^t(x)\}\Lambda_j,
\notag\\
L_P^t(x,y)
&=\max_{1\le j\le J}
A_{j,P}^t(x)\gamma_{j,P}^t(y,x),
&
U_P^t(x,y)
&=\min_{1\le j\le J}
B_{j,P}^t(x)\gamma_{j,P}^t(y,x).
\label{eq:supp_finite_iv_envelopes}
\end{align}

\begin{proposition}[Compatibility and bounds with a finite instrument]
\label{prop:supp_finite_iv_bounds}
Suppose Assumptions~\ref{ass:positivity} and~\ref{ass:abs_cont} hold, extended to all instrument categories, with $z_{j_0}$ as the reference category.
A compatible full-data extension satisfying the restrictions for $Y^t$ exists if and only if the following conditions hold for $P_X$-almost every $x$:
\begin{equation}
\begin{gathered}
L_P^t(x,y)\le U_P^t(x,y)
\quad\text{for }
P(Y\mid T=t,X=x,Z=z_{j_0})\text{-almost every }y,\\
\E_P\{L_P^t(x,Y)\mid T=t,X=x,Z=z_{j_0}\}
\le1\le
\E_P\{U_P^t(x,Y)\mid T=t,X=x,Z=z_{j_0}\}.
\end{gathered}
\label{eq:supp_finite_iv_compatibility}
\end{equation}
Here $L_P^t$ and $U_P^t$ bound the common potential-outcome density relative to the reference outcome density, rather than the missing-versus-observed density ratio used in the binary-IV formulation.
The feasible common density ratios are precisely the functions $\rho$ such that $L_P^t\le\rho\le U_P^t$ and $\E_P(\rho\mid T=t,X,Z=z_{j_0})=1$.

The feasible conditional distributions of $Y^t$ and the resulting sharp bounds do not depend on the choice of reference category.
For an atomless reference outcome law, the sharp conditional means are obtained by the same threshold allocation as in the binary-IV analysis, with $(L_P,U_P)$ replaced by $(L_P^t,U_P^t)$.
For a prespecified threshold class $\widetilde{\mc Q}$, define
\begin{align}
\rho_{P,q}^{t-}(y,x)
&=U_P^t(x,y)I\{y\le q(x)\}
+L_P^t(x,y)I\{y>q(x)\},\notag\\
\rho_{P,q}^{t+}(y,x)
&=L_P^t(x,y)I\{y\le q(x)\}
+U_P^t(x,y)I\{y>q(x)\},\notag\\
\theta_{P,q}^{tb}(x)
&=q(x)+\E_P\!\left[
\rho_{P,q}^{tb}(Y,x)\{Y-q(x)\}
\mid T=t,X=x,Z=z_{j_0}
\right],\notag\\
\nu_P^{tb}(q)&=\E_P\{\theta_{P,q}^{tb}(X)\}.
\label{eq:supp_finite_iv_fixed_q}
\end{align}
Then the restricted-class bounds are $\widetilde\mu^{t-}(P)=\sup_{q\in\widetilde{\mc Q}}\nu_P^{t-}(q)$ and $\widetilde\mu^{t+}(P)=\inf_{q\in\widetilde{\mc Q}}\nu_P^{t+}(q)$.
The resulting restricted-class interval contains the sharp identified interval, as in Corollary~\ref{cor:restricted_classes}.
\end{proposition}

Changing the reference category rescales $L_P^t$, $U_P^t$, and every
feasible $\rho$ by the same Radon--Nikodym derivative while changing the
integrating measure inversely. Therefore the exact bounds are algebraically
equivalent for every reference category; choosing the largest category in a
sample can improve numerical stability but cannot tighten the population
bounds. A fixed additive smooth approximation of
equation~\eqref{eq:supp_finite_iv_envelopes}, however, need not be reference
invariant.

We next record the fixed-threshold influence function. Let
$a_{U,q}^{t-}=I\{Y\le q(X)\}$,
$a_{U,q}^{t+}=I\{Y>q(X)\}$, and $a_{L,q}^{tb}=1-a_{U,q}^{tb}$.
Assume temporarily that the maximizing and minimizing categories in
equation~\eqref{eq:supp_finite_iv_envelopes} are unique, and denote their
indicators by
\[
\omega_{L,j,P}^t
=I\{A_{j,P}^t\gamma_{j,P}^t=L_P^t\},
\qquad
\omega_{U,j,P}^t
=I\{B_{j,P}^t\gamma_{j,P}^t=U_P^t\}.
\]
Define
\begin{align*}
\dot L_{\gamma_j,P}^t
&=\omega_{L,j,P}^tA_{j,P}^t,
&
\dot U_{\gamma_j,P}^t
&=\omega_{U,j,P}^tB_{j,P}^t,
&&j\ne j_0,\\
\dot L_{\pi_j,P}^t
&=\omega_{L,j,P}^t\gamma_{j,P}^t(1-\Lambda_j^{-1}),
&
\dot U_{\pi_j,P}^t
&=\omega_{U,j,P}^t\gamma_{j,P}^t(1-\Lambda_j),
&&1\le j\le J.
\end{align*}
Put
\begin{align*}
D_{\gamma_j,P,q}^{tb}
&=\{Y-q(X)\}
\left(a_{U,q}^{tb}\dot U_{\gamma_j,P}^t
+a_{L,q}^{tb}\dot L_{\gamma_j,P}^t\right),
\qquad j\ne j_0,\\
D_{\pi_j,P,q}^{tb}
&=\{Y-q(X)\}
\left(a_{U,q}^{tb}\dot U_{\pi_j,P}^t
+a_{L,q}^{tb}\dot L_{\pi_j,P}^t\right),
\qquad 1\le j\le J,\\
H_{j_0,P,q}^{tb}
&=\rho_{P,q}^{tb}(Y,X)\{Y-q(X)\}
-\sum_{j\ne j_0}\gamma_{j,P}^t(Y,X)
D_{\gamma_j,P,q}^{tb}(Y,X),\\
H_{j,P,q}^{tb}&=D_{\gamma_j,P,q}^{tb},\qquad j\ne j_0,\\
\overline H_{j,P,q}^{tb}(x)
&=\E_P\{H_{j,P,q}^{tb}(Y,x)\mid T=t,X=x,Z=z_j\},\\
C_{j,P,q}^{tb}(x)
&=\E_P\{D_{\pi_j,P,q}^{tb}(Y,x)
\mid T=t,X=x,Z=z_{j_0}\}.
\end{align*}

\begin{proposition}[Fixed-threshold influence function with a finite instrument]
\label{prop:supp_finite_iv_eif}
Suppose the conditions of Proposition~\ref{prop:supp_finite_iv_bounds} hold, the outcome domain $\mc Y$ is bounded, and there are no ties in the maximum and minimum defining the envelopes, almost surely under $P(X,Y\mid T=t,Z=z_{j_0})$.
For every fixed bounded measurable threshold function $q$, $\nu_P^{tb}(q)$ is pathwise differentiable, with efficient influence function
\begin{align}
\phi_{P,q}^{tb}(O)
={}&\theta_{P,q}^{tb}(X)-\nu_P^{tb}(q)\notag\\
&+\sum_{j=1}^J\frac{I(Z=z_j)}{e_{j,P}(X)}
\left[
C_{j,P,q}^{tb}(X)\{D_t-\pi_{j,P}^t(X)\}
\right.\notag\\[-2mm]
&\hspace{39mm}\left.
+\frac{D_t}{\pi_{j,P}^t(X)}
\{H_{j,P,q}^{tb}(Y,X)-\overline H_{j,P,q}^{tb}(X)\}
\right].
\label{eq:supp_finite_iv_eif}
\end{align}
\end{proposition}

The optimizer and cross-fitting results in the main text extend using equation~\eqref{eq:supp_finite_iv_eif}, under the corresponding regularity conditions.
For exact envelopes, the margin condition controls the distance between the largest and second-largest arguments in the maximum and between the smallest and second-smallest arguments in the minimum.
For smoothed envelopes, the indicators selecting the active arguments are replaced by the corresponding smooth derivative weights.

If $Y$ also has finite support, the bounds can be expressed directly in terms of $a_{j,k,P}^t(x)=P(Y=y_k\mid T=t,X=x,Z=z_j)$, without density ratios.
For each outcome category $k$, intersect the probability and sensitivity intervals contributed by all instrument categories:
\[
\ell_{k,P}^t
=\max_{1\le j\le J}
\left\{
\pi_{j,P}^t a_{j,k,P}^t,
\left[\pi_{j,P}^t+\frac{1-\pi_{j,P}^t}{\Lambda_j}\right]a_{j,k,P}^t
\right\},
\]
\[
u_{k,P}^t
=\min_{1\le j\le J}
\left\{
\pi_{j,P}^t a_{j,k,P}^t+1-\pi_{j,P}^t,
\left[\pi_{j,P}^t+(1-\pi_{j,P}^t)\Lambda_j\right]a_{j,k,P}^t
\right\}.
\]
Propositions~\ref{prop:supp_finite_outcome_bounds} and
\ref{prop:supp_finite_outcome_eif} then apply with the sums over the two
binary-IV strata replaced by sums over $j=1,\ldots,J$. This combined
extension is reference-free even before optimization.

\subsection{Binary-outcome bounds under joint exchangeability}
\label{subsec:supp_binary_joint_exchangeability}

Suppose $\mathcal Y=\{0,1\}$. The IV assumptions used above require each potential outcome
to be independent of the instrument conditional on $X$.
Here we additionally impose
\begin{equation}
(Y^0,Y^1)\perp Z\mid X.
\label{eq:supp_joint_exchangeability}
\end{equation}
This restriction strengthens the separate marginal independence
conditions by requiring the joint distribution of $(Y^0,Y^1)$
to agree across instrument strata.
It holds in a randomized trial with nonadherence under
individual-level exclusion, $Y^{t,0}=Y^{t,1}$ almost surely
for each $t$: holding treatment fixed, changing instrument
assignment does not change any individual's potential outcome.
However, it need not follow from randomization under the weaker
distributional exclusion restriction maintained here.
We therefore retain the other IV-MSM assumptions and impose
\eqref{eq:supp_joint_exchangeability} as an additional restriction
on the treatment-specific potential outcomes $Y^0$ and $Y^1$.
The resulting identification problem admits a finite linear
program, following the approach underlying the Balke--Pearl
bounds \citep{balke1997bounds}, with additional constraints
representing the marginal sensitivity model.

For $z,t,y\in\{0,1\}$, define the observed probabilities
\[
p_{zty,P}(x)
:=P(T=t,Y=y\mid X=x,Z=z),
\qquad
\pi_{z,P}^t(x):=\sum_{y=0}^1p_{zty,P}(x),
\]
and
\[
a_{z,P}^t(y,x)
:=\frac{p_{zty,P}(x)}{\pi_{z,P}^t(x)}.
\]
Thus $a_{z,P}^t(y,x)$ is the observed conditional probability
of outcome $y$ in treatment arm $t$ and instrument stratum $z$.
We use the same finite sensitivity parameter $\Lambda_z\ge1$
for both potential outcomes within stratum $z$.

At each $x$, the optimization variables are sixteen
nonnegative numbers $w_{ztuv}(x)$, interpreted as
\[
Q(T=t,Y^0=u,Y^1=v\mid X=x,Z=z),
\qquad z,t,u,v\in\{0,1\}.
\]
Write $y_0(u,v)=u$ and $y_1(u,v)=v$.
Let $\mathcal W_P(x)$ contain the vectors satisfying the
following constraints.

First, consistency with the observed distribution requires
\begin{equation}
\sum_{\substack{u,v\in\{0,1\}\\y_t(u,v)=y}}
w_{ztuv}(x)
=p_{zty,P}(x),
\qquad z,t,y\in\{0,1\}.
\label{eq:supp_joint_observed}
\end{equation}
Second, joint exchangeability requires
\begin{equation}
\sum_{t=0}^1w_{0tuv}(x)
=
\sum_{t=0}^1w_{1tuv}(x)
=:r_{uv}(x),
\qquad u,v\in\{0,1\}.
\label{eq:supp_joint_common}
\end{equation}
Here $r_{uv}(x)$ is the common conditional probability of
$(Y^0,Y^1)=(u,v)$.
Third, the sensitivity restrictions require
\begin{equation}
\begin{aligned}
\frac{\{1-\pi_{z,P}^t(x)\}a_{z,P}^t(y,x)}{\Lambda_z}
\le
\sum_{\substack{u,v\in\{0,1\}\\y_t(u,v)=y}}
w_{z,1-t,u,v}(x)\le
\{1-\pi_{z,P}^t(x)\}a_{z,P}^t(y,x)\Lambda_z,
\quad z,t,y\in\{0,1\}.
\end{aligned}
\label{eq:supp_joint_msm}
\end{equation}
The middle expression is the conditional probability of
$T=1-t$ and $Y^t=y$, given $(X,Z)=(x,z)$.
Dividing by $1-\pi_{z,P}^t(x)$ gives the missing-outcome
probability constrained by the MSM.
All constraints are linear in $w$.

\begin{proposition}[Binary bounds under joint exchangeability]
\label{prop:supp_binary_joint_bounds}
Suppose Assumption~\ref{ass:positivity} holds.
A full-data extension of $P$ satisfying Assumptions~\ref{ass:consist}, \ref{ass:iv_excl}, \ref{ass:abs_cont}, and \ref{ass:msm_iv_dual}, with the potential-outcome restrictions imposed for both $Y^1$ and $Y^0$ by treatment relabeling, together with \eqref{eq:supp_joint_exchangeability}, exists if and only if $\mathcal W_P(x)$ is nonempty for $P_X$-almost every $x$.

For a fixed function $c:\{0,1\}^2\to\mathbb R$, define
\[
\begin{aligned}
m_{c,P}^-(x)
&:=\min_{w\in\mathcal W_P(x)}
\sum_{u,v\in\{0,1\}}c(u,v)r_{uv}(x),\\
m_{c,P}^+(x)
&:=\max_{w\in\mathcal W_P(x)}
\sum_{u,v\in\{0,1\}}c(u,v)r_{uv}(x).
\end{aligned}
\]
When the model is compatible, the sharp identified interval for
$\E_Q\{c(Y^0,Y^1)\}$ is
\[
\left[
\E_P\{m_{c,P}^-(X)\},
\E_P\{m_{c,P}^+(X)\}
\right].
\]
Taking $c(u,v)=u$, $v$, and $v-u$ gives the bounds for
$\mu^0(Q)$, $\mu^1(Q)$, and the ATE, respectively.
\end{proposition}

The additional joint restriction can prevent separately
optimized potential-outcome means from being attained
simultaneously. The sharp ATE bounds must therefore be obtained
by optimizing $c(u,v)=v-u$ directly, rather than by subtracting
the separate mean bounds.
Likewise, intersecting the final IV-MSM and Balke--Pearl ATE
intervals need not give the sharp interval under their combined
restrictions.

Joint exchangeability need not tighten the bounds. For binary
outcomes, the following result identifies the sensitivity range
over which the additional restriction leaves identification
unchanged.

\begin{proposition}[Sensitivity threshold for additional tightening]
\label{prop:supp_joint_exchangeability_cutoff}
Suppose outcomes are binary and the IV-MSM restrictions apply
to both potential outcomes.
If $1\le\Lambda_0,\Lambda_1\le3$, adding
\eqref{eq:supp_joint_exchangeability} leaves the feasible pairs
of conditional potential-outcome marginal distributions
unchanged. Consequently, it leaves compatibility and the sharp
population bounds for $\mu^0$, $\mu^1$, and the ATE unchanged.

The constant three is sharp: for every common sensitivity
parameter $\Lambda>3$, there exists an observed law compatible
with the strengthened model for which joint exchangeability
strictly narrows the sharp ATE interval.
\end{proposition}

To illustrate the second assertion, fix any $\Lambda>3$ and
consider a population without covariates, with
$P(Z=0)=P(Z=1)=1/2$ and observed conditional probabilities
\[
\begin{array}{c|cccc}
& T=0,Y=0 & T=0,Y=1 & T=1,Y=0 & T=1,Y=1\\
\hline
Z=0 & d & \Lambda d & \Lambda d & d\\
Z=1 & \Lambda d & d & \Lambda d & d
\end{array}
\qquad
d=\frac{1}{2(\Lambda+1)}.
\]
At $\Lambda_0=\Lambda_1=\Lambda$, the original IV-MSM gives
the sharp ATE interval
\[
\left[-\frac{\Lambda-1}{2\Lambda},\,0\right],
\]
whereas adding joint exchangeability gives
\[
\left[
-\frac{\Lambda-1}{2\Lambda},
\,\frac{3-\Lambda}{2(\Lambda+1)}
\right].
\]
These expressions are derived in the proof of Proposition~\ref{prop:supp_joint_exchangeability_cutoff}.
The latter is strictly narrower and excludes zero for every
$\Lambda>3$.
For example, at $\Lambda=3.5$, the intervals are
$[-0.35714,\,0]$ and $[-0.35714,\,-0.05556]$, respectively.
Thus, three is the exact universal cutoff: tightening is
impossible at or below this value and possible above it,
although it need not occur for every observed law.

This extension concerns population identification.
Estimation and inference under the additional joint restriction
require a separate analysis.

%% file: supp_experiments.tex

\section{Additional numerical experiments}
\label{sec:supp_experiments}

We provide implementation details and additional results for the  main-text experiments, followed by a scalar-covariate experiment exploring a broader range of $(a,b)$ combinations.
We also examine smoothed bounds, finite ordered outcomes, and finite discrete instruments.

\subsection{Implementation and additional results for the main-text experiments}
\label{subsec:supp_main_experiments}

We use five-fold cross-fitting, with nuisance estimation and tuning performed within each outer-training sample.
Instrument and treatment probabilities are estimated using a Super Learner combining main-effect and interaction logistic models, penalized logistic regression, random forests, and gradient-boosted trees.
The Super Learner combines the candidate predictions using weights chosen to minimize the three-fold cross-validated negative Bernoulli log-likelihood.
For continuous outcomes, the conditional mean regressions entering the influence functions use linear models, generalized additive models, and random forests, with three-fold cross-validation and a one-standard-error selection rule.

The across-instrument density ratios $\gamma_t$ are estimated by regularized moment matching in a reproducing kernel Hilbert space, using Gaussian Nystr\"om features.
Positive predictions are normalized to have empirical mean one in the training reference stratum.
Standardization of $(Y,X)$ uses the outer-training sample only.
Let $M$ denote the median squared distance between standardized training observations and $n_t$ the training-sample size in treatment arm $t$.
For the continuous-outcome experiments reported here, we jointly select a bandwidth $c\sqrt M$, with $c\in\{0.25,0.5,1,2,4\}$, and a penalty $\lambda=C/\sqrt{n_t}$, with
$C\in\{10^{-1.5},10^{-1},10^{-0.5},1,10^{0.5},10\}$,
by minimizing the three-fold cross-validated Riesz loss.
For binary outcomes, the kernel variance is fixed at $M/2$, and $C$ is selected by averaging the Riesz validation loss over two independent three-fold partitions.
The alternative spline/Riesz learner selects the complexity of a growing sparse tensor-spline basis and its regularization parameter using two independent three-fold partitions.

For continuous outcomes, the main threshold class consists of additive natural splines with interior knots $(-1.2,-0.4,0.4,1.2)$ and boundary knots $(-2,2)$ for each covariate, together with an intercept.
Thresholds are fitted by weighted check-loss optimization.
The alternative class uses 60 fixed radial-basis functions with bandwidth 2 and an intercept.
Each class defines its own restricted-class population bounds.
For binary outcomes, we estimate the baseline conditional success probabilities and combine them with the fitted $\gamma_t$ values to obtain the comparison-stratum probabilities, then evaluate the explicit bound formulas.

Population references are computed from the known generating law by numerical integration and, for continuous IV-MSM bounds, optimization over the specified threshold class.
Centered coverage is the proportion of replications in which a two-sided confidence interval contains its corresponding population bound.
Outward coverage is the proportion in which the confidence set contains the entire method-specific population interval.
Error bars in the coverage plots are exact 95\% binomial Monte Carlo intervals.

\begin{figure}[!htb]
\centering
\includegraphics[width=\linewidth]{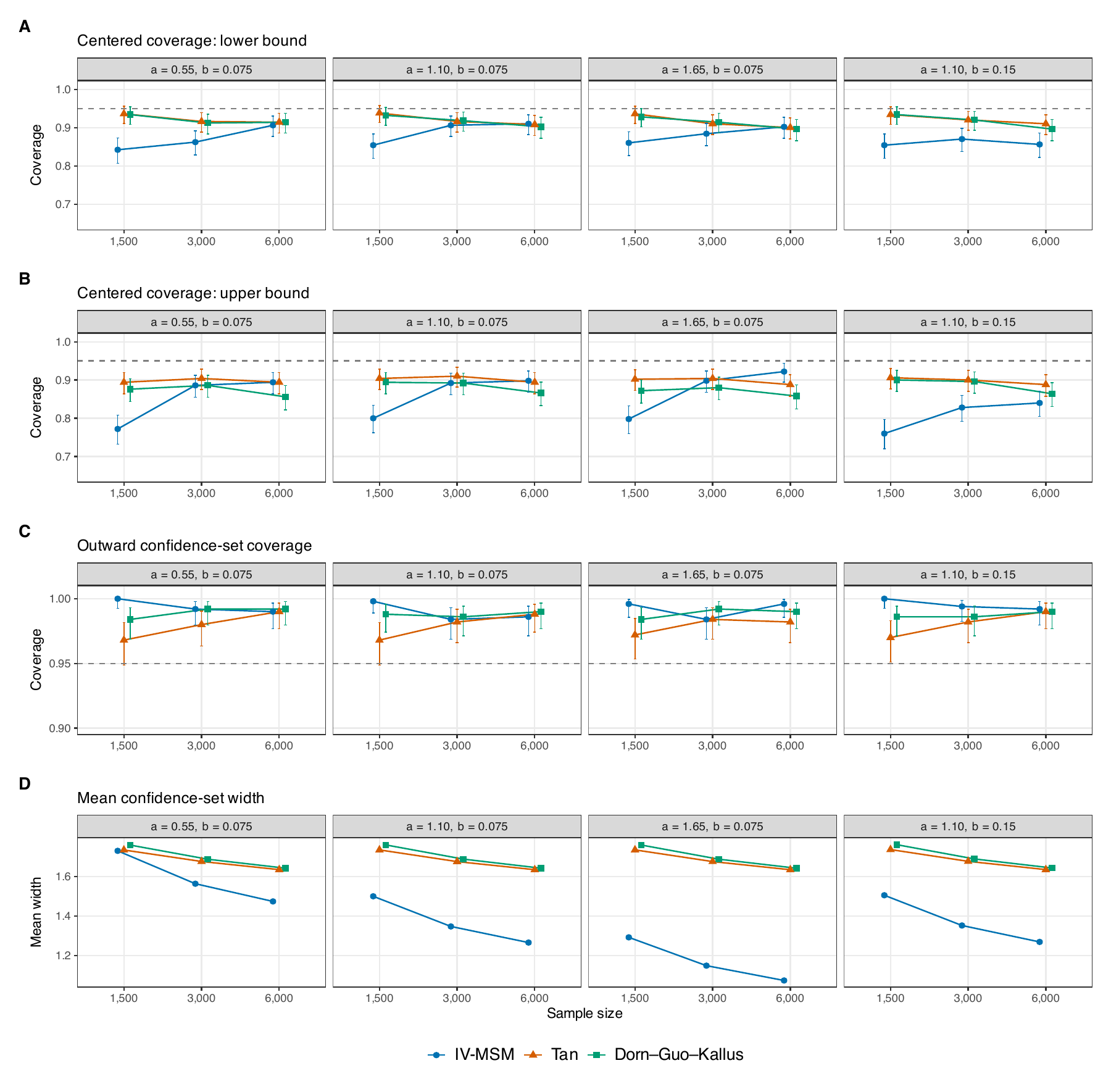}
\caption{Continuous-outcome results across sample sizes at $\Lambda=3$,
with 500 replications per sample size and $(a,b)$ combination.
Columns identify $(a,b)$ settings.
Panels show centered coverage for the lower (A) and upper (B) ATE bounds,
outward coverage (C), and mean confidence-set width (D).
The IV-MSM uses the RKHS/Riesz learner and natural-spline threshold class;
conventional-MSM procedures adjust for $X$.
Coverage is evaluated against method-specific population bounds.
Error bars show exact 95\% binomial Monte Carlo intervals;
horizontal dotted lines mark 0.95.}
\label{fig:supp_stress_continuous_sample_size}
\end{figure}

\begin{figure}[!htb]
\centering
\includegraphics[width=\linewidth]{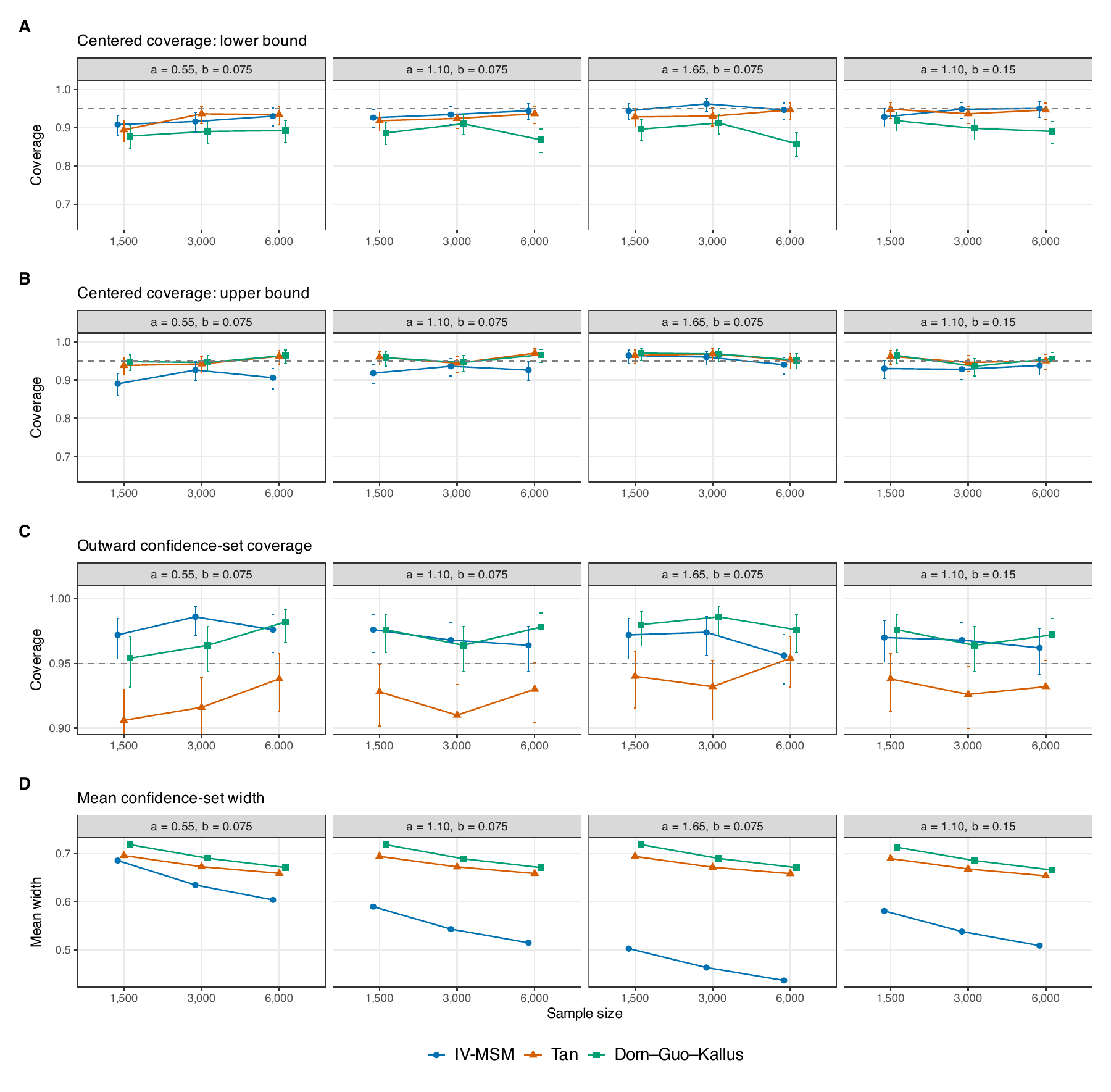}
\caption{Binary-outcome results across sample sizes at $\Lambda=3$,
with 500 replications per sample size and $(a,b)$ combination.
Columns identify $(a,b)$ settings.
Panels show centered coverage for the lower (A) and upper (B) ATE bounds,
outward coverage (C), and mean confidence-set width (D).
The IV-MSM uses the RKHS/Riesz learner;
conventional-MSM procedures adjust for $X$.
Coverage is evaluated against method-specific population bounds.
Error bars show exact 95\% binomial Monte Carlo intervals;
horizontal dotted lines mark 0.95.}
\label{fig:supp_stress_binary_sample_size}
\end{figure}

Figures~\ref{fig:supp_stress_continuous_sample_size}
and~\ref{fig:supp_stress_binary_sample_size} compare the IV-MSM with the conventional-MSM procedures adjusting for $X$ at
$n\in\{1500,3000,6000\}$, using 500 replications for each outcome type, sample size, and $(a,b)$ combination.
We fix $\Lambda=3$, which is compatible in all four settings.
Increasing sample size reduces mean confidence-set widths, although centered coverage does not improve uniformly.
For continuous outcomes, IV-MSM upper-bound coverage improves relative to $n=1500$ but remains below the nominal level.
IV-MSM outward coverage is generally conservative, and its confidence sets are narrower than those of the conventional-MSM procedures.

\begin{figure}[!htb]
\centering
\includegraphics[width=\linewidth]{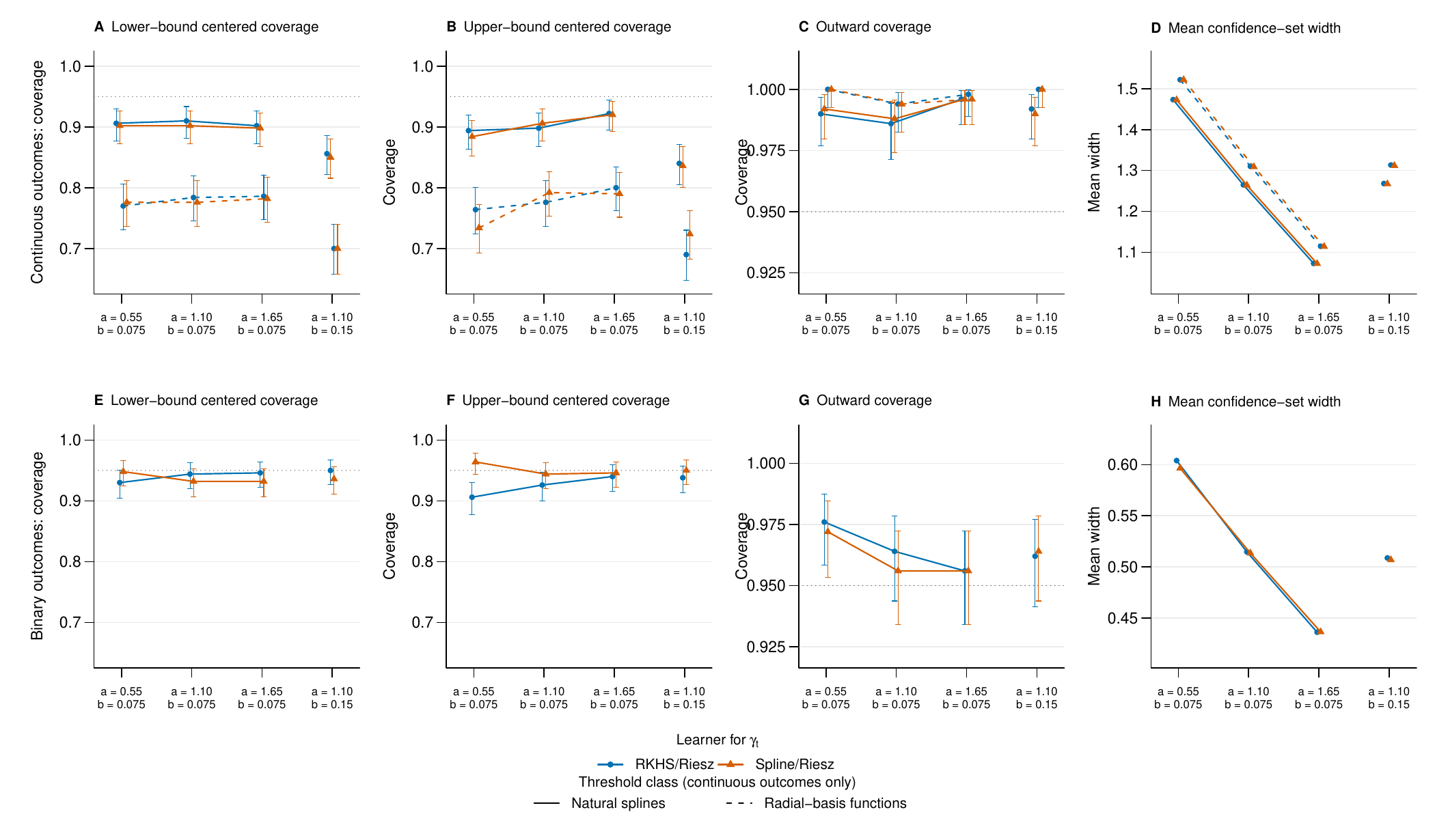}
\caption{Learner and threshold-class comparisons at $n=6000$ and
$\Lambda=3$, with 500 replications per outcome type and $(a,b)$ combination.
Rows distinguish continuous (top) and binary (bottom) outcomes.
Columns show centered coverage for the lower and upper ATE bounds,
outward coverage, and mean confidence-set width.
Colors and symbols distinguish learners for $\gamma_t$.
For continuous outcomes, solid and dashed lines distinguish
natural-spline and radial-basis function threshold classes,
each evaluated against its own population bounds.
Lines connect settings with $b=0.075$.
Error bars show exact 95\% binomial Monte Carlo intervals;
horizontal dotted lines mark 0.95.}
\label{fig:supp_stress_implementations}
\end{figure}

Figure~\ref{fig:supp_stress_implementations} compares the learners for $\gamma_t$ at $n=6000$ and $\Lambda=3$ across all four settings, including both threshold classes for continuous outcomes.
Holding the continuous threshold class fixed, the learners give similar coverage and mean widths.
The radial-basis function class yields lower centered coverage and wider confidence sets than the natural-spline class, despite high outward coverage.
For binary outcomes, the spline/Riesz learner generally improves upper-bound centered coverage, particularly at $(a,b)=(0.55,0.075)$; outward coverage and mean widths are similar for the two learners.
Changing the learner for $\gamma_t$ leaves the population target unchanged, whereas changing the threshold class changes the restricted-class population bounds.

\begin{figure}[!htb]
\centering
\includegraphics[width=\linewidth]{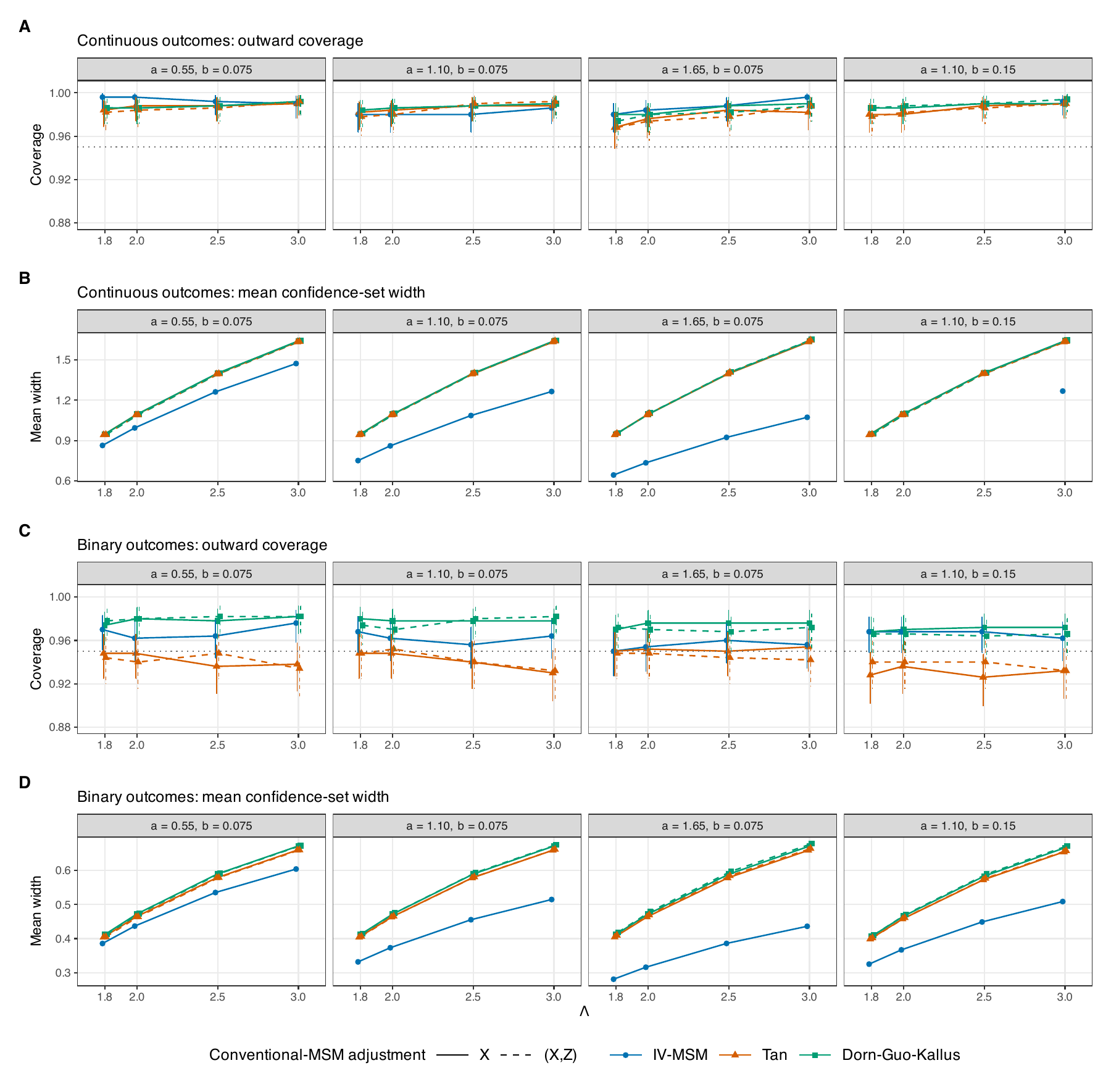}
\caption{Comparison with conventional-MSM procedures adjusting for $X$ and $(X,Z)$,
with $n=6000$ and 500 replications per outcome type and $(a,b)$ combination.
Columns identify $(a,b)$ settings.
Panels show outward coverage and mean confidence-set width for
continuous (A--B) and binary (C--D) outcomes.
For conventional-MSM procedures, solid and dashed lines indicate
adjustment for $X$ and $(X,Z)$, respectively.
Coverage is evaluated against method-specific population intervals;
incompatible IV-MSM combinations are omitted.
Error bars show exact 95\% binomial Monte Carlo intervals;
horizontal dotted lines mark 0.95.}
\label{fig:supp_stress_adjustment_sets}
\end{figure}

Figure~\ref{fig:supp_stress_adjustment_sets} compares conventional-MSM procedures adjusting for $X$ or $(X,Z)$, together with the IV-MSM, at $n=6000$ across the evaluated sensitivity values.
Additional adjustment for $Z$ changes outward coverage relatively little and slightly increases mean confidence-set widths.
The IV-MSM confidence sets remain narrower in every compatible comparison.
The conventional MSM adjusting for $(X,Z)$ imposes the same stratum-specific sensitivity restrictions as the IV-MSM but does not impose the IV assumptions.
At the population level, adding these assumptions therefore yields a sharp identified interval contained in that of the conventional MSM adjusting for $(X,Z)$.

\subsection{Scalar-covariate experiment}
\label{subsec:supp_scalar_full}

We use the same form of observed-data specification as in the main continuous-outcome experiment, but with a scalar covariate and $g_T(x)=0.5x$, $g_Y(x)=0.65x$, and $h(x)=1$.
For each $i=1,\dots,n$, we generate $X_i$ from a standard-normal distribution truncated to $[-2,2]$ and independently generate $Z_i\sim\operatorname{Bernoulli}(1/2)$.
Treatment and outcomes are then generated as in Section~\ref{sec:numerical_experiments} in the main text, using these functions.
We consider all combinations of $a\in\{0,0.55,1.10,1.65\}$ and $b\in\{0,0.025,0.075,0.15\}$, giving 16 observed-data laws.
For each design, we use 500 replications of $n=6000$ observations.

We evaluate the bound estimators for $\Lambda_0=\Lambda_1=\Lambda\in\{1.8,2,2.5,3,3.5\}$.
All these design--sensitivity combinations are compatible with the IV-MSM.
The setting $a=b=0$ provides an exact-envelope-tie benchmark, outside the no-tie conditions of the general inference results.
Our one-step estimator uses five-fold cross-fitting, a fixed natural-spline threshold class, and a direct spline/Riesz learner for $\gamma_t$.
The conventional-MSM comparisons adjust for either $X$ or $(X,Z)$.

\begin{landscape}
\begin{figure}
\centering
\includegraphics[width=\linewidth,height=.80\textheight,keepaspectratio]
{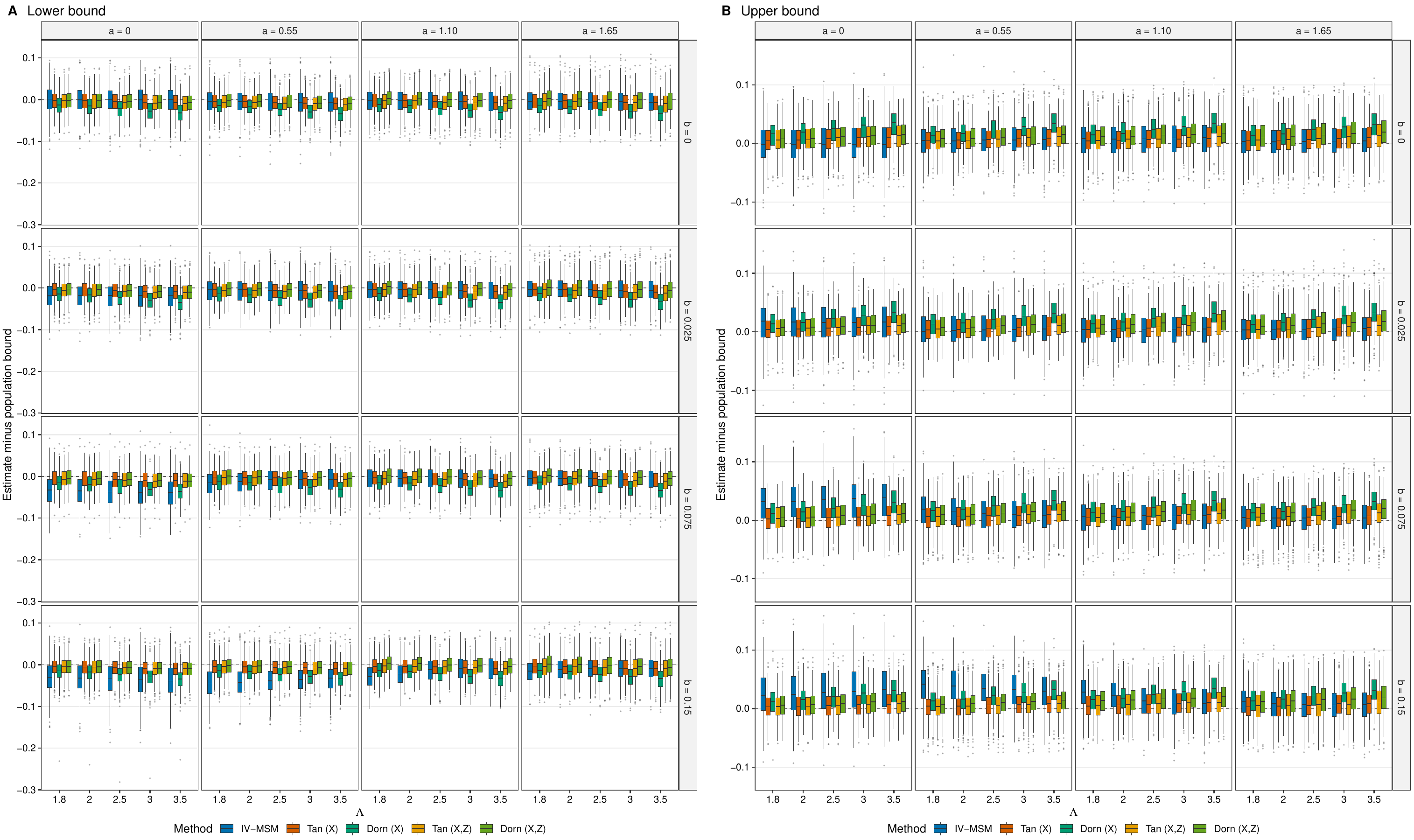}
\caption{ATE Bound estimation errors in the scalar-covariate experiment:
lower bound (A) and upper bound (B). Errors are relative to each method's population target. Facets index all 16 $(a,b)$ combinations.
Sample size $n=6000$; 500 repetitions per setting, apart from the single
failed Tan $(X,Z)$ lower-bound fit described in the text.}
\label{fig:supp_scalar_errors}
\end{figure}
\end{landscape}

\begin{figure}
\centering
\includegraphics[width=\linewidth,height=.84\textheight,keepaspectratio]
{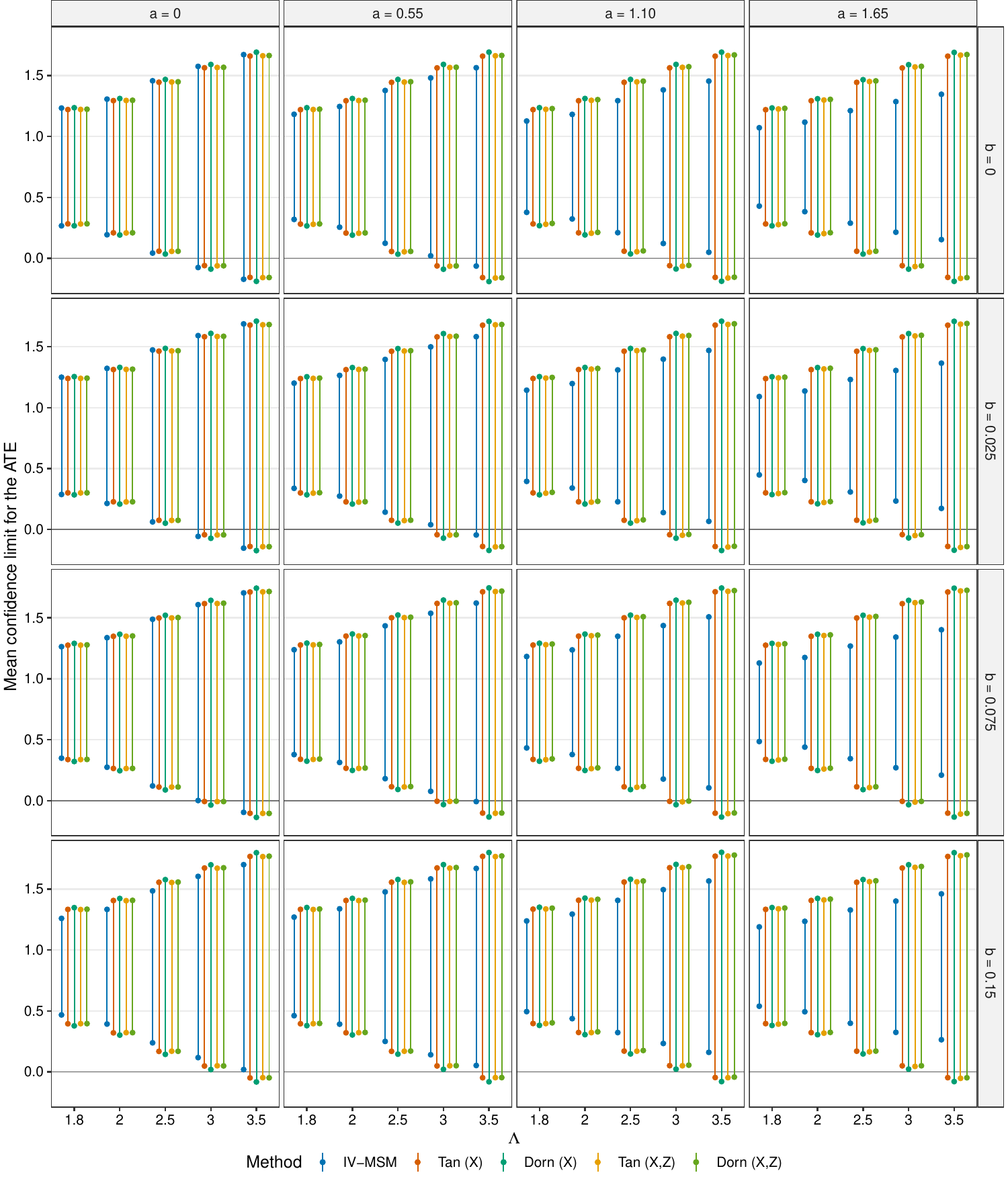}
\caption{Mean lower and upper limits of the 95\% outward confidence sets
in the scalar-covariate experiment. Each method is evaluated against its
own population interval. Sample size $n=6000$; 500 repetitions per setting,
with 499 complete intervals for the single Tan $(X,Z)$ failure setting.}
\label{fig:supp_scalar_mean_sets}
\end{figure}

\begin{landscape}
\begin{figure}
\centering
\includegraphics[width=\linewidth,height=.80\textheight,keepaspectratio]
{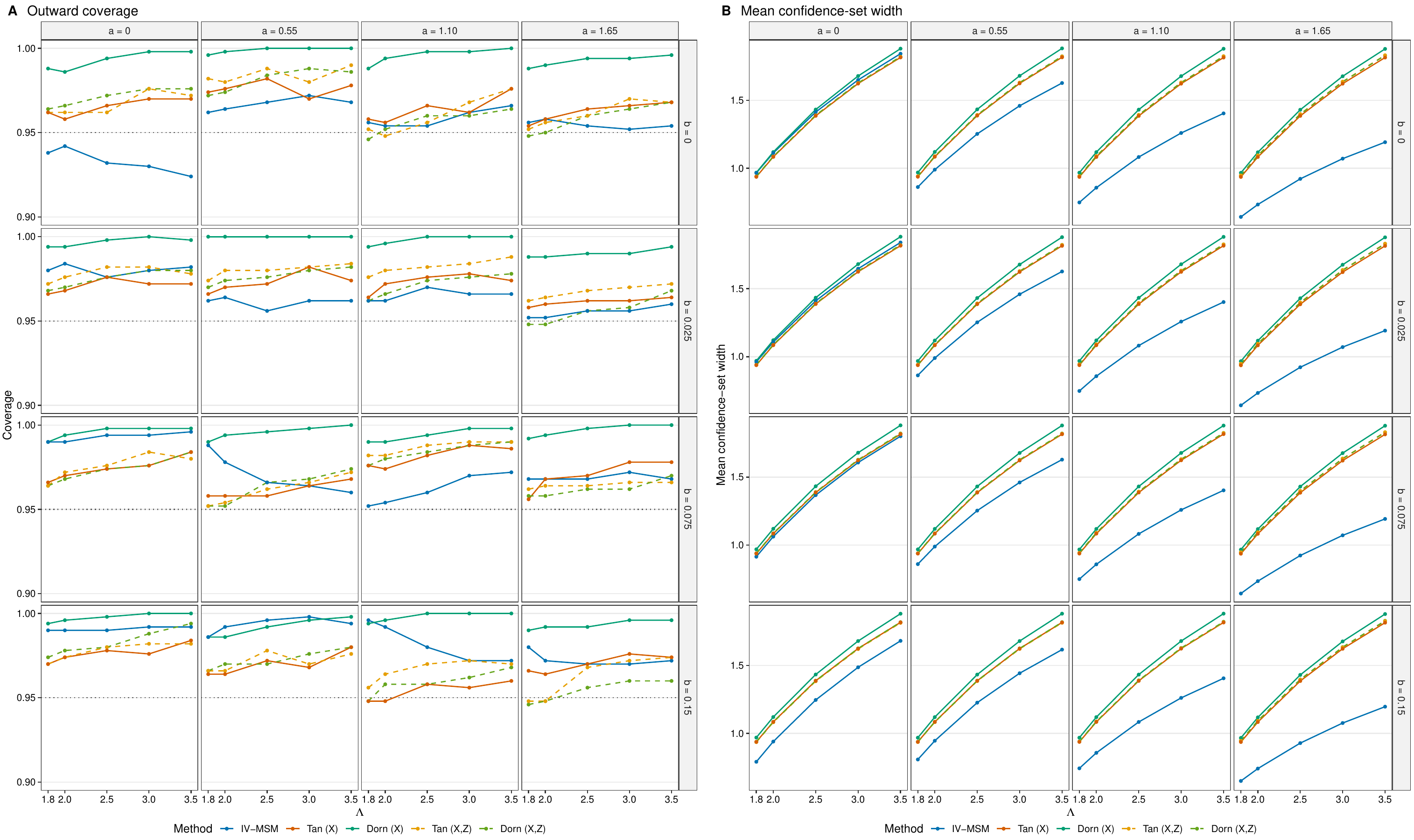}
\caption{Coverage (A) and mean width (B) of the 95\% outward
confidence sets in the scalar-covariate experiment. Coverage refers to
containment of the method-specific population interval at each sensitivity
value. Sample size $n=6000$; 500 repetitions per setting, with the exception
described in the text.}
\label{fig:supp_scalar_outward}
\end{figure}
\end{landscape}

Figures~\ref{fig:supp_scalar_errors}
and~\ref{fig:supp_scalar_mean_sets} show bound estimation
errors and mean outward confidence limits, respectively.
Increasing IV relevance generally narrows the IV-MSM confidence
sets relative to the conventional-MSM procedures
(Figure~\ref{fig:supp_scalar_outward}).
Relative to Tan's procedure adjusting for $X$, mean widths
are approximately 8--14\%, 20--23\%, and 31--34\% smaller
at $a=0.55$, $1.10$, and $1.65$, respectively.
These comparisons involve method-specific population targets.

\begin{landscape}
\begin{figure}
\centering
\includegraphics[width=\linewidth,height=.80\textheight,keepaspectratio]
{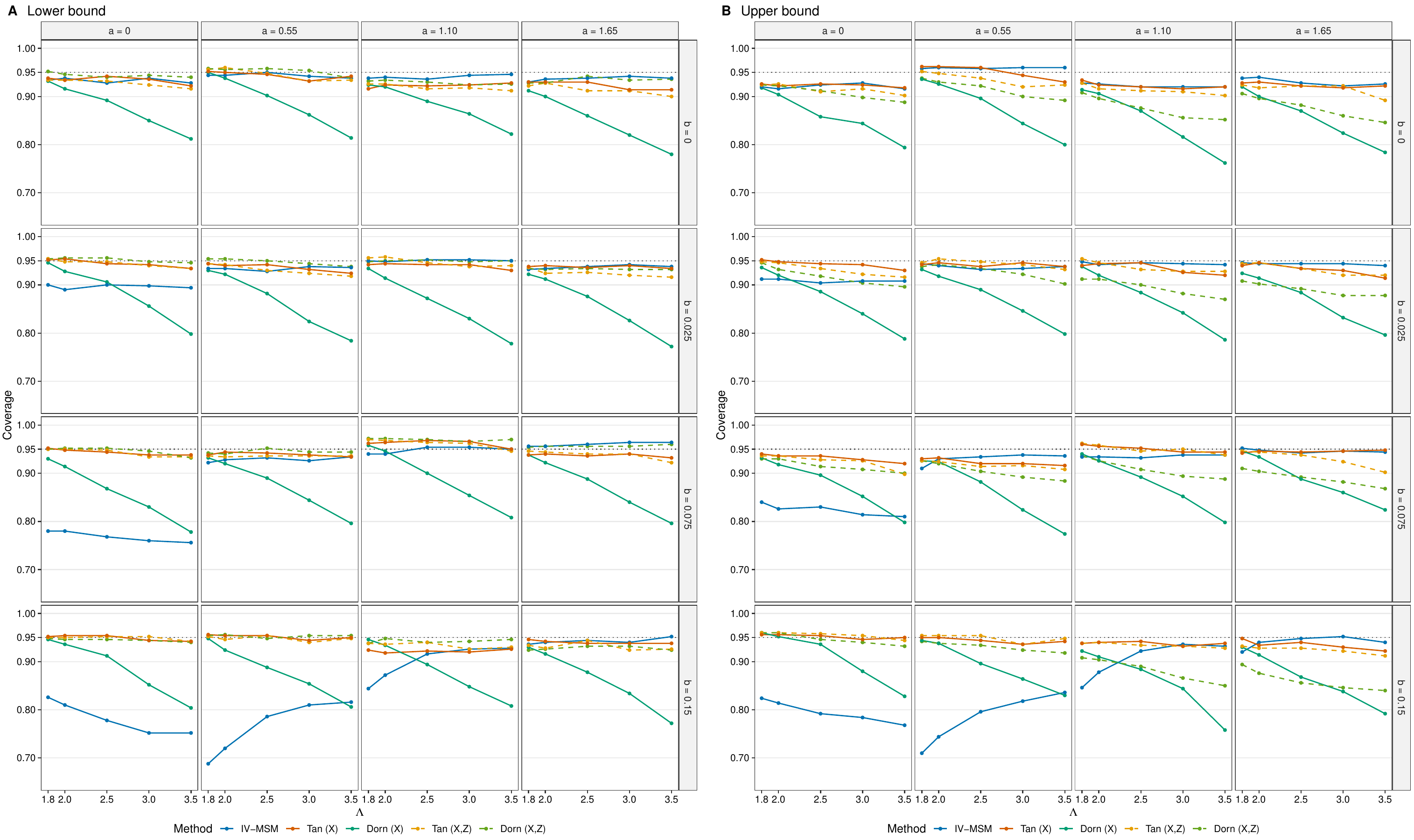}
\caption{Coverage of centered 95\% confidence intervals for the lower
(A) and upper (B) population ATE bounds in the scalar-covariate
experiment. Sample size $n=6000$; 500 repetitions per setting, apart
from the single failed lower-bound fit described in the text.
Targets are method-specific.}
\label{fig:supp_scalar_centered}
\end{figure}
\end{landscape}

Across the evaluated combinations, IV-MSM outward coverage
ranges from 92.4\% to 99.8\%, with the lowest value occurring
in the exact-tie setting at $\Lambda=3.5$.
Centered coverage ranges from 68.8\% to 96.4\% for the lower
bound and from 71.0\% to 96.0\% for the upper bound
(Figure~\ref{fig:supp_scalar_centered}).
Thus, good outward coverage can coexist with substantially
lower centered coverage.

\begin{figure}
\centering
\includegraphics[width=\linewidth,height=.84\textheight,keepaspectratio]
{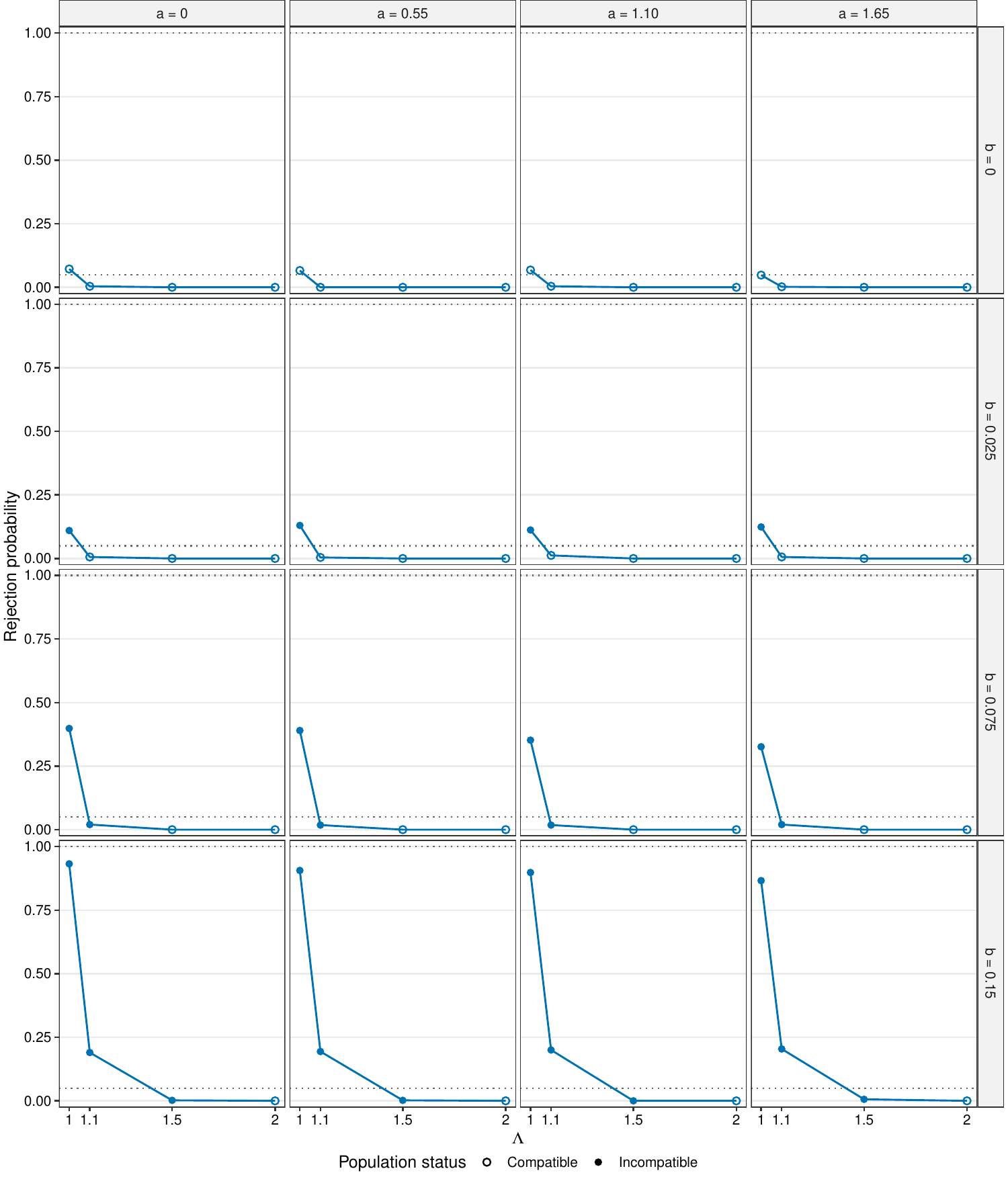}
\caption{Rejection rates of the nominal $5\%$ compatibility test in the
scalar-covariate experiment, using the spline-based density-ratio learner.
Sample size $n=6000$; 500 repetitions per setting. Open and filled circles
indicate population-compatible and incompatible specifications,
respectively. The dotted and dashed lines mark $0.05$ and $1$.}
\label{fig:supp_scalar_compatibility}
\end{figure}

\begin{figure}[!htb]
\centering
\includegraphics[width=\linewidth]
{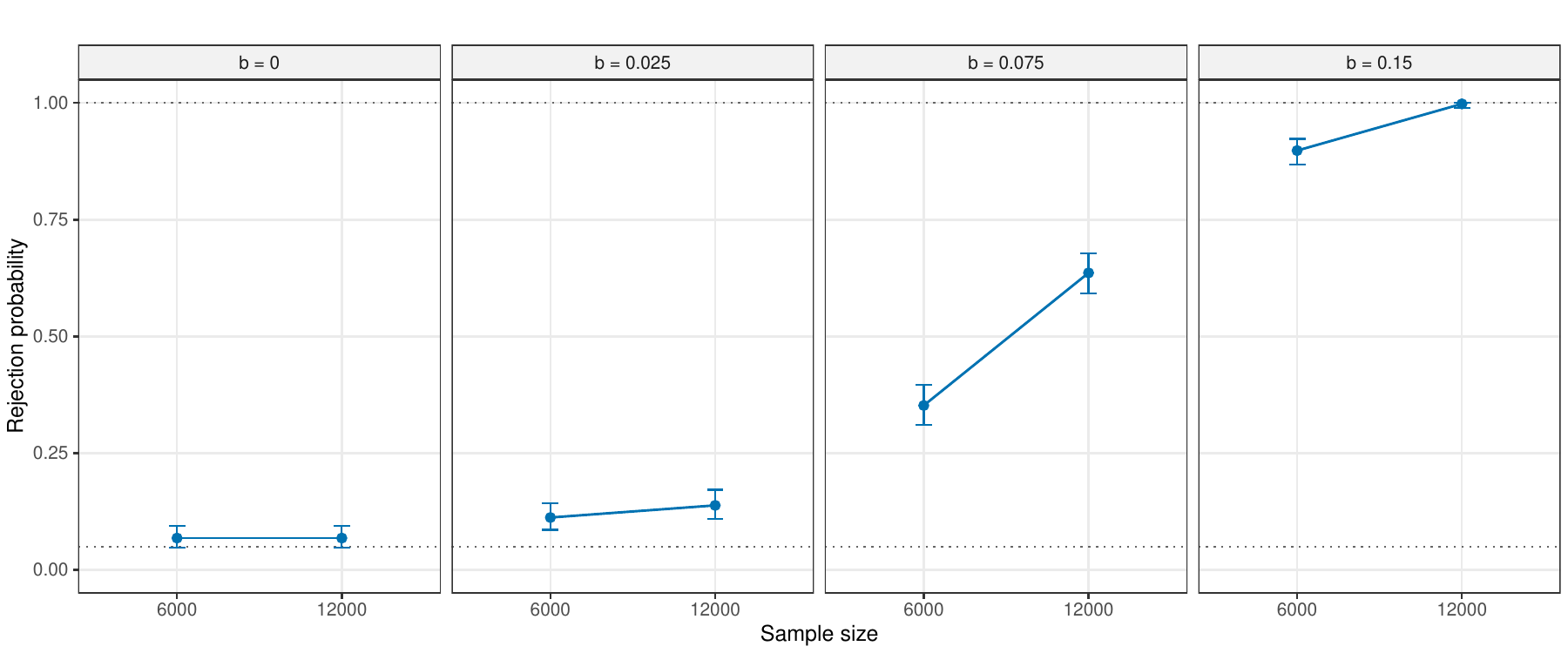}
\caption{Rejection rates at $n=6000$ and $12000$, restricted to
$a=1.10$ and $\Lambda=1$, using the spline-based density-ratio learner.
Each setting uses 500 repetitions; error bars are exact $95\%$ binomial
Monte Carlo confidence intervals. The specification is compatible when
$b=0$ and incompatible otherwise. The dotted and dashed lines mark
$0.05$ and $1$.}
\label{fig:supp_scalar_compatibility_sample_size}
\end{figure}

We evaluate the compatibility test at $\Lambda\in\{1,1.1,1.5,2\}$ (Figure~\ref{fig:supp_scalar_compatibility}).
At $(a,b)=(1.10,0.15)$, the smallest compatible value is
approximately $\Lambda=1.759$, but rejection rates at
$\Lambda=1,1.1,1.5$ are 89.8\%, 20.0\%,
and 0\%, respectively.
We also compare sample sizes $n=6000$ and $12000$ at
$a=1.10$ and $\Lambda=1$, using 500 replications per setting
(Figure~\ref{fig:supp_scalar_compatibility_sample_size}).
Power increases from 35.2\% to 63.6\% at $b=0.075$
and from 89.8\% to 99.8\% at $b=0.15$, but remains low at $b=0.025$.

\subsection{Smoothed bounds}
\label{subsec:supp_smoothed_experiments}

We examine smoothing in a design with envelope ties on a set of positive probability.
For each $i=1,\dots,n$, we generate $X_i\in\{1,\ldots,7\}$ with $P(X_i=x)=1/9$ for $x=1,\ldots,6$ and $P(X_i=7)=1/3$, and independently generate $Z_i\sim\operatorname{Bernoulli}(1/2)$.
Conditional on $(X_i,Z_i)$, we generate $T_i\sim\operatorname{Bernoulli}(p_i)$, where
\[
p_i=
\begin{cases}
0.9,& X_i=7,\ Z_i=1,\\
0.1,& \text{otherwise}.
\end{cases}
\]
We set $Y_i=0.75T_i+R_i$, where
$R_i\sim\operatorname{Unif}[0,1]$ independently of $(X_i,Z_i,T_i)$.
Thus $\gamma_t(y,x)=1$ for both treatment arms.

We use $\Lambda_0=\Lambda_1=2$, $n=6000$, and 500 replications.
In the first six strata, the treatment probabilities are equal across instrument strata, so $\Gamma_{1/2,P}(y,x)=1/2$ and $\Gamma_{2,P}(y,x)=2$.
The two arguments defining $L_P$ therefore coincide, as do those defining $U_P$.
These strata together contain two-thirds of the population.

The threshold class contains all functions on the seven strata.
We compare $\varepsilon\in\{0,0.01,0.02\}$, using the same observations, folds, and fitted nuisances across smoothing values.
The primary targets are the lower and upper bounds for $\mu^1$.
Each procedure is evaluated against its corresponding population bounds, accounting for the change in target induced by smoothing.

\begin{figure}[!htb]
\centering
\includegraphics[width=\linewidth]{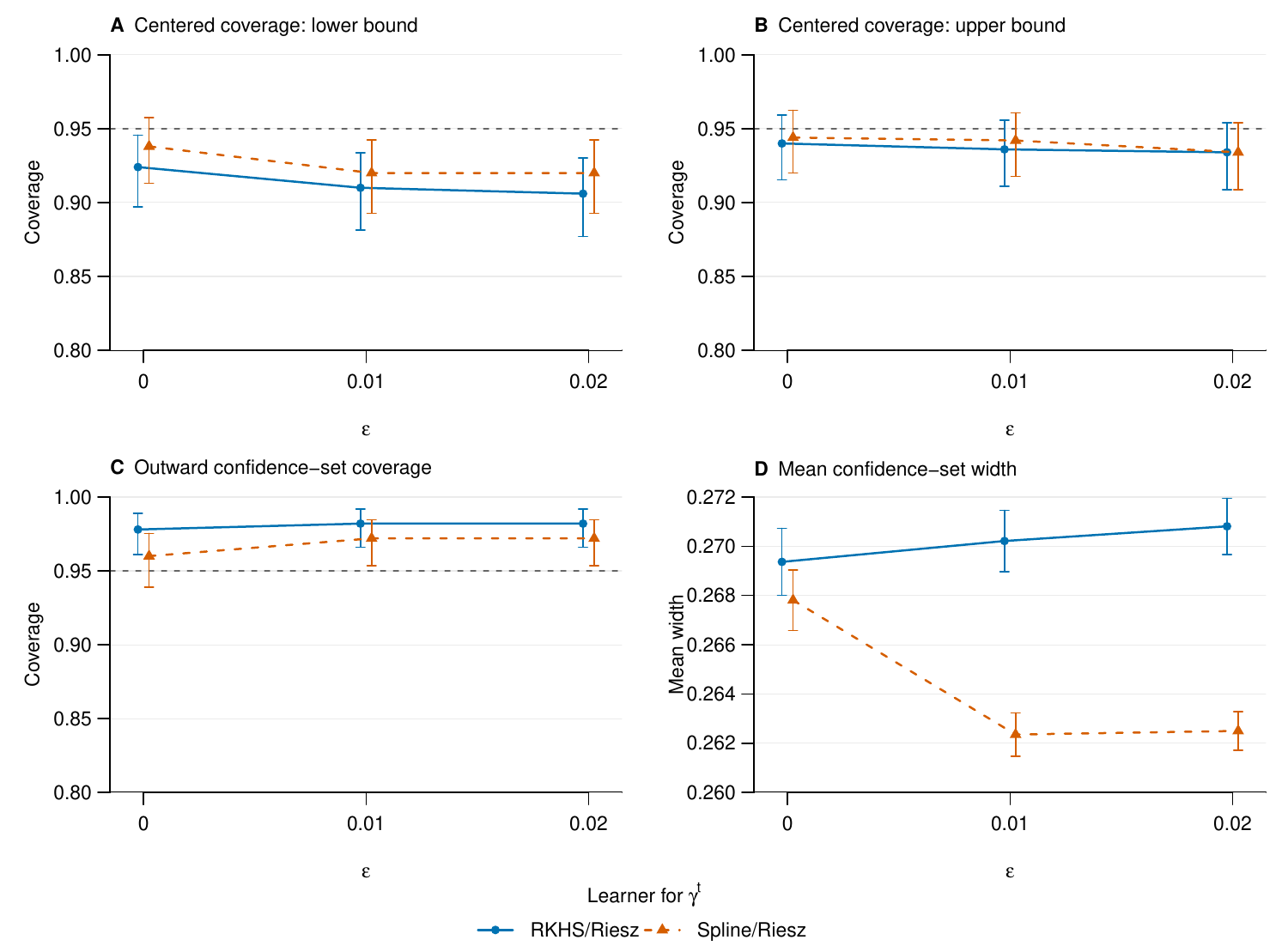}
\caption{Smoothed-bound results for $\mu^1$, with $n=6000$,
$\Lambda=2$, and 500 replications.
Panels show centered coverage for the lower (A) and upper (B)
bounds, outward coverage (C), and mean confidence-set width (D).
Colors and symbols distinguish learners for $\gamma_t$.
Coverage is evaluated against the corresponding population
bounds at each $\varepsilon$.
Error bars show 95\% Monte Carlo intervals; horizontal reference
lines mark 0.95.}
\label{fig:supp_smoothed_results}
\end{figure}

Figure~\ref{fig:supp_smoothed_results} shows centered coverage for the lower and upper bounds, outward coverage, and mean confidence-set width across smoothing values.
Outward coverage remains above 95\% for both learners, ranging from 97.8--98.2\% for RKHS/Riesz and 96.0--97.2\% for spline/Riesz.
Smoothing does not improve centered coverage: lower-bound coverage decreases from 92.4\% to 90.6\% for RKHS/Riesz and from 93.8\% to 92.0\% for spline/Riesz, while upper-bound coverage changes relatively little.
Mean confidence-set widths also change little, increasing slightly for RKHS/Riesz and decreasing slightly for spline/Riesz.
Thus, removing envelope ties does not by itself improve centered coverage in this experiment.

\subsection{Finite ordered outcomes}
\label{subsec:supp_finite_outcome_experiments}

For each $i=1,\dots,n$, we generate $(X_i,Z_i,T_i)$ as in the main experiment, with $a=1.10$.
For $K\in\{5,10,20\}$, define $y_k=-2+4(k-1)/(K-1)$, $k=1,\ldots,K$.
Conditional on $(X_i,Z_i,T_i)$, we generate $Y_i\in\{y_1,\ldots,y_K\}$ with probabilities
\[
P(Y_i=y_k\mid X_i,Z_i,T_i)
\propto
\exp\left[
-\frac{\{y_k-g_Y(X_i)-0.75T_i\}^2}{2}
+Z_i(2T_i-1)b h(X_i)y_k
\right],
\qquad b=0.04,
\]
where the probabilities are normalized to sum to one over $k=1,\ldots,K$.
These are distinct finite-support distributions rather than coarsenings of a common continuous outcome.
For each value of $K$, we use 500 replications of $n=6000$ observations and evaluate the bound estimators at $\Lambda_0=\Lambda_1=2$.

The one-step estimator uses the finite-support bound formulas in Section~\ref{subsec:supp_finite_outcomes}.
Within each treatment arm and the baseline instrument stratum, we estimate $P(Y=y_k\mid Y\ge y_k,T=t,X,Z=0)$, $k=1,\ldots,K-1$, using binary Super Learners and recover the category probabilities from these conditional probabilities.
We combine them with the RKHS or spline/Riesz density-ratio estimates to obtain the comparison-stratum probabilities.

\begin{figure}[!htb]
\centering
\includegraphics[width=\linewidth]
{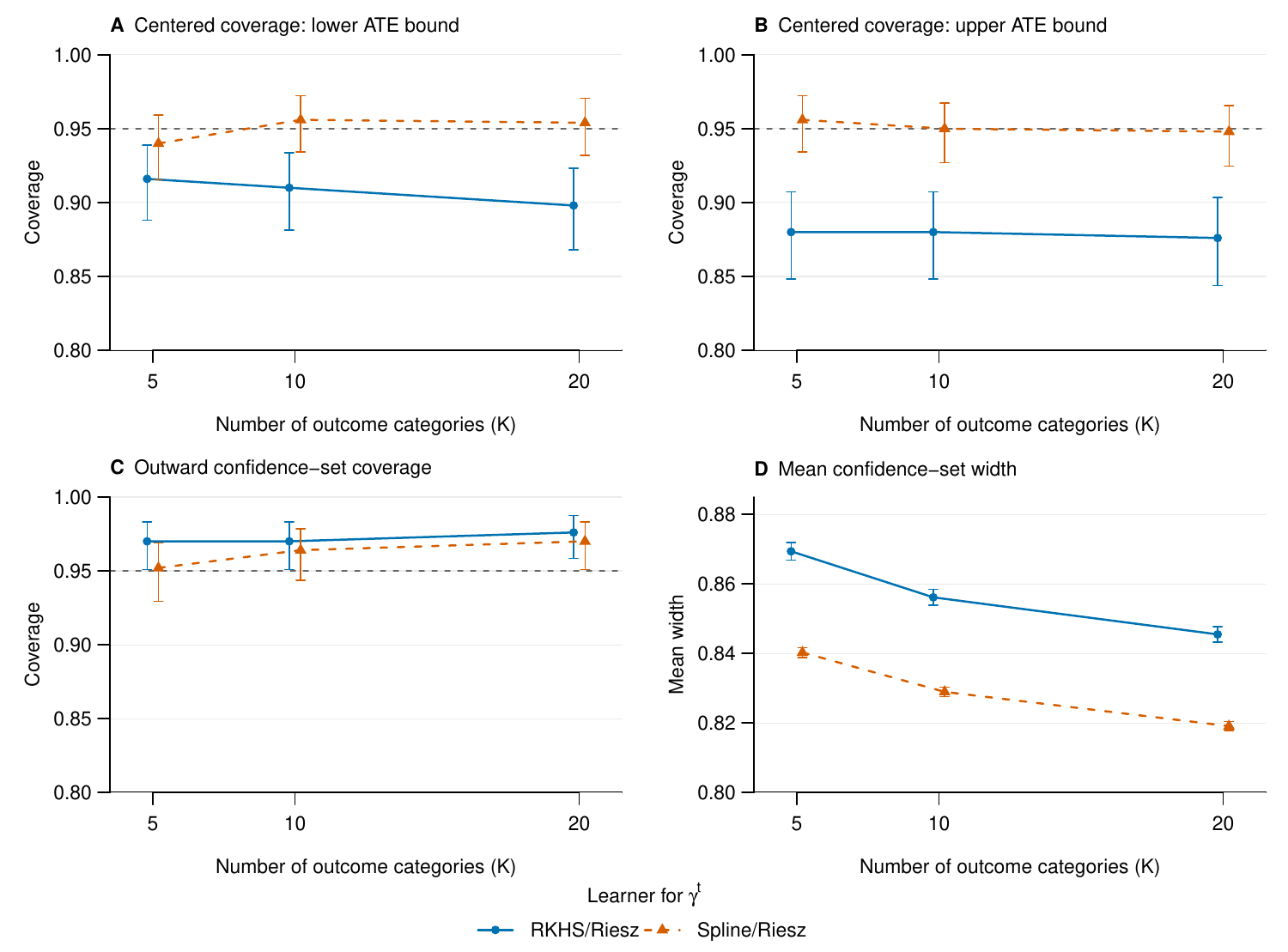}
\caption{Finite ordered-outcome results, with $n=6000$,
$\Lambda=2$, and 500 replications per setting.
Panels show centered coverage for the lower (A) and upper (B)
ATE bounds, outward coverage (C), and mean confidence-set width (D).
Colors and symbols distinguish learners for $\gamma_t$.
Each value of $K$ defines a distinct outcome distribution.
Error bars show 95\% Monte Carlo intervals; horizontal reference
lines mark 0.95.}
\label{fig:supp_ordered_outcomes_results}
\end{figure}

Figure~\ref{fig:supp_ordered_outcomes_results} summarizes centered coverage for the lower and upper ATE bounds, outward coverage, and mean confidence-set width.
The spline/Riesz learner yields centered coverage closer to the nominal level than RKHS/Riesz: coverage ranges from 94.0--95.6\% across both bounds, compared with 89.8--91.6\% for the lower bound and 87.6--88.0\% for the upper bound using RKHS/Riesz.
Both implementations attain outward coverage above 95\%.
The spline/Riesz implementation also produces slightly narrower confidence sets.
Coverage remains broadly stable as the number of outcome categories increases, with a slight decrease in lower-bound centered coverage for RKHS/Riesz.
Mean confidence-set widths decrease for both learners.
Because the outcome distribution changes with $K$, these comparisons do not isolate an effect of the number of categories.

\subsection{Finite discrete instruments}
\label{subsec:supp_finite_iv_experiments}

For each $i=1,\dots,n$, we generate the five-dimensional covariate vector $X_i$ as in the main experiment and independently generate $Z_i^\star$ uniformly on $\{1,\ldots,10\}$.
Writing $v_z=(z-1)/9$, we generate $T_i\mid X_i,Z_i^\star\sim\operatorname{Bernoulli}(p_i)$, where
\[
\log\frac{p_i}{1-p_i}
=g_T(X_i)+1.10(v_{Z_i^\star}-1/2).
\]
We set $Y_i=g_Y(X_i)+0.75T_i+R_i$, where, conditional on $(X_i,Z_i^\star,T_i)$, $R_i$ is drawn from $N\{v_{Z_i^\star}(2T_i-1)0.075h(X_i),1\}$ truncated to $[-2,2]$. For binary outcomes, we retain the same distribution of $(X_i,Z_i^\star,T_i,R_i)$ and set
\[
Y_i=I\{g_Y(X_i)+0.75T_i+R_i>0\}.
\]
Thus, the binary experiment dichotomizes the continuous outcome at zero.

We construct observed instruments with $J\in\{3,6,10\}$ categories by grouping the values of $Z_i^\star$.
For $J=3$, the groups are $\{1,2,3\}$, $\{4,5,6,7\}$, and $\{8,9,10\}$; for $J=6$, they are
$\{1,2\}$, $\{3\}$, $\{4,5\}$, $\{6,7\}$, $\{8\}$, and $\{9,10\}$.
For $J=10$, we retain $Z_i^\star$.
These groupings are nested, and the same generated observations are used for all three versions.
We use 500 replications of $n=6000$ observations and evaluate the finite-instrument one-step estimator at a common sensitivity parameter $\Lambda=2$ across instrument categories. For continuous outcomes, we use the fixed additive natural-spline threshold class from the main experiment; binary outcomes require no threshold estimation.
For both outcome types, we include an oracle-$\gamma_t$ comparison using the true across-instrument density ratios.
The oracle-primitive comparison additionally uses the true instrument and treatment probabilities and, for binary outcomes, the true conditional outcome probabilities.
For continuous outcomes, thresholds and auxiliary conditional
means remain estimated in both oracle comparisons.

\begin{figure}[!htb]
\centering
\includegraphics[width=\linewidth]{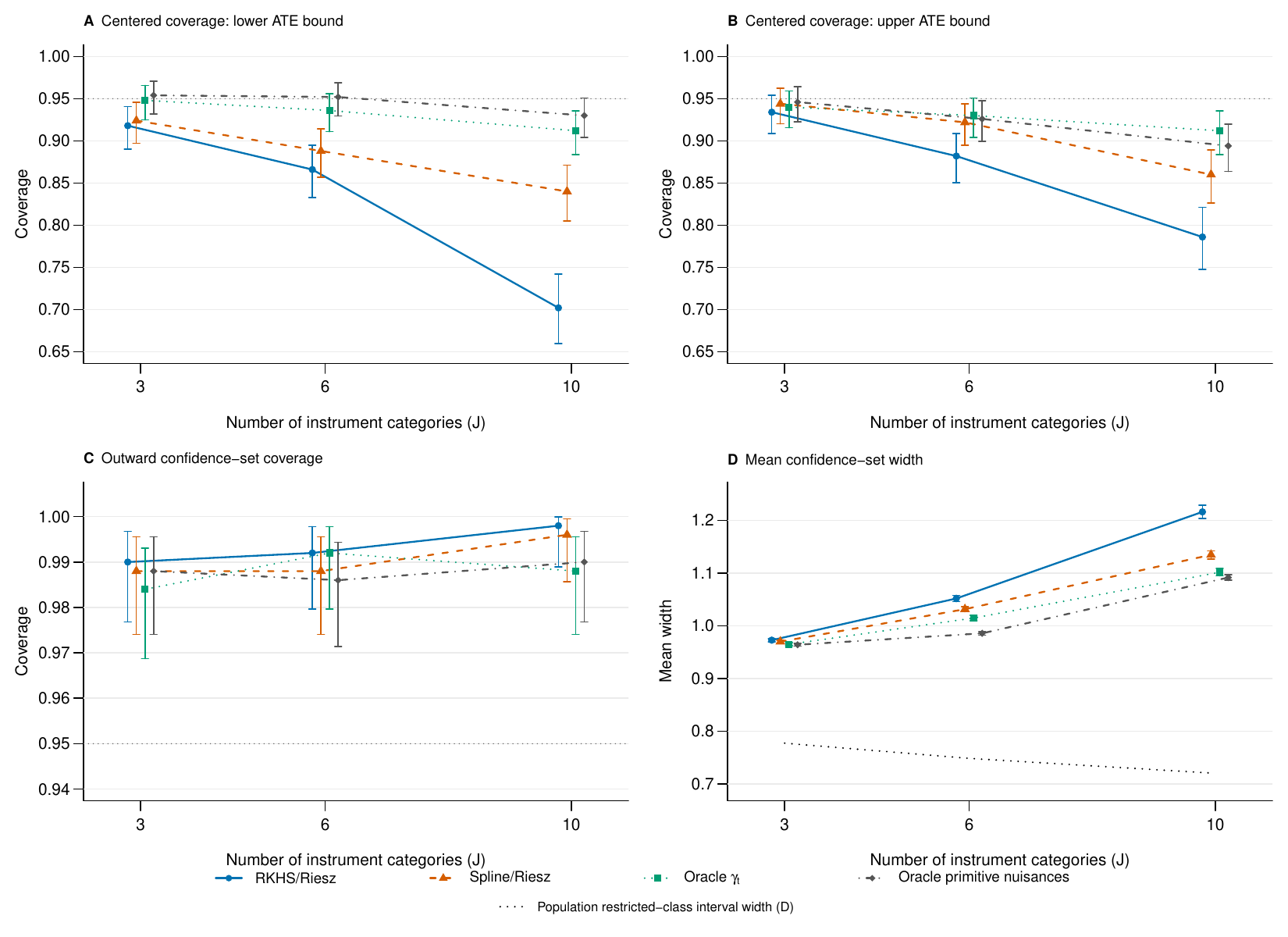}
\caption{Continuous-outcome results with finite discrete instruments,
with $n=6000$, $\Lambda=2$, and 500 replications per resolution.
Panels show centered coverage for the lower (A) and upper (B) ATE
bounds, outward coverage (C), and mean confidence-set width (D).
Colors and symbols distinguish the four implementations.
The black dotted curve in D shows the population restricted-class
interval width.
Error bars show 95\% Monte Carlo intervals; horizontal reference
lines mark 0.95.}
\label{fig:supp_finite_iv_results}
\end{figure}

\begin{figure}[!htb]
\centering
\includegraphics[width=\linewidth]{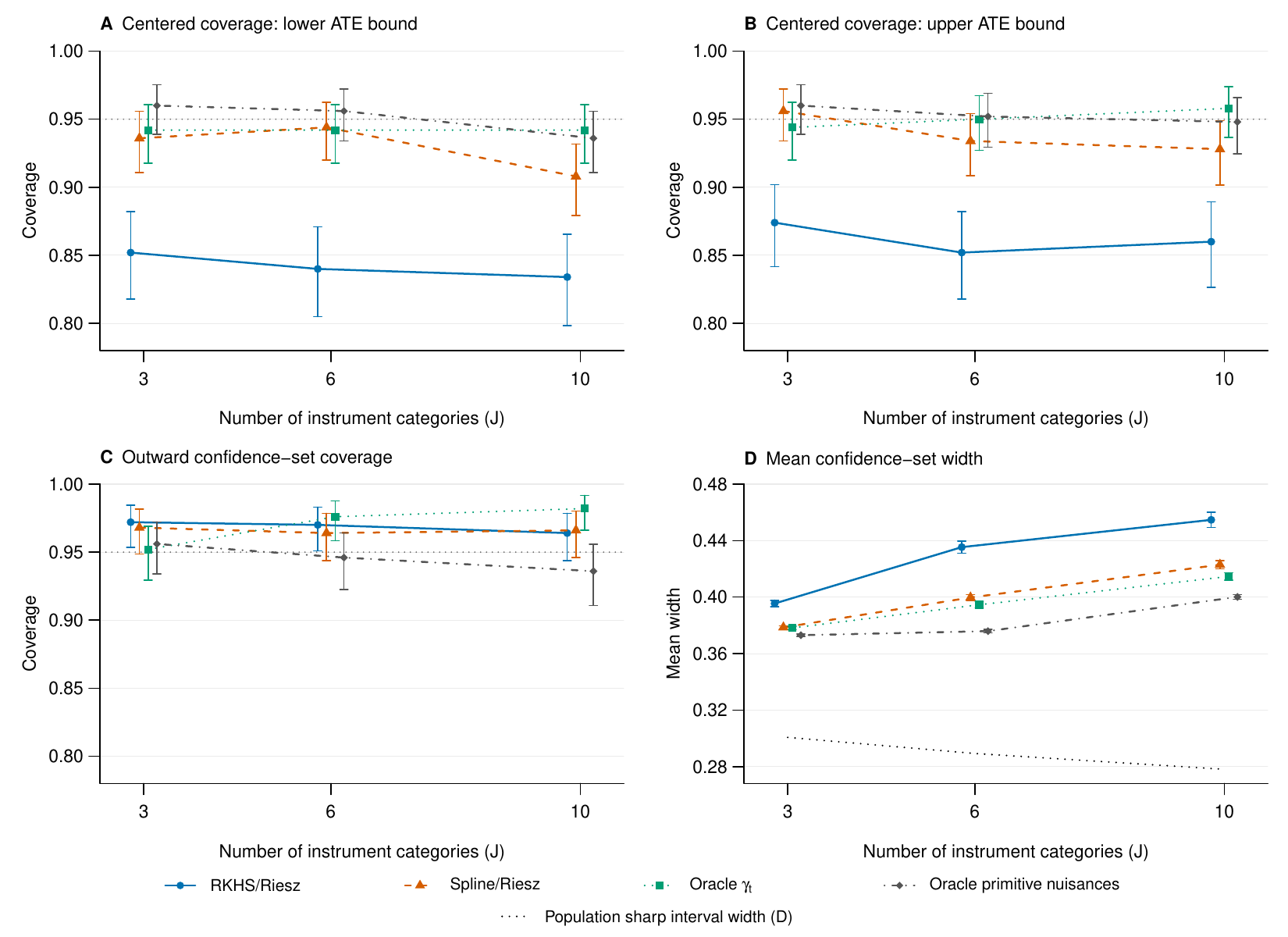}
\caption{Binary-outcome results with finite discrete instruments,
with $n=6000$, $\Lambda=2$, and 500 replications per resolution.
Panels show centered coverage for the lower (A) and upper (B) ATE
bounds, outward coverage (C), and mean confidence-set width (D).
Colors and symbols distinguish the four implementations.
The black dotted curve in D shows the population interval width.
Error bars show 95\% Monte Carlo intervals; horizontal reference
lines mark 0.95.}
\label{fig:supp_finite_iv_binary_results}
\end{figure}

Figures~\ref{fig:supp_finite_iv_results}
and~\ref{fig:supp_finite_iv_binary_results} show centered bound
coverage, outward coverage, and mean confidence-set width for
continuous and binary outcomes, respectively. For continuous outcomes, centered coverage deteriorates as $J$ increases, particularly for RKHS/Riesz, while outward coverage remains conservative.
The oracle comparisons substantially improve centered coverage, but some undercoverage persists even with the true instrument and treatment probabilities and across-instrument density ratios.
Thus, estimation of $\gamma_t$ contributes to, but does not fully
explain, the deterioration.

For binary outcomes, spline/Riesz gives centered coverage closer to nominal and narrower confidence sets than RKHS/Riesz.
Both oracle implementations yield centered coverage near 95\% across instrument resolutions.
Outward coverage remains near or above nominal, except for modest undercoverage in some all-oracle comparisons.

For both outcome types, the population intervals narrow as instrument resolution increases, but mean confidence-set widths increase, including in the oracle comparisons.
Thus, additional identifying information does not necessarily translate into narrower estimated confidence sets at a fixed sample size.

%% file: supp_application.tex
\section{Oregon Health Insurance Experiment: implementation and additional results}
\label{sec:supp_oregon_application}

\subsection{Implementation}

The covariate vector contains centered age, sex, prior emergency-department use, a two-person-household indicator, and seven lottery-draw indicators.
We use five-fold cross-fitting, with nuisance estimation and tuning performed within each training sample.

Instrument and treatment probabilities and baseline conditional outcome probabilities are estimated using a Super Learner combining main-effect and interaction logistic models, penalized logistic regression, random forests, gradient-boosted trees, generalized additive models, and highly adaptive lasso.
Combination weights minimize the three-fold cross-validated negative Bernoulli log-likelihood.
The application library therefore adds generalized additive models and highly adaptive lasso to the library used in the numerical experiments.
Instrument and treatment probabilities are truncated to $[0.05,0.95]$, and outcome probabilities to $[10^{-4},1-10^{-4}]$.

We estimate $\gamma_t$ using the same RKHS procedure and tuning rule as in the binary-outcome numerical experiments (Section~\ref{subsec:supp_main_experiments}).
Baseline conditional outcome probabilities and the fitted $\gamma_t$ values are used to reconstruct the
comparison-stratum probabilities.

For the compatibility test, witness functions are learned on each training sample and evaluated on the corresponding held-out observations.
We calibrate the test using 999 Gaussian multiplier draws, without adjusting the nuisance estimates to
satisfy the restrictions being tested.
For bound estimation, we adjust fitted probabilities when necessary to satisfy the compatibility restrictions.

We calibrate the outward bands on the 81-point grid
$\{1.20,1.21,\ldots,2.00\}$ using 200,000 Gaussian draws and the estimated joint covariance of the centered score contributions.
The bands are extended between grid points using Section~\ref{sec:supp_binary_bands}.
Calibration is performed separately for each method and smoothing value; coverage is not simultaneous across analyses.

\subsection{Smoothed IV-MSM bounds and additional conventional-MSM comparisons}

We compare the exact-envelope IV-MSM analysis with smoothing at $\varepsilon=0.01$, retaining the same folds and fitted nuisance functions.
Smoothing is applied to both the envelopes and the final binary bound maxima and minima.
We also evaluate the conventional-MSM estimators adjusting for $(X,Z)$ instead of $X$ alone.
Tan's estimates use the authors' \texttt{RCALsa} implementation; the Dorn--Guo--Kallus estimates use the authors' binary-outcome implementation with generalized-random-forest nuisance fits.

\begin{figure}[!htb]
\centering
\includegraphics[width=\linewidth]{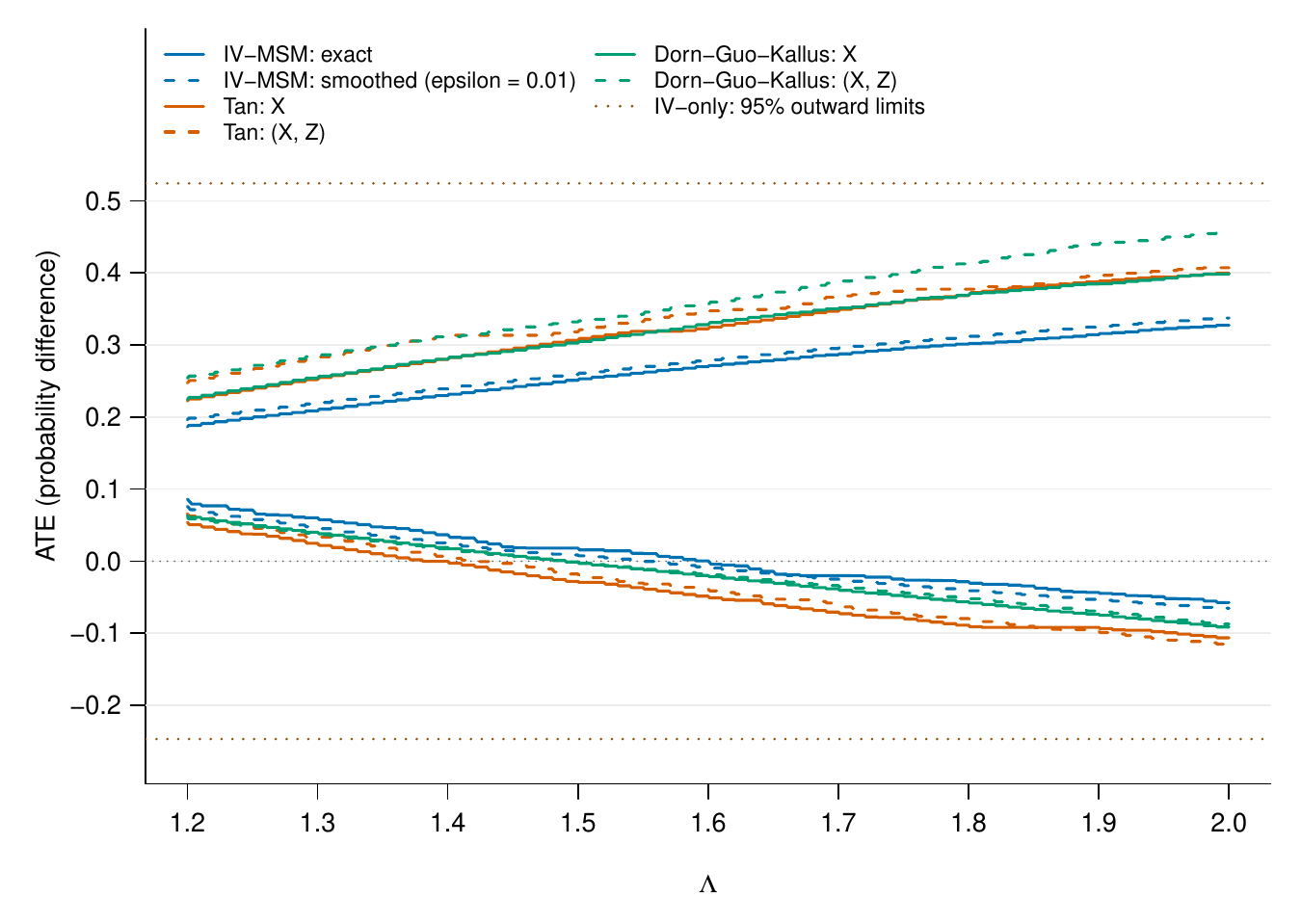}
\caption{Oregon application: 95\% simultaneous outward
confidence bands for the ATE over $\Lambda\in[1.2,2]$.
Colors distinguish methods. For the IV-MSM, solid and dashed
curves correspond to exact envelopes and smoothing with
$\varepsilon=0.01$, respectively. For the conventional-MSM
procedures, solid and dashed curves correspond to adjustment
for $X$ and $(X,Z)$, respectively.
Each band is calibrated for its method-specific population
bounds. Brown horizontal dotted lines show the IV-only
95\% outward confidence limits.
The gray horizontal dotted line marks zero.}
\label{fig:supp_oregon_additional}
\end{figure}

Figure~\ref{fig:supp_oregon_additional} compares the resulting 95\% simultaneous outward bands and includes the IV-only outward confidence limits as a reference.
Smoothing slightly widens the IV-MSM band: its lower limit first reaches zero at approximately $\Lambda=1.551$, compared with $\Lambda=1.600$ for the exact-envelope analysis.
Additionally adjusting for $Z$ generally widens the conventional-MSM confidence sets, which remain wider than those of the IV-MSM.
The conventional MSM adjusting for $(X,Z)$ imposes the same stratum-specific sensitivity restrictions as the IV-MSM but does not impose the IV assumptions.
At the population level, the IV-MSM therefore yields a sharp identified interval contained in that of the conventional MSM adjusting for $(X,Z)$.

%% file: supp_main_proofs.tex
\section{Proofs of Results in the Main Text}\label{sec:proofs_identification}

\begin{proof}[Proof of Proposition~\ref{prop:comp_msm_iv}]
Fix a full-data extension $Q$ satisfying the restrictions pertaining to $Y^1$.
For brevity, let $F_z(\cdot\mid x)$ denote the conditional distribution $P(Y\mid T=1,X=x,Z=z)$, for $z\in\{0,1\}$.
Consistency and composition (Assumption~\ref{ass:consist}) implies that $F_z$ is also the conditional law of $Y^1$ under $Q$ given $(T=1,X=x,Z=z)$.
By the definition of $\lambda^1_{z,Q}$ and the law of total probability, the conditional law $G_z$ of $Y^1$ given $(X=x,Z=z)$ is
\begin{equation}
dG_z(y\mid x)
=\big[\pi_{z,P}(x)+\{1-\pi_{z,P}(x)\}\lambda^1_{z,Q}(y,x)\big]
dF_z(y\mid x).
\label{eq:proof_mixture_Gz}
\end{equation}

We next show that $G_0=G_1$.
For every measurable set $C$ and $z\in\{0,1\}$, the relation $Y^1=Y^{1,Z}$ (Assumption~\ref{ass:consist}), instrument exchangeability (Assumption~\ref{ass:iv_exchan}), and distributional exclusion (Assumption~\ref{ass:iv_excl}) give
\[
\begin{aligned}
Q(Y^1\in C\mid X,Z=z)
&=Q(Y^{1,z}\in C\mid X,Z=z)\\
&=Q(Y^{1,z}\in C\mid X)
=Q(Y^1\in C\mid X).
\end{aligned}
\]
Thus $G_z(\cdot\mid x)$ does not depend on $z$. 
Moreover, by definition of $\gamma_{1,P}$, $dF_1(y\mid x)=\gamma_{1,P}(y,x)dF_0(y\mid x)$.
Substituting this identity into equation~\eqref{eq:proof_mixture_Gz} and equating the densities of $G_0$ and $G_1$ with respect to $F_0$ yields
\[
\pi_{0,P}(x)+\{1-\pi_{0,P}(x)\}\lambda^1_{0,Q}(y,x)
=\gamma_{1,P}(y,x)\left[\pi_{1,P}(x)
+\{1-\pi_{1,P}(x)\}\lambda^1_{1,Q}(y,x)\right]
\]
for $P_X$-almost every $x$ and $F_0(\cdot\mid x)$-almost every $y$.
Treatment positivity (Assumption~\ref{ass:positivity}) permits division by $1-\pi_{0,P}(x)$, and rearrangement gives
\[
\lambda^1_{0,Q}(Y^1,X)
=A_P(X,Y^1)\lambda^1_{1,Q}(Y^1,X)+B_P(X,Y^1).
\]
By equation~\eqref{eq:proof_mixture_Gz}, $G_0(\cdot\mid x)$ and $F_0(\cdot\mid x)$ have the same null sets, since $\pi_{0,P}(x)+\{1-\pi_{0,P}(x)\}\lambda^1_{0,Q}(y,x)>0$.
Because $G_0(\cdot\mid x)=Q(Y^1\mid X=x)$, the identity also holds at $(Y^1,X)$ $Q$-almost surely.

Compatibility also forces $A_P>0$ $F_0(\cdot\mid X)$-almost surely. 
Indeed, $A_P=0$ is equivalent to $\gamma_{1,P}=0$ because treatment probabilities are bounded away from one. 
On that set the affine identity gives
\[
\lambda^1_{0,Q}=B_P
=-\frac{\pi_{0,P}(x)}{1-\pi_{0,P}(x)}<0,
\]
contradicting the nonnegativity of a Radon--Nikodym derivative (and, for finite $\Lambda_0$, the bound $\lambda^1_{0,Q}\ge\Lambda_0^{-1}>0$).
Since $A_P>0$, the comparison-stratum sensitivity restriction in Assumption~\ref{ass:msm_iv_dual} implies
\[
A_P\Lambda_1^{-1}+B_P
\le \lambda^1_{0,Q}
\le A_P\Lambda_1+B_P.
\]
Intersecting this interval with the baseline restriction $[\Lambda_0^{-1},\Lambda_0]$ gives $L_P\le\lambda^1_{0,Q}\le U_P$.
\end{proof}

\begin{proof}[Proof of Corollary~\ref{coro:falsification_dual}]
Let $F_0(\cdot\mid x)$ denote the conditional distribution $P(Y\mid T=1,X=x,Z=0)$, as in the preceding proof.
By Proposition~\ref{prop:comp_msm_iv}, compatibility requires $L_P(x,y)\le U_P(x,y)$ for $P_X$-almost every $x$ and $F_0(\cdot\mid x)$-almost every $y$.
By the definitions of the
envelopes, this implies
\[
\Lambda_0^{-1}\le A_P(x,y)\Lambda_1+B_P(x,y),
\qquad
A_P(x,y)\Lambda_1^{-1}+B_P(x,y)\le\Lambda_0.
\]
Substituting the definitions of $A_P$ and $B_P$ and multiplying by $1-\pi_{0,P}(x)>0$ gives
\[
\begin{aligned}
\pi_{0,P}(x)+\{1-\pi_{0,P}(x)\}/\Lambda_0
&\le
\gamma_{1,P}(y,x)
\big[\pi_{1,P}(x)+\{1-\pi_{1,P}(x)\}\Lambda_1\big],\\
\gamma_{1,P}(y,x)
\big[\pi_{1,P}(x)+\{1-\pi_{1,P}(x)\}/\Lambda_1\big]
&\le
\pi_{0,P}(x)+\{1-\pi_{0,P}(x)\}\Lambda_0.
\end{aligned}
\]
The bracketed factors are strictly positive by treatment positivity.
Dividing by these factors yields
\[
\frac{\pi_{0,P}(x)+\{1-\pi_{0,P}(x)\}/\Lambda_0}
     {\pi_{1,P}(x)+\{1-\pi_{1,P}(x)\}\Lambda_1}
\le \gamma_{1,P}(y,x)\le
\frac{\pi_{0,P}(x)+\{1-\pi_{0,P}(x)\}\Lambda_0}
     {\pi_{1,P}(x)+\{1-\pi_{1,P}(x)\}/\Lambda_1}.
\]

Finally, Proposition~\ref{prop:comp_msm_iv} gives $L_P(x,y)\le\lambda^1_{0,Q}(y,x)\le U_P(x,y)$.
Integrating with respect to $F_0(\cdot\mid x)$ and using
\[
\int \lambda^1_{0,Q}(y,x)\,dF_0(y\mid x)=1
\]
yields $\widetilde L_P(x)\le1\le\widetilde U_P(x)$ for $P_X$-almost every $x$, completing the proof.
\end{proof}

\begin{proof}[Proof of Proposition~\ref{prop:sharpness_msm_iv}]
We prove the forward mapping by constructing the required conditional distributions. Fix $\lambda\in\mc E_P(\Lambda_0,\Lambda_1)$ and retain the notation $F_z$ from the proof of Proposition~\ref{prop:comp_msm_iv}.
To construct the full-data extension, we must obtain valid conditional distributions for $Y^1$ among untreated individuals in both instrument strata.
We begin with the baseline instrument stratum. Using $\lambda$, define a candidate conditional distribution $H_0$ for $Y^1\mid T=0,X=x,Z=0$ by
\[
dH_0(y\mid x)=\lambda(y,x)dF_0(y\mid x).
\]
The conditional normalization defining $\mc E_P$ gives $\int dH_0(y\mid x)=1$ for $P_X$-almost every $x$.
Since $\lambda\ge 0$, $H_0$ is therefore a conditional probability distribution.
The resulting completed conditional distribution of $Y^1$ in the baseline instrument stratum is
\begin{equation}
\begin{aligned}
dG_0(y\mid x)
&=\pi_{0,P}(x)dF_0(y\mid x)
+\{1-\pi_{0,P}(x)\}dH_0(y\mid x)\\
&=\big[\pi_{0,P}(x)+\{1-\pi_{0,P}(x)\}\lambda(y,x)\big]dF_0(y\mid x).
\end{aligned}
\label{eq:proof_common_G}
\end{equation}
As established in the proof of Proposition~\ref{prop:comp_msm_iv}, instrument exchangeability and distributional exclusion require the completed conditional distribution of $Y^1$ to agree across the two instrument strata.
Consequently, the candidate $H_1$ for the corresponding unobserved conditional distribution in the comparison stratum must satisfy
\[
dG_0(y\mid x)
=\pi_{1,P}(x)dF_1(y\mid x)
+\{1-\pi_{1,P}(x)\}dH_1(y\mid x).
\]
Solving this equality for $H_1$ gives
\begin{equation}
dH_1(y\mid x)
=\frac{dG_0(y\mid x)-\pi_{1,P}(x)dF_1(y\mid x)}
{1-\pi_{1,P}(x)}.
\label{eq:proof_H1_definition}
\end{equation}
We now verify that this candidate is a conditional probability distribution.
Since $P\in\mc M_1(\Lambda_0,\Lambda_1)$, the preceding compatibility proof gives $A_P(x,y)>0$ for $P_X$-almost every $x$ and $F_0(\cdot\mid x)$-almost every $y$.
Using $dF_1(y\mid x)=\gamma_{1,P}(y,x)\,dF_0(y\mid x)$ and the definitions of $A_P$ and $B_P$ gives
\begin{equation}
dH_1(y\mid x)
=\frac{\lambda(y,x)-B_P(x,y)}{A_P(x,y)}dF_1(y\mid x).
\label{eq:proof_H1_ratio}
\end{equation}
The restrictions $L_P\le\lambda\le U_P$ and $A_P>0$ imply
\[
\Lambda_1^{-1}
\le\frac{\lambda-B_P}{A_P}\le\Lambda_1,
\]
so $H_1$ is nonnegative and satisfies the comparison-stratum sensitivity bound.
Integrating equation~\eqref{eq:proof_H1_definition} gives
\[
\int dH_1(y\mid x)
=\frac{1-\pi_{1,P}(x)}{1-\pi_{1,P}(x)}=1.
\]
Hence $H_1$ is a conditional probability distribution. The displayed formulas are measurable in $x$ and therefore define valid conditional distributions.

We can now define the joint law. 
Give $(X,Z,T)$ its distribution under $P$, and conditionally on $(T=t,X=x,Z=z)$ let $Y^1$ have law $F_z$ if $t=1$ and $H_z$ if $t=0$.  Equation~\eqref{eq:proof_common_G} and the definition of $H_1$ show that
\[
Q(Y^1\in C\mid X=x,Z=0)
=G_0(C\mid x)
=Q(Y^1\in C\mid X=x,Z=1)
\]
for every measurable $C$.
Thus $Y^1\indep_Q Z\mid X$. 
Set $Y^{1,0}=Y^{1,1}=Y^1$.
This choice satisfies the relation $Y^1=Y^{1,Z}$ in Assumption~\ref{ass:consist}, as well as distributional exclusion (Assumption~\ref{ass:iv_excl}) and instrument exchangeability (Assumption~\ref{ass:iv_exchan}) for treatment level one.
It also gives
\[
\lambda^1_{0,Q}(Y^1,X)=\lambda(Y^1,X),\qquad
\lambda^1_{1,Q}(Y^1,X)=\frac{\lambda(Y^1,X)-B_P(X,Y^1)}{A_P(X,Y^1)}.
\]
The two sensitivity restrictions follow respectively from $\Lambda_0^{-1}\le L_P\le\lambda\le U_P\le\Lambda_0$ and equation~\eqref{eq:proof_H1_ratio}.

It remains to complete the law while preserving the observed outcome. 
When $T=1$, set the observed $Y$ equal to $Y^1$ and assign any conditional distribution to $Y^0$.
When $T=0$, draw the observed $Y$ from $P(Y\mid T=0,X,Z)$, set $Y^0=Y$, and couple this draw with $Y^1$ by any conditional distribution, for example by making them conditionally independent given $(T,X,Z)$.
Finally set $Y^{0,0}=Y^{0,1}=Y^0$. 
The resulting law satisfies $Y=(1-T)Y^0+TY^1$ and induces exactly $P$ as the law of $(Y,T,X,Z)$.
The proposition imposes only the restrictions pertaining to $Y^1$, so the arbitrary completion of the counterfactual $Y^0$ among treated units creates no additional obligation.
This proves that the constructed $Q$ is a valid full-data extension of $P$ satisfying all restrictions pertaining to $Y^1$.

Under this extension, the common conditional mean of $Y^1$ given $X$ is, by equation~\eqref{eq:proof_common_G},
\[
\int y\left[\pi_{0,P}(X)
+\{1-\pi_{0,P}(X)\}\lambda(y,X)\right]dF_0(y\mid X).
\]
Taking expectation over $X$ and rewriting the conditional integrals by inverse weighting in the $Z=0$ stratum gives
\[
\begin{aligned}
\mu^1(Q)
&=\E_P\left[
\frac{I(Z=0)}{e_P(X)}
\left\{TY+T\frac{1-\pi_{0,P}(X)}{\pi_{0,P}(X)}
\lambda(Y,X)Y\right\}
\right]\\
&=R_P(\lambda),
\end{aligned}
\]
proving part~(i).

For the converse, let $Q$ be any extension satisfying the restrictions pertaining to $Y^1$ and choose a jointly measurable version
\[
\lambda(y,x)
=\frac{dQ(Y^1\mid T=0,X=x,Z=0)}
       {dQ(Y^1\mid T=1,X=x,Z=0)}(y).
\]
Outcome overlap (Assumption~\ref{ass:abs_cont}) guarantees existence, consistency and composition identify the denominator with $F_0$, and the Radon--Nikodym property gives
$\int\lambda(y,x)dF_0(y\mid x)=1$.  Equivalently,
\[
\E_P\left\{
\frac{T\lambda(Y,X)}{\pi_{0,P}(X)}\,\middle|\,X,Z=0
\right\}=1.
\]
The baseline sensitivity restriction and Proposition~\ref{prop:comp_msm_iv} give $L_P\le\lambda\le U_P$, so $\lambda\in\mc E_P(\Lambda_0,\Lambda_1)$.
The same mixture calculation used above yields $\mu^1(Q)=R_P(\lambda)$.
This proves part~(ii).
\end{proof}

\begin{proof}[Proof of Theorem~\ref{thm:sharp_bounds_reps}]
Retain the notation $F_0(\cdot\mid x)$ from the preceding proofs.
All conditional statements below hold for $P_X$-almost every $x$.
Compatibility gives $\widetilde L_P(x)\le1\le\widetilde U_P(x)$.

First suppose $\widetilde L_P(x)<1<\widetilde U_P(x)$.
Since $F_0(\cdot\mid x)$ is atomless, the function
\[
q\longmapsto
\int_{-\infty}^{q}\Delta_P(x,y)\,dF_0(y\mid x)
\]
is continuous and nondecreasing, with limits zero and $\widetilde U_P(x)-\widetilde L_P(x)$.
Consequently, finite thresholds can be chosen to satisfy
\[
\begin{aligned}
\int_{-\infty}^{q_P^-(x)}
\Delta_P(x,y)\,dF_0(y\mid x)
&=1-\widetilde L_P(x),\\
\int_{-\infty}^{q_P^+(x)}
\Delta_P(x,y)\,dF_0(y\mid x)
&=\widetilde U_P(x)-1.
\end{aligned}
\]
Choosing the corresponding generalized inverses gives measurable functions of $x$.
Define
\[
\begin{aligned}
\lambda_P^-(y,x)
&=L_P(x,y)+\Delta_P(x,y)I\{y\le q_P^-(x)\},\\
\lambda_P^+(y,x)
&=U_P(x,y)-\Delta_P(x,y)I\{y\le q_P^+(x)\}.
\end{aligned}
\]
Both ratios lie between the envelopes and integrate to one under $F_0(\cdot\mid x)$.

To establish optimality, let $\lambda$ be any feasible density ratio.
For the lower bound, normalization gives
\[
\begin{aligned}
\int y\{\lambda(y,x)-\lambda_P^-(y,x)\}\,dF_0(y\mid x)=
\int\{y-q_P^-(x)\}
\{\lambda(y,x)-\lambda_P^-(y,x)\}\,dF_0(y\mid x)
\ge0.
\end{aligned}
\]
Indeed, below the threshold, $\lambda_P^-=U_P$, so both factors in the integrand are nonpositive; above it, $\lambda_P^-=L_P$, so both are nonnegative.
For the upper bound, the same argument gives
\[
\int y\{\lambda(y,x)-\lambda_P^+(y,x)\}\,dF_0(y\mid x)
=
\int\{y-q_P^+(x)\}
\{\lambda(y,x)-\lambda_P^+(y,x)\}\,dF_0(y\mid x)
\le0,
\]
because $\lambda_P^+=L_P$ below the threshold and $\lambda_P^+=U_P$ above it.

At a normalization boundary, no threshold construction is needed.
If $\widetilde L_P(x)=1$, feasibility and normalization force $\lambda=L_P$ $F_0(\cdot\mid x)$-almost surely.
If $\widetilde U_P(x)=1$, they force $\lambda=U_P$. When both equal one, $L_P=U_P$ almost surely.
Define both extremizers by these forced values on the boundary.
The resulting measurable functions belong to $\mc E_P$ and minimize and maximize the conditional objective, respectively.
Integrating over $X$ and applying Proposition~\ref{prop:sharpness_msm_iv} establishes that $R_P(\lambda_P^-)$ and $R_P(\lambda_P^+)$ are the sharp bounds.
Their integrability follows from $\Lambda_0<\infty$, treatment and instrument positivity, and $\E_P|Y|<\infty$.

We next derive the check-loss representations.
For every $\nu\in[0,1]$ and finite $y,q$,
\[
yI\{y\le q\}
=y\nu-\rho_\nu(y,q)-q\{\nu-I(y\le q)\}.
\]
On the set of strict normalization inequalities, the defining equations for the thresholds imply
\[
\begin{aligned}
\int\Delta_P(x,y)
\{1-\tau_P(x)-I(y\le q_P^-(x))\}\,dF_0(y\mid x)&=0,\\
\int\Delta_P(x,y)
\{\tau_P(x)-I(y\le q_P^+(x))\}\,dF_0(y\mid x)&=0.
\end{aligned}
\]
Thus, multiplying the check-loss identity by $\Delta_P$ and integrating eliminates its last term. Substitution into the conditional objectives gives
\[
\begin{aligned}
\int y\lambda_P^-(y,x)\,dF_0(y\mid x)
&=
\int y\{\tau_P(x)L_P(x,y)
+[1-\tau_P(x)]U_P(x,y)\}\,dF_0(y\mid x)\\
&\quad-
\int\Delta_P(x,y)
\rho_{1-\tau_P(x)}(y,q_P^-(x))\,dF_0(y\mid x),\\
\int y\lambda_P^+(y,x)\,dF_0(y\mid x)
&=
\int y\{\tau_P(x)L_P(x,y)
+[1-\tau_P(x)]U_P(x,y)\}\,dF_0(y\mid x)\\
&\quad+
\int\Delta_P(x,y)
\rho_{\tau_P(x)}(y,q_P^+(x))\,dF_0(y\mid x).
\end{aligned}
\]
The thresholds are the corresponding weighted conditional quantiles and therefore minimize these weighted check losses.
Multiplying the preceding identities by $1-\pi_{0,P}(x)$, adding $\pi_{0,P}(x)\int y\,dF_0(y\mid x)$, and integrating over $X$ yields the functionals in~\eqref{eq:fixed_threshold_objectives} evaluated at $q_P^-$ and $q_P^+$, respectively.
The check loss is subtracted in $\nu_P^{1-}(q)$ and added in $\nu_P^{1+}(q)$, while all remaining terms are independent of $q$.
Thus, minimizing the respective check losses maximizes $\nu_P^{1-}(q)$ and minimizes $\nu_P^{1+}(q)$.

For completeness, consider the normalization boundaries.
If $\widetilde L_P(x)=1<\widetilde U_P(x)$, then $\tau_P(x)=1$.
The check-loss term in $\nu_P^{1-}(q)$ is therefore $\rho_0(y,q)=(q-y)_+$, which decreases to zero as $q\to-\infty$.
The term in $\nu_P^{1+}(q)$ is $\rho_1(y,q)=(y-q)_+$, which decreases to zero as $q\to+\infty$.
If $\widetilde L_P(x)<\widetilde U_P(x)=1$, then $\tau_P(x)=0$, so these two check losses and the corresponding threshold directions are reversed.
In either case, the conditional expressions converge to $\int yL_P(x,y)\,dF_0(y\mid x)$ when $\widetilde L_P(x)=1$, or to $\int yU_P(x,y)\,dF_0(y\mid x)$ when $\widetilde U_P(x)=1$.
If both envelope averages equal one, $\Delta_P(x,Y)=0$ almost surely under $F_0(\cdot\mid x)$, and the threshold does not affect the value.

To approximate the boundary values using finite thresholds, construct a sequence indexed by positive integers $m$, with $m\to\infty$.
For each $m$, retain the optimal thresholds where $\widetilde L_P(x)<1<\widetilde U_P(x)$.
Where $\widetilde L_P(x)=\widetilde U_P(x)=1$, the envelopes coincide almost surely and the threshold values do not affect the functionals, so set both thresholds to zero.
Where $\widetilde L_P(x)=1<\widetilde U_P(x)$, set the lower threshold to $-m$ and the upper threshold to $m$; where $\widetilde L_P(x)<\widetilde U_P(x)=1$, reverse these choices.
Each member of this sequence consists of measurable real-valued threshold functions and therefore belongs to $\mc Q$.
The weighted check-loss terms on the boundary sets are bounded by $\Lambda_0|Y|I\{|Y|>m\}$, so the corresponding population functionals converge to the sharp bounds as $m\to\infty$ by dominated convergence, using $\E_P|Y|<\infty$ and the uniform bounds on the inverse probability weights implied by positivity.
Since the check losses are nonnegative and are minimized by the weighted quantiles where the normalization inequalities are strict, no other threshold can improve upon these endpoint values.
Therefore
\[
\mu^{1-}(P)=\sup_{q\in\mc Q}\nu_P^{1-}(q),
\qquad
\mu^{1+}(P)=\inf_{q\in\mc Q}\nu_P^{1+}(q).
\]

If, almost surely, either $\widetilde L_P(X)=\widetilde U_P(X)=1$ or $\widetilde L_P(X)<1<\widetilde U_P(X)$, the finite thresholds constructed above attain these extrema, completing the proof.
\end{proof}

\begin{proof}[Proof of Corollary~\ref{cor:sharp_ate}]
The bounds for the potential-outcome means imply that every compatible extension satisfies
\[
\mu^{1-}(P)-\mu^{0+}(P)
\le\psi(Q)\le
\mu^{1+}(P)-\mu^{0-}(P).
\]

To attain the lower bound, apply the construction in Proposition~\ref{prop:sharpness_msm_iv} to the minimizing density ratio for $Y^1$ and, after relabeling treatment, to the maximizing density ratio for $Y^0$.
These constructions specify the missing distribution of $Y^1$ among untreated individuals and that of $Y^0$ among treated individuals, respectively.
They can therefore be combined: draw $(Y,T,X,Z)$ from $P$, set $Y^T=Y$, and draw the missing potential outcome from its constructed conditional distribution given $(T,X,Z)$.
Set $Y^{t,0}=Y^{t,1}=Y^t$ for both treatment levels.
This preserves each construction's joint distribution of $(Y^t,T,X,Z)$ and hence its IV, outcome-overlap, and sensitivity restrictions.
The resulting extension attains $\mu^{1-}(P)-\mu^{0+}(P)$.
Reversing the choices of extremizing density ratios attains $\mu^{1+}(P)-\mu^{0-}(P)$.

Finally, any mixture $Q_s=sQ^-+(1-s)Q^+$, $s\in[0,1]$, preserves the observed-data law $P$ and the consistency restrictions.
Both extensions satisfy the exclusion and exchangeability restrictions.
Because they share the same distribution of $(X,Z)$, conditioning on these variables leaves the mixture weights unchanged, so the conditional distributional equalities expressing these restrictions are preserved under $Q_s$.
The missing potential-outcome distributions are mixed while their observed counterparts remain fixed, so the resulting density ratios are convex combinations of the original ratios and satisfy the same sensitivity bounds and outcome-overlap conditions.
Thus $Q_s$ is compatible and
\[
\psi(Q_s)=s\psi(Q^-)+(1-s)\psi(Q^+).
\]
Every value between the two endpoints is therefore attainable.
\end{proof}

\begin{proof}[Proof of Corollary~\ref{cor:restricted_classes}]
Since $\widetilde{\mc Q}\subseteq\mc Q$, Theorem~\ref{thm:sharp_bounds_reps} gives
\[
\widetilde\mu^{1-}(P)
=\sup_{q\in\widetilde{\mc Q}}\nu_P^{1-}(q)
\le
\sup_{q\in\mc Q}\nu_P^{1-}(q)
=\mu^{1-}(P)
\]
and
\[
\widetilde\mu^{1+}(P)
=\inf_{q\in\widetilde{\mc Q}}\nu_P^{1+}(q)
\ge
\inf_{q\in\mc Q}\nu_P^{1+}(q)
=\mu^{1+}(P).
\]
Together with $\mu^{1-}(P)\le\mu^{1+}(P)$, these inequalities give the asserted ordering.
\end{proof}

\begin{proof}[IV-MSM bounds contained within MSM bounds adjusting for $X$]
Let $Q$ be a full-data extension satisfying the IV-MSM
restrictions pertaining to $Y^1$, with
$\Lambda_0=\Lambda_1=\Lambda<\infty$.
As established in the proof of
Proposition~\ref{prop:comp_msm_iv},
$Y^1\indep_Q Z\mid X$.
Define
\[
g_{z,Q}(y,x):=Q(T=1\mid Y^1=y,X=x,Z=z).
\]
By Bayes' rule, the within-stratum sensitivity restrictions
are equivalent to
\[
\Lambda^{-1}
\le
\frac{\operatorname{odds}\{g_{z,Q}(y,x)\}}
     {\operatorname{odds}\{\pi_{z,P}(x)\}}
\le \Lambda.
\]
Equivalently,
\[
f_\Lambda\{\pi_{z,P}(x)\}
\le g_{z,Q}(y,x)\le
h_\Lambda\{\pi_{z,P}(x)\},
\]
where
\[
f_\Lambda(p):=\frac{p}{p+\Lambda(1-p)},
\qquad
h_\Lambda(p):=\frac{\Lambda p}{1-p+\Lambda p}.
\]
These inequalities hold for $Q_{Y^1,X}$-almost every $(y,x)$
in both instrument strata.

Let
\[
\widetilde\pi_P(x):=P(T=1\mid X=x),
\qquad
\widetilde g_Q(y,x):=Q(T=1\mid Y^1=y,X=x).
\]
The law of total probability and
$Y^1\indep_Q Z\mid X$ give
\[
\begin{aligned}
\widetilde\pi_P(x)
&=e_P(x)\pi_{0,P}(x)
  +\{1-e_P(x)\}\pi_{1,P}(x),\\
\widetilde g_Q(y,x)
&=e_P(x)g_{0,Q}(y,x)
  +\{1-e_P(x)\}g_{1,Q}(y,x).
\end{aligned}
\]
For $\Lambda\ge1$, $f_\Lambda$ is convex and $h_\Lambda$
is concave on $[0,1]$. Jensen's inequality therefore yields
\[
\begin{aligned}
f_\Lambda\{\widetilde\pi_P(x)\}
&\le e_P(x)f_\Lambda\{\pi_{0,P}(x)\}
  +\{1-e_P(x)\}f_\Lambda\{\pi_{1,P}(x)\}\\
&\le \widetilde g_Q(y,x)\\
&\le e_P(x)h_\Lambda\{\pi_{0,P}(x)\}
  +\{1-e_P(x)\}h_\Lambda\{\pi_{1,P}(x)\}\\
&\le h_\Lambda\{\widetilde\pi_P(x)\}.
\end{aligned}
\]
Transforming back to odds gives
\[
\Lambda^{-1}
\le
\frac{\operatorname{odds}\{Q(T=1\mid Y^1,X)\}}
     {\operatorname{odds}\{\widetilde\pi_P(X)\}}
\le \Lambda,
\qquad Q\text{-almost surely}.
\]
Thus every full-data extension satisfying the IV-MSM
restrictions pertaining to $Y^1$ also satisfies the
conventional MSM adjusting for $X$ alone at the same
$\Lambda$. Its feasible values of $\mu^1(Q)$ therefore
belong to the conventional-MSM identified interval,
proving the claimed containment.

The corresponding result for $\mu^0(Q)$ follows by
relabeling treatment. When the restrictions hold for both
potential outcomes, the same full-data extension satisfies
both conventional-MSM restrictions, establishing containment
of the sharp ATE interval as well.
\end{proof}

\begin{proof}[Proof of Theorem~\ref{thm:eif_fixed_q}]
Fix $b$ and $q^b$, and consider a regular submodel $P_s$ through $P$ with score $S$.
A dot denotes differentiation at $s=0$ in this proof.
Let $F_z(\cdot\mid x)=P(Y\mid T=1,X=x,Z=z)$, and let
$s_z(y,x)$ be its conditional score. These scores have conditional mean zero.
Differentiating the density-ratio identity gives
\[
\dot\gamma_{1,P}(y,x)
=\gamma_{1,P}(y,x)\{s_1(y,x)-s_0(y,x)\}.
\]
The threshold and its allocation indicators are held fixed along the submodel.
Away from the envelope ties, the derivatives of $L_P$ and $U_P$ are the active-term derivatives specified in the main text.
Thus the derivative of $\lambda_{P,q}^b(y,x)\{y-q^b(x)\}$ with respect to $\zeta\in\{\gamma_1,\pi_0,\pi_1\}$ is $D_{P,q,\zeta}^b(y,x)$.

Holding $q^b$ fixed, differentiation of the integrand and its
conditional distribution gives
\[
\begin{aligned}
\dot\theta_{P,q}^b(x)
={}&\int \lambda_{P,q}^b(y,x)\{y-q^b(x)\}
s_0(y,x)\,dF_0(y\mid x)\\
&+\int D_{P,q,\gamma_1}^b(y,x)
\dot\gamma_{1,P}(y,x)\,dF_0(y\mid x)+\sum_{z=0}^1
\overline D_{P,q,\pi_z}^b(x)\dot\pi_{z,P}(x).
\end{aligned}
\]
Substituting
$\dot\gamma_{1,P}=\gamma_{1,P}(s_1-s_0)$,
using $dF_1=\gamma_{1,P}\,dF_0$, and collecting the terms
involving $s_0$ yields
\[
\begin{aligned}
\dot\theta_{P,q}^b(x)
={}&\int\left[\lambda_{P,q}^b(y,x)\{y-q^b(x)\}
-\gamma_{1,P}(y,x)D_{P,q,\gamma_1}^b(y,x)\right]
s_0(y,x)\,dF_0(y\mid x)\\
&+\int D_{P,q,\gamma_1}^b(y,x)s_1(y,x)\,dF_1(y\mid x)
+\sum_{z=0}^1\overline D_{P,q,\pi_z}^b(x)\dot\pi_{z,P}(x).
\end{aligned}
\]
Also $\dot m_{0,P}(x)=\int ys_0(y,x)\,dF_0(y\mid x)$.
Applying the product rule to $M_{P,q}^b=\pi_{0,P}m_{0,P}+(1-\pi_{0,P})\theta_{P,q}^b$ therefore gives
\[
\dot M_{P,q}^b(x)
=\sum_{z=0}^1\left[
\int H_{z,P,q}^b(y,x)s_z(y,x)\,dF_z(y\mid x)
+C_{z,P,q}^b(x)\dot\pi_{z,P}(x)\right].
\]
In particular, differentiating the outer factor $\pi_{0,P}$ contributes $m_{0,P}-\theta_{P,q}^b$ to $C_{0,P,q}^b$.

The preceding calculation differentiates each envelope using the argument that determines it at $P$.
We now show that changes in the determining argument introduce only an $o(|s|)$ additional error after integration. We give the argument for the lower envelope.

Its contribution to $M_{P,q}^b(x)$ is
\[
\int
a_{L,q}^b(y,x)\{y-q^b(x)\}
(1-\pi_{0,P}(x))L_P(x,y)\,dF_0(y\mid x).
\]
Using $\gamma_{1,P}\,dF_0=dF_1$, the weighted measure in this integral satisfies
\[
(1-\pi_{0,P})L_P\,dF_0
=
\max\left\{
\frac{1-\pi_{0,P}}{\Lambda_0}\,dF_0,\,
\left[\pi_{1,P}
+\frac{1-\pi_{1,P}}{\Lambda_1}\right]\,dF_1
-\pi_{0,P}\,dF_0
\right\}.
\]
The maximum acts pointwise on the densities of the two measures inside the braces.

The population objective also averages over $X$, so these measures are multiplied by $dP_X(x)$.
Write $F_{z,P_s}$ for $F_z$ evaluated under $P_s$, and express all joint measures below as densities relative to one common dominating measure held fixed along the submodel.
Let $g_s$ be the density of the signed measure
\[
\left\{
\left[\pi_{1,P_s}(x)
+\frac{1-\pi_{1,P_s}(x)}{\Lambda_1}\right]
dF_{1,P_s}(y\mid x)
-
\left[\pi_{0,P_s}(x)
+\frac{1-\pi_{0,P_s}(x)}{\Lambda_0}\right]
dF_{0,P_s}(y\mid x)
\right\}dP_{s,X}(x).
\]
This is the difference between the two measures compared in the maximum, including the averaging over $X$.
Since $\max(u,v)=u+(v-u)_+$, the weighted lower-envelope density equals the density of $(1-\pi_{0,P_s})\Lambda_0^{-1}dF_{0,P_s}dP_{s,X}$ plus $(g_s)_+$.
We therefore need to justify differentiating $(g_s)_+$.

Write $\|\cdot\|_1$ for integrated absolute value with respect to the common dominating measure.
An $o_{L_1}(|s|)$ remainder denotes a density whose $L_1$ norm is $o(|s|)$.
For each $z\in\{0,1\}$, the relevant distributional expansion is
\[
\begin{aligned}
dF_{z,P_s}(y\mid x)dP_{s,X}(x)
={}&
\left[
1+s\left\{
s_z(y,x)+\E_P[S(O)\mid X=x]
\right\}
\right]
dF_z(y\mid x)dP_X(x)
+o_{L_1}(|s|).
\end{aligned}
\]
To obtain this expansion, note that the joint measure $dF_{z,P_s}dP_{s,X}$ is obtained from the observed-data law by reweighting with
$I(T=1,Z=z)/P_s(T=1,Z=z\mid X)$.
The logarithmic derivative of $P_s(T=1,Z=z\mid X)$ at $s=0$ is $\E_P[S(O)\mid T=1,X,Z=z]-\E_P[S(O)\mid X]$.
Inverse-probability weighting subtracts this quantity from the observed-data score, so the score of $dF_{z,P_s}(y\mid x)dP_{s,X}(x)$ is $s_z(y,x)+\E_P[S(O)\mid X=x]$.
Regularity of the submodel and positivity give the expansion with the stated integrated remainder.

Applying the product rule to the propensity coefficients in the definition of $g_s$ now gives
\[
\|g_s-g_0-s\dot g_0\|_1=o(|s|),
\]
where $\dot g_0$ is the density of
\[
\begin{aligned}
&\Bigg[
\left(1-\frac1{\Lambda_1}\right)\dot\pi_{1,P}(x)
+
\left\{\pi_{1,P}(x)
+\frac{1-\pi_{1,P}(x)}{\Lambda_1}\right\}
\left\{s_1(y,x)+\E_P[S(O)\mid X=x]\right\}
\Bigg]
dF_1(y\mid x)dP_X(x)\\
&\quad-
\Bigg[
\left(1-\frac1{\Lambda_0}\right)\dot\pi_{0,P}(x)
+
\left\{\pi_{0,P}(x)
+\frac{1-\pi_{0,P}(x)}{\Lambda_0}\right\}
\left\{s_0(y,x)+\E_P[S(O)\mid X=x]\right\}
\Bigg]
dF_0(y\mid x)dP_X(x).
\end{aligned}
\]
The derivative is integrable by positivity and
square integrability of the submodel score.

At $P$, the signed measure defining $g_0$ equals
\[
(1-\pi_{0,P}(x))
\{\Gamma_{\Lambda_1^{-1},P}(x,y)-\Lambda_0^{-1}\}
\,dF_0(y\mid x)dP_X(x).
\]
The lower-envelope no-tie condition therefore implies that
$\{g_0=0\}$ has zero probability under
$P(X,Y\mid T=1,Z=0)$.
Positivity and outcome overlap imply that this set also
has zero mass under both $dF_0(y\mid x)dP_X(x)$ and
$dF_1(y\mid x)dP_X(x)$.
The explicit expression for the derivative shows that both
measures defining $\dot g_0$ vanish on this set.
Consequently, $\dot g_0=0$ almost everywhere on $\{g_0=0\}$.

The inequality $|u_+-v_+|\le |u-v|$ gives
\[
\|(g_s)_+-(g_0+s\dot g_0)_+\|_1
\le
\|g_s-g_0-s\dot g_0\|_1
=o(|s|).
\]
Thus replacing $(g_s)_+$ by $(g_0+s\dot g_0)_+$ introduces an integrated error of $o(|s|)$.
Whenever $|s\dot g_0|<|g_0|$, the quantities $g_0+s\dot g_0$ and $g_0$ have the same sign, giving the exact identity
\[
(g_0+s\dot g_0)_+
=(g_0)_++sI(g_0>0)\dot g_0.
\]
On this region, therefore, taking the positive part of the linear approximation introduces no further error.
Otherwise, its absolute value is at most $|s\dot g_0|$.
It follows that
\[
\begin{aligned}
\|(g_s)_+-(g_0)_+-sI(g_0>0)\dot g_0\|_1\le
\|g_s-g_0-s\dot g_0\|_1
+
|s|\left\|
\dot g_0 I\{|g_0|\le |s\dot g_0|\}
\right\|_1
=o(|s|).
\end{aligned}
\]
For the last equality, wherever $g_0\ne0$, the indicator tends to zero as $s\to0$.
Where $g_0=0$, its multiplier $\dot g_0$ is zero.
The product therefore tends to zero almost everywhere and is bounded in absolute value by the integrable function $|\dot g_0|$.
Dominated convergence makes its $L_1$ norm tend to zero.

Adding the ordinary product-rule expansion of $(1-\pi_{0,P_s})\Lambda_0^{-1}dF_{0,P_s}dP_{s,X}$ establishes the required derivative of the weighted lower envelope, with an integrated absolute remainder of $o(|s|)$.
The multiplier $a_{L,q}^b(y,x)\{y-q^b(x)\}$ is bounded because the outcome domain and the fixed threshold function are bounded.
Multiplication by this factor preserves the remainder order.

The upper-envelope argument is analogous.
Thus changes in which argument determines an envelope add no first-order term beyond the derivatives calculated above.
We now express the derivative of $\nu_P^{1b}(q^b)$ as an expectation against the observed-data score $S(O)$ to identify its efficient influence function.

Write $e_{0,P}=e_P$ and $e_{1,P}=1-e_P$.
The conditional-score identities
\[
\begin{aligned}
\dot\pi_{z,P}(x)
&=\E_P[\{T-\pi_{z,P}(x)\}S(O)\mid X=x,Z=z],\\
\int H_{z,P,q}^b s_z\,dF_z
&=\E_P[\{H_{z,P,q}^b-\overline H_{z,P,q}^b\}S(O)
\mid T=1,X=x,Z=z]
\end{aligned}
\]
show, by inverse weighting with $e_{z,P}$ and $\pi_{z,P}$, that the residual terms in~\eqref{eq:eif_fixed_q} represent $\E_P\big\{\dot M_{P,q}^b(X)\big\}$.
The remaining term $M_{P,q}^b(X)-\nu_P^{1b}(q^b)$ represents perturbation of the distribution of $X$.
Consequently,
\[
\left.\frac{d}{ds}\nu_{P_s}^{1b}(q^b)\right|_{s=0}
=\E_P\{\phi_{P,q}^{1b}(O)S(O)\}.
\]
The proposed influence function is centered by conditional expectation and is square integrable by positivity, the finite sensitivity parameters, compatibility, and boundedness of the outcome domain and the fixed threshold function.
In the nonparametric observed-data model it is therefore the canonical gradient, proving part~\textup{(i)}.

For part~\textup{(ii)}, hold $\widetilde q_P^b$ fixed and write
\[
\begin{aligned}
\widetilde\mu^{1b}(P_s)-\widetilde\mu^{1b}(P)
={}&\nu_{P_s}^{1b}(\widetilde q_P^b)-\nu_P^{1b}(\widetilde q_P^b)\\
&+\widetilde\mu^{1b}(P_s)-\nu_{P_s}^{1b}(\widetilde q_P^b).
\end{aligned}
\]
The last difference is $o(|s|)$ by~\eqref{eq:first_order_optimizer}.
Part~\textup{(i)} identifies the derivative of the first difference, which proves the claimed differentiability and efficient influence function.
\end{proof}

\begin{proof}[Relaxing the boundedness conditions of
Theorem~\ref{thm:eif_fixed_q}]
The boundedness requirements can be replaced by local moment
conditions. Fix $q^b\in\widetilde{\mc Q}$ and retain the other
assumptions of Theorem~\ref{thm:eif_fixed_q}.
Consider regular submodels $\{P_s\}$ through $P$ such that,
for some $\delta>0$, Assumption~\ref{ass:positivity} holds
with a common positivity constant for $|s|<\delta$ and
\begin{equation}
\sup_{|s|<\delta}
\E_{P_s}\!\left[Y^2+\{q^b(X)\}^2\right]<\infty.
\label{eq:supp_unbounded_moments}
\end{equation}
This condition allows both $Y$ and $q^b(X)$ to be unbounded,
but controls their second moments along the submodel.

Under these conditions, the distributional expansions in
the preceding proof remain valid after weighting their
absolute remainders by $1+|y|+|q^b(x)|$.
To obtain this conclusion, apply quadratic-mean
differentiability to the nonnegative propensity-weighted
outcome measures appearing in the envelope representations.
Cauchy--Schwarz and
equation~\eqref{eq:supp_unbounded_moments} control the
square-root expansion remainders; truncating the
square-integrable scores controls the remaining terms.
Taking differences gives, in the notation of the preceding
proof,
\[
\left\|
\{1+|y|+|q^b(x)|\}
\{g_s-g_0-s\dot g_0\}
\right\|_1=o(|s|).
\]
Moreover,
$\{1+|y|+|q^b(x)|\}|\dot g_0|$ is integrable.
The same positive-part inequality and dominated-convergence
argument therefore control changes in the determining
envelope argument, now including the unbounded residual
$y-q^b(x)$.

Consequently, along every such submodel,
\[
\left.\frac{d}{ds}\nu_{P_s}^{1b}(q^b)\right|_{s=0}
=
\E_P\{\phi_{P,q}^{1b}(O)S(O)\},
\]
with the same square-integrable influence function.
These submodels include bounded-score perturbations, so
the canonical-gradient expression is unchanged.
Part~\textup{(ii)} extends in the same way when the selected
optimizer satisfies equation~\eqref{eq:supp_unbounded_moments}
and condition~\eqref{eq:first_order_optimizer} holds along
these submodels.
\end{proof}

\begin{proof}[Verification of equation~\eqref{eq:linked_M}]
Changing measure from $F_1$ to $F_0$ gives
\[
\overline H_{1,P,q}^b(x)
=\left\{1-\pi_{0,P}(x)\right\}
\int\gamma_{1,P}(y,x)D_{P,q,\gamma_1}^b(y,x)\,dF_0(y\mid x).
\]
This cancels the corresponding negative term in
$\overline H_{0,P,q}^b$. Their sum is
$\pi_{0,P}m_{0,P}+(1-\pi_{0,P})(\theta_{P,q}^b-q^b)$;
adding $(1-\pi_{0,P})q^b$ gives $M_{P,q}^b$.
\end{proof}

\begin{proof}[Proof of Theorem~\ref{thm:restricted_crossfit_inference}]
Let $\E_{n,k}$ denote the empirical measure in validation fold $k$.
Within this proof abbreviate
\[
\widehat\Phi_k=\Phi^{1b}(\cdot;\widehat\eta_{-k},\widehat q_{-k}^b),
\quad \Phi_k^0=\Phi^{1b}(\cdot;\eta_P,\widehat q_{-k}^b),
\quad \Phi_\infty=\Phi^{1b}(\cdot;\eta_P,q_\infty^b).
\]
Since $\E_P\left(\Phi_k^0\right)=\nu_P^{1b}(\widehat q_{-k}^b)$, direct subtraction gives
\[
\begin{aligned}
\widehat{\widetilde\mu}^{1b}-\overline\nu_n^{1b}
={}&(\E_n-\E_P)\Phi_\infty+R_n^{1b}\\
&+\sum_k\frac{n_k}{n}(\E_{n,k}-\E_P)(\widehat\Phi_k-\Phi_k^0)
+\sum_k\frac{n_k}{n}(\E_{n,k}-\E_P)(\Phi_k^0-\Phi_\infty).
\end{aligned}
\]
We control the empirical-process terms using the standard cross-fitting argument.
Conditional on the training sample for fold $k$, its validation observations are i.i.d. under $P$ and independent of the fitted functions.
The conditional variance of each centered validation-fold average is at most the squared $L_2(P)$ norm of the function difference being averaged, divided by $n_k$.
Conditional Chebyshev's inequality, the nuisance-substitution condition, and influence-function stabilization therefore make both sums $o_p(n^{-1/2})$.
This argument is applied separately to each of the fixed number of folds.
The negligible-drift condition yields
\begin{equation}
\widehat{\widetilde\mu}^{1b}-\overline\nu_n^{1b}
=\frac1n\sum_{i=1}^n\phi_{P,q_\infty^b}^{1b}(O_i)+o_p(n^{-1/2}).
\label{eq:proof_crossfit_expansion}
\end{equation}
The central limit theorem proves the first conclusion.

For the lower bound,
$\overline\nu_n^{1-}-\widetilde\mu^{1-}(P)
=-r_{\widetilde{\mc Q},n}^{1-}$; for the upper bound the difference
is $r_{\widetilde{\mc Q},n}^{1+}$.
Negligible regret therefore gives the second conclusion with the same
influence function and variance.

Under the additional optimizer conditions, Theorem~\ref{thm:eif_fixed_q}\textup{(ii)} identifies $\phi_{P,q_\infty^b}^{1b}$ as the canonical gradient of the restricted-class bound.
Together with the preceding asymptotic linear expansion, this establishes regularity and efficiency by standard results for pathwise differentiable parameters.
Applying the expansions jointly to the four bounds and using the multivariate central limit theorem gives the joint conclusions and those for their ATE contrasts.
\end{proof}

\begin{proof}[Consistency of the variance estimators]
With the notation of the preceding proof, $\max_k\|\widehat\Phi_k-\Phi_\infty\|_{P,2}=o_p(1)$.
Conditional Markov's inequality in each fold gives
\[
\frac1n\sum_{i=1}^n
\{\widehat\Phi_{k(i)}(O_i)-\Phi_\infty(O_i)\}^2=o_p(1).
\]
Cauchy--Schwarz and the law of large numbers then imply convergence of the empirical first and second moments to those of $\Phi_\infty$.
Thus the empirical variance consistently estimates $\operatorname{Var}_P(\phi_{P,q_\infty^b}^{1b})$.
The same argument applied to the vector of four contributions gives consistency of every empirical covariance.
In particular, using the row-level ATE contrasts in~\eqref{eq:ate_bound_variance} consistently estimates the variance of each difference, including its covariance term.
\end{proof}

\begin{proof}[Proof of Corollary~\ref{cor:outward_confidence_set}]
The joint cross-fitted expansion and standard-error consistency give
\[
\frac{\widehat{\widetilde\psi}^b-\overline\psi_n^b}
{\widehat{\mathrm{se}}_b}\rightsquigarrow N(0,1).
\]
Since $\overline\psi_n^-\le\widetilde\psi^-(P)$,
\[
\{L_{n,\alpha}\le\overline\psi_n^-\}
\subseteq\{L_{n,\alpha}\le\widetilde\psi^-(P)\}.
\]
The event on the left has limiting probability $1-\alpha$.
The upper claim follows from
$\{\overline\psi_n^+\le U_{n,\alpha}\}
\subseteq\{\widetilde\psi^+(P)\le U_{n,\alpha}\}$.
Using $\alpha/2$ for each bound and applying the union bound proves coverage of the entire restricted-class interval.
Its containment of the sharp interval proves the corresponding sharp-interval coverage.
The argument for each potential-outcome interval is identical.
\end{proof}

\begin{proof}[Local coverage of one-sided confidence limits and outward confidence sets]
Fix a law $P$ satisfying the conditions for the first conclusion of Theorem~\ref{thm:restricted_crossfit_inference}, with $0<\sigma_{1b}^2<\infty$ and $\sqrt n\,\widehat{\mathrm{se}}_{1b}\to_P\sigma_{1b}$.
Let $\{P_s\}$ be a differentiable-in-quadratic-mean parametric submodel through $P$, with score $S\in L_2^0(P)$, and set $P_n=P_{h/\sqrt n}$ for fixed $h\in\mathbb R$.
Consider paths in the domain of the fixed-threshold functionals along which Theorem~\ref{thm:eif_fixed_q}\textup{(i)} gives their derivatives.

If the limiting threshold $q_\infty^b$ in Assumption~\ref{ass:sharp_if_stabilization} is not optimal, choose a fixed $q^\star\in\widetilde{\mc Q}$ with a strictly better objective value at $P$: larger when $b=-$ and smaller when $b=+$.
Under the conditions of Theorem~\ref{thm:eif_fixed_q}\textup{(i)}, the fixed-threshold functional $\nu_P^{1b}(q^\star)$ is pathwise differentiable at $P$ and therefore continuous along the regular submodel, i.e., $\nu_{P_s}^{1b}(q^\star)\longrightarrow\nu_P^{1b}(q^\star)$ as $s\to0$.

Define
\[
L_{n,\alpha}^{1}
=\widehat{\widetilde\mu}^{1-}
-z_{1-\alpha}\widehat{\mathrm{se}}_{1-},
\qquad
U_{n,\alpha}^{1}
=\widehat{\widetilde\mu}^{1+}
+z_{1-\alpha}\widehat{\mathrm{se}}_{1+}.
\]
We show that
\begin{equation}
\begin{aligned}
\liminf_{n\to\infty}P_n^{\otimes n}
\{L_{n,\alpha}^{1}\le\widetilde\mu^{1-}(P_n)\}
&\ge1-\alpha,\\
\liminf_{n\to\infty}P_n^{\otimes n}
\{\widetilde\mu^{1+}(P_n)\le U_{n,\alpha}^{1}\}
&\ge1-\alpha.
\end{aligned}
\label{eq:supp_local_one_sided}
\end{equation}

Write $\phi_b=\phi_{P,q_\infty^b}^{1b}$ and use $\overline\nu_{n,P}^{1b}$ for the fold-average population value $\overline\nu_n^{1b}$ evaluated under the base law $P$.
The cross-fitted expansion gives
\[
\widehat{\widetilde\mu}^{1b}
=\overline\nu_{n,P}^{1b}
+\frac1n\sum_{i=1}^n\phi_b(O_i)+o_P(n^{-1/2}).
\]
Taking expectations in the influence-function stabilization condition and applying Cauchy--Schwarz also gives $\overline\nu_{n,P}^{1b}\to_P\nu_P^{1b}(q_\infty^b)$.
Contiguity of $P_n^{\otimes n}$ to $P^{\otimes n}$ transfers these remainder and consistency statements, including standard-error consistency, to sampling under $P_n$.

First suppose $q_\infty^-$ is optimal at $P$.
For every realized training sample,
\[
\overline\nu_{n,P}^{1-}
\le\widetilde\mu^{1-}(P)
=\nu_P^{1-}(q_\infty^-).
\]
Consequently,
\[
\widehat{\widetilde\mu}^{1-}
\le A_n^-+o_{P_n}(n^{-1/2}),
\qquad
A_n^-=\nu_P^{1-}(q_\infty^-)
+\frac1n\sum_{i=1}^n\phi_-(O_i).
\]
The inequality compares population objectives under $P$ and holds regardless of the distribution generating the training sample.
Le Cam's third lemma gives, under $P_n^{\otimes n}$,
\[
\frac1{\sqrt n}\sum_{i=1}^n\phi_-(O_i)
\rightsquigarrow
N\{h\E_P(\phi_-S),\sigma_{1-}^2\}.
\]
Separately, fixed-threshold differentiability gives
\[
\sqrt n\{\nu_{P_n}^{1-}(q_\infty^-)
-\nu_P^{1-}(q_\infty^-)\}
=h\E_P(\phi_-S)+o(1).
\]
Thus the shift in the influence-function sum equals the first-order change in the fixed-threshold target.
Subtracting that target cancels the shift, so
\[
\sqrt n\{A_n^--\nu_{P_n}^{1-}(q_\infty^-)\}
\rightsquigarrow N(0,\sigma_{1-}^2)
\qquad\text{under }P_n^{\otimes n}.
\]
Together with the upper comparison for the estimator and standard-error consistency, this implies
\[
\liminf_{n\to\infty}P_n^{\otimes n}
\{L_{n,\alpha}^{1}\le\nu_{P_n}^{1-}(q_\infty^-)\}
\ge1-\alpha.
\]
Moreover,
\[
\begin{aligned}
P_n^{\otimes n}
\{L_{n,\alpha}^{1}\le\widetilde\mu^{1-}(P_n)\}
&\ge
P_n^{\otimes n}
\{L_{n,\alpha}^{1}\le\nu_{P_n}^{1-}(q_\infty^-)\},
\end{aligned}
\]
because $\nu_{P_n}^{1-}(q_\infty^-)\le\widetilde\mu^{1-}(P_n)$.
This proves the lower-limit claim when the limiting threshold is optimal.

If $q_\infty^-$ is not optimal, choose the fixed $q^\star$ specified above, with $\nu_P^{1-}(q^\star)>\nu_P^{1-}(q_\infty^-)$.
Stabilization and contiguity give $L_{n,\alpha}^{1}\to\nu_P^{1-}(q_\infty^-)$ in probability under $P_n^{\otimes n}$, whereas
\[
\widetilde\mu^{1-}(P_n)
\ge\nu_{P_n}^{1-}(q^\star)
\longrightarrow\nu_P^{1-}(q^\star)
>\nu_P^{1-}(q_\infty^-).
\]
The positive gap implies that coverage tends to one.

For the upper bound, reverse the inequalities.
If $q_\infty^+$ is optimal at $P$, define
$A_n^+=\nu_P^{1+}(q_\infty^+)+n^{-1}\sum_i\phi_+(O_i)$.
Then $\widehat{\widetilde\mu}^{1+}\ge A_n^++o_{P_n}(n^{-1/2})$
and $\widetilde\mu^{1+}(P_n)\le\nu_{P_n}^{1+}(q_\infty^+)$.
The same mean-shift cancellation proves the upper-limit claim.
A nonoptimal limiting threshold again leaves a positive outward gap, giving coverage tending to one.
This establishes~\eqref{eq:supp_local_one_sided} without
condition~\eqref{eq:first_order_optimizer}.

Relabeling treatment gives the corresponding conclusions for $Y^0$.
For the lower ATE bound, apply the argument to the fixed-threshold contrast $\nu_P^{1-}(q_1)-\nu_P^{0+}(q_0)$, whose supremum over the two threshold classes is $\widetilde\psi^-(P)$.
For the upper bound, use $\nu_P^{1+}(q_1)-\nu_P^{0-}(q_0)$ and its infimum.
Under the corresponding conditions, the gradients are the differences of the potential-outcome gradients, with variances including their covariance.
The same argument therefore gives one-sided local coverage for the ATE bounds.
Allocating error probability $\alpha/2$ to each bound and applying the union bound gives local coverage at least $1-\alpha$ for the outward confidence set.
At compatible local laws, this set also covers the sharp identified interval by containment.
\end{proof}

\begin{proof}[Proof of Proposition~\ref{prop:binary_bounds}]
Fix $x$ and suppose first that $0<a_{0,P}(x)<1$.
By the definition of $\vartheta_Q^1(x)$, the two density-ratio constraints are
\[
a_{0,P}L_P(x,1)\le\vartheta_Q^1(x)\le a_{0,P}U_P(x,1),
\qquad
(1-a_{0,P})L_P(x,0)\le1-\vartheta_Q^1(x)\le(1-a_{0,P})U_P(x,0).
\]
Combining these inequalities, $\vartheta_Q^1(x)$ must be at least the larger of the two lower limits and at most the smaller of the two upper limits.
By the definitions of $\vartheta_P^{1-}$ and $\vartheta_P^{1+}$, this is equivalent to
\[
\vartheta_P^{1-}(x)
\le \vartheta_Q^1(x)
\le \vartheta_P^{1+}(x).
\]
Each value in this interval defines a nonnegative normalized density ratio at the two outcome values.
If $a_{0,P}(x)=0$ or $1$, outcome overlap instead forces $\vartheta_Q^1(x)=a_{0,P}(x)$.

The conditional mean is $\pi_{0,P}(x)a_{0,P}(x)+\{1-\pi_{0,P}(x)\}\vartheta_Q^1(x)$, an increasing affine function of $\vartheta_Q^1$.
Its extrema are attained at the stated measurable bound functions.
Proposition~\ref{prop:sharpness_msm_iv} supplies the corresponding full-data extensions, and averaging over $X$ gives the asserted bounds.
Convex combinations of the feasible missing probabilities attain every intermediate mean.
The treatment-relabeling and joint construction in Corollary~\ref{cor:sharp_ate} give the ATE bounds.
\end{proof}

\begin{proof}[Binary specialization of the one-step inference argument]
Use $K=2$ and $(y_1,y_2)=(0,1)$ in the finite-support representation.
Its conditional mean $M_P^{tb}(X)$ is the binary sharp bound.
Under the branch regularity of
Proposition~\ref{prop:supp_finite_outcome_eif}, its centered gradient
is the specialization of~\eqref{eq:supp_finite_outcome_eif}.
For $t=1$, for example, the outcome-residual term in stratum $z$ is
\[
\frac{I(Z=z)T}{e_{z,P}(X)\pi_{z,P}(X)}
\frac{\partial M_P^{1b}(X)}{\partial a_{z,P}}
\{Y-a_{z,P}(X)\}.
\]
The derivative with respect to $a_{z,P}$ varies the success probability
and its complementary failure probability together. This is equivalent
to the two-category sum in the finite-support formula.

There is no learned threshold or optimization regret in this
representation. For cross-fitting, subtract the true uncentered gradient
from its fitted version in each validation fold. If this difference is
$o_p(1)$ in $L_2(P)$ and its fold-average expectation is
$o_p(n^{-1/2})$, the argument proving~\eqref{eq:proof_crossfit_expansion}
gives asymptotic linearity centered at $\mu^{tb}(P)$.
The variance, joint-contrast, and confidence-limit arguments then apply
without threshold stabilization or a first-order optimizer condition.

To justify negligible drift for exact binary bounds, the margin
condition must control all branch switches in the bound map, including
the additional maximum and minimum defining $\vartheta_P^{1b}$.
For bounded fitted probabilities, the smooth-branch Taylor remainder is
quadratic in their $L_2(P)$ errors. If the maximal uniform error in the
competing branch arguments is $a_n=o_p(1)$ and the probability of being
within $v$ of any relevant tie is $O(v^\kappa)$, the switching remainder
is $O_p(a_n^{1+\kappa})$. This follows from the maximum/minimum
inequality used in the nuisance-drift proof above. Thus quadratic and
switching remainders of order $o_p(n^{-1/2})$, together with the
propensity--regression product conditions and $L_2(P)$ consistency of
the fitted uncentered gradients, supply the binary inference conditions.
No conclusion at a positive-probability endpoint tie is inferred from
cross-fitting alone.
\end{proof}

\begin{proof}[The IV-only limit under nondegenerate conditional binary laws]
For $0<a_{z,P}(x)<1$ in both strata, the common success probability
must lie in the intersection, over $z$, of
\[
\begin{split}
\big[\max\{[\pi_{z,P}+(1-\pi_{z,P})/\Lambda_z]a_{z,P},
1-[\pi_{z,P}+(1-\pi_{z,P})\Lambda_z](1-a_{z,P})\},\\
\min\{[\pi_{z,P}+(1-\pi_{z,P})\Lambda_z]a_{z,P},
1-[\pi_{z,P}+(1-\pi_{z,P})/\Lambda_z](1-a_{z,P})\}\big].
\end{split}
\]
As both sensitivity parameters diverge, the lower bound converges
to $\max_z\pi_{z,P}a_{z,P}$ and the upper bound to
$\min_z\{\pi_{z,P}a_{z,P}+1-\pi_{z,P}\}$.
Dominated convergence applies to these conditional probabilities.
At degenerate outcome laws, the separate outcome-overlap assumption
continues to force zero missing mass at an unobserved outcome value;
the unrestricted IV-only limit need not follow there.
\end{proof}

%% file: supp_additional_proofs.tex

\section{Proofs of Results in Appendices~\ref{sec:supp_extensions_regularity} and~\ref{sec:supp_finite_extensions}}
\label{sec:supp_extension_proofs}

\begin{proof}[The four witnessed compatibility inequalities]
Fix a treatment arm $t$, a covariate value $x$, and an allocation $r$.
Change of measure gives
\[
\E_P\{\gamma_{t,P}(Y,X)r(Y,X)\mid T=t,X,Z=0\}
=m_{1,t,P}^r(X).
\]
Multiplying the lower and upper density-ratio inequalities by $r\ge0$
and integrating under the baseline conditional outcome law yields
the two pointwise moments in~\eqref{eq:supp_compatibility_moment_bases}.

For every $0\le r\le1$,
\[
\begin{aligned}
\max\{\Lambda_0^{-1},c_{t,P}^{\ell}\gamma_{t,P}+b_{t,P}\}
&\ge\Lambda_0^{-1}
+r(c_{t,P}^{\ell}\gamma_{t,P}+b_{t,P}-\Lambda_0^{-1}),\\
\min\{\Lambda_0,c_{t,P}^{u}\gamma_{t,P}+b_{t,P}\}
&\le\Lambda_0
+r(c_{t,P}^{u}\gamma_{t,P}+b_{t,P}-\Lambda_0).
\end{aligned}
\]
Integrating the first inequality and using $\widetilde L_{t,P}\le1$
gives $\theta_{t,P}^{\mathrm{norm},-}\le0$.
Integrating the second and using $\widetilde U_{t,P}\ge1$ gives
$\theta_{t,P}^{\mathrm{norm},+}\le0$.
Multiplying each conditional inequality by $h(X)\ge0$ and averaging
establishes $\Theta_{t,j}(P;w)\le0$ under the null.
These arguments use expectations only and apply equally to discrete
and continuous outcome laws.
\end{proof}

\begin{proof}[Proof of Lemma~\ref{lem:supp_compatibility_fixed_witness_if}]
Hold $w=(r,h)$ fixed along a regular submodel with score $S$.
The conditional-score identities give
\[
\begin{aligned}
\dot m_{z,t,P}^r(x)
&=\E_P[\{r(Y,X)-m_{z,t,P}^r(x)\}S(O)
\mid T=t,X=x,Z=z],\\
\dot\pi_{z,P}^t(x)
&=\E_P[\{D_t-\pi_{z,P}^t(x)\}S(O)\mid X=x,Z=z].
\end{aligned}
\]
The chain rule differentiates the fixed-witness conditional moment
through its four arguments $(m_0,m_1,p_0,p_1)$.
Inverse weighting with $e_{z,P}\pi_{z,P}^t$ and $e_{z,P}$ embeds
the two displayed derivatives in the observed-data score inner product.
Perturbation of the distribution of $X$ contributes
$h(X)\theta_{t,j,P}(X;r)-\Theta_{t,j}(P;w)$.
These are exactly the terms in~\eqref{eq:supp_compatibility_uncentered_if}
after centering. All residual terms have conditional mean zero;
hence the uncentered function has expectation $\Theta_{t,j}(P;w)$.
Bounded witnesses and positivity ensure square integrability and
bounded derivatives of the coefficient maps for the fixed finite
sensitivity pair. Thus the centered function is the canonical gradient.
\end{proof}

\begin{proof}[Proof of Theorem~\ref{thm:learned_witness_compatibility}]
For each coordinate $(t,j)$, put
\[
\overline\Theta_{n,t,j}
=\sum_k\frac{n_k}{n}
\Theta_{t,j}(P;\widehat w_{-k,t,j}).
\]
Every summand is nonpositive under the null for every realized
training-sample witness, so $\overline\Theta_{n,t,j}\le0$.
Define the centered studentized maximum
\[
\max_{t,j}
\frac{\sqrt n\{
\overline\Phi_{t,j}^{\mathrm{comp}}-\overline\Theta_{n,t,j}\}}
{\widehat\sigma_{t,j}}.
\]
The observed statistic $T_n$ is bounded above by this maximum.
The negligible nuisance remainders and the assumed joint centered
Gaussian multiplier approximation calibrate this maximum at level
$\alpha$. Continuity of its limiting distribution follows from
the finite number of nondegenerate Gaussian coordinates.
Consequently, comparison with the centered maximum gives
\[
\limsup_{n\to\infty}P\{T_n>c_{1-\alpha}^*\}\le\alpha.
\]
Here multiplier validity is used as an assumption of the theorem;
training-fold measurability and bounded moments alone would not
establish it when selected witnesses fail to stabilize.
\end{proof}

\begin{proof}[Consistency against detected fixed alternatives]
Suppose that under the alternative the selection rule yields a
coordinate $(t,j)$ with
$\overline\Theta_{n,t,j}\ge c>0$ with probability tending to one,
and that the corresponding centered expansion and variance conditions
continue to hold. Its studentized statistic diverges to $+\infty$,
whereas the multiplier critical value is $O_p(1)$, since it is the
maximum of a fixed number of standardized Gaussian coordinates.
The rejection probability therefore tends to one.
For a fixed finite candidate collection, uniform consistency of the
training estimates and selection of a maximizer of the witnessed
moment guarantee this positive separation whenever the collection
contains a witness with a positive population moment. Mere presence
of such a witness, without a corresponding selection guarantee,
does not establish consistency.
\end{proof}

\begin{proof}[Proof of Proposition~\ref{prop:supp_binary_bound_shape}]
We first consider the bounds for $\mu^1(Q)$.
Fix $x$ and set $\Lambda_0=\Lambda_1=\Lambda$.
Substituting the envelope definitions into
$\vartheta_P^{1-}(x)$ and $\vartheta_P^{1+}(x)$ gives
\[
\begin{aligned}
&\pi_{0,P}(x)a_{0,P}(x)
+\{1-\pi_{0,P}(x)\}\vartheta_P^{1-}(x)\\
&\quad=
\max_{z\in\{0,1\}}\max\left\{
\left[\pi_{z,P}(x)
+\frac{1-\pi_{z,P}(x)}{\Lambda}\right]a_{z,P}(x),\,
1-\left[\pi_{z,P}(x)
+\{1-\pi_{z,P}(x)\}\Lambda\right]\{1-a_{z,P}(x)\}
\right\},\\[1ex]
&\pi_{0,P}(x)a_{0,P}(x)
+\{1-\pi_{0,P}(x)\}\vartheta_P^{1+}(x)\\
&\quad=
\min_{z\in\{0,1\}}\min\left\{
\left[\pi_{z,P}(x)
+\{1-\pi_{z,P}(x)\}\Lambda\right]a_{z,P}(x),\,
1-\left[\pi_{z,P}(x)
+\frac{1-\pi_{z,P}(x)}{\Lambda}\right]\{1-a_{z,P}(x)\}
\right\}.
\end{aligned}
\]
These are the sharp conditional mean bounds.
On the compatible interval $\mathcal I$, the identities also hold when $a_{0,P}(x)$ equals zero or one, using the conventions in Section~\ref{sec:binary_bounds}.

For each $z$, the two candidates defining the lower bound are nonincreasing in $\Lambda$, whereas those defining the upper bound are nondecreasing.
The absolute values of their derivatives are
\[
\frac{\{1-\pi_{z,P}(x)\}a_{z,P}(x)}{\Lambda^2},
\;
\{1-\pi_{z,P}(x)\}\{1-a_{z,P}(x)\},
\;
\{1-\pi_{z,P}(x)\}a_{z,P}(x),
\;
\frac{\{1-\pi_{z,P}(x)\}\{1-a_{z,P}(x)\}}{\Lambda^2},
\]
respectively.
Each is at most one for $\Lambda\ge1$.
Taking a finite maximum or minimum preserves a common Lipschitz constant.
Thus the conditional lower bound is nonincreasing, the conditional upper bound is
nondecreasing, and both are $1$-Lipschitz.
Averaging over $X$ preserves these properties, proving them for $\mu^{1-}$ and $\mu^{1+}$.
Relabeling treatment gives the same conclusions for $\mu^{0-}$ and $\mu^{0+}$.

Since $\psi^-=\mu^{1-}-\mu^{0+}$, the lower ATE bound is nonincreasing. Similarly,
$\psi^+=\mu^{1+}-\mu^{0-}$ is nondecreasing.
For either contrast, the triangle inequality bounds its change by the sum of the changes in its two mean endpoints.
Each of those changes is at most $|\Lambda'-\Lambda|$,
so
\[
|\psi^b(\Lambda')-\psi^b(\Lambda)|
\le 2|\Lambda'-\Lambda|,
\qquad
\Lambda,\Lambda'\in\mathcal I,\quad b\in\{-,+\}.
\]
This proves the asserted Lipschitz bounds and monotonicities.
\end{proof}

\begin{proof}[Proof of Proposition~\ref{prop:supp_binary_continuous_bands}]
The joint expansions, positive variances, and consistent standard errors imply
\[
\max_{1\le j\le J}\left\{
\frac{\widehat\psi_j^--\psi^-(\lambda_j)}
{\widehat{\mathrm{se}}_j^-},
\frac{\psi^+(\lambda_j)-\widehat\psi_j^+}
{\widehat{\mathrm{se}}_j^+}
\right\}
\rightsquigarrow
\max_j\{G_j^-,-G_j^+\}.
\]
By the definitions of $\ell_j$ and $u_j$, the maximum of the standardized estimation errors is at most $\widehat c_{1-\alpha}$ if and only if
\[
\ell_j\le\psi^-(\lambda_j),
\qquad
\psi^+(\lambda_j)\le u_j
\quad\text{for every }j=1,\ldots,J.
\]
Consistent critical-value estimation and continuity of
the limiting distribution at its critical quantile
therefore imply that this grid-coverage event has
probability tending to $1-\alpha$.

On this event, fix any $\Lambda\in\mathcal I$.
For every grid point $\lambda_j$,
\[
\begin{aligned}
\ell_j-2(\Lambda-\lambda_j)_+
&\le
\psi^-(\lambda_j)-2(\Lambda-\lambda_j)_+
\le\psi^-(\Lambda),\\
\psi^+(\Lambda)
&\le
\psi^+(\lambda_j)+2(\Lambda-\lambda_j)_+
\le u_j+2(\Lambda-\lambda_j)_+.
\end{aligned}
\]
The inequalities involving $\ell_j$ and $u_j$ follow
from grid coverage.
The inequalities comparing population bounds follow
from the Lipschitz property when $\lambda_j\le\Lambda$
and monotonicity when $\lambda_j>\Lambda$.

Taking the maximum over $j$ in the first line gives
$L_n(\Lambda)\le\psi^-(\Lambda)$.
Taking the minimum over $j$ in the second line gives
$\psi^+(\Lambda)\le U_n(\Lambda)$.
Compatibility also gives
$\psi^-(\Lambda)\le\psi^+(\Lambda)$. Thus the grid-coverage event is contained in the event of coverage throughout $\mathcal I$.
Consequently,
\[
\begin{aligned}
P\!\left\{
L_n(\Lambda)\le\psi^-(\Lambda)\le\psi^+(\Lambda)\le U_n(\Lambda)
\;\forall\;\Lambda\in\mathcal I
\right\}\ge
P\!\left\{
\ell_j\le\psi^-(\lambda_j),\
\psi^+(\lambda_j)\le u_j
\;\forall\;j=1,\ldots,J
\right\}.
\end{aligned}
\]
The probability on the right converges to $1-\alpha$, which proves the asserted limit-inferior bound.
\end{proof}

\begin{proof}[Proof of Proposition~\ref{prop:supp_finite_outcome_bounds}]
Fix $t\in\{0,1\}$ and a covariate value $x$.
For brevity, omit the superscript $t$, the subscript $P$ or $Q$, and the argument $x$ in the calculations below.

We first establish the compatibility conditions.
If $h_k$ is the common potential-outcome probability at $y_k$, the mixture identity forces
\[
\vartheta_{z,k}=\frac{h_k-\pi_z a_{z,k}}{1-\pi_z}.
\]
The probability and sensitivity requirements on these missing probabilities, across instrument strata, are equivalent to $\ell_k\le h_k\le u_k$ for every $k$.
Moreover, $\sum_k h_k=1$ implies $\sum_k\vartheta_{z,k}=1$.
For finite sensitivity parameters, $a_{z,k}=0$ forces $h_k=0$ and $\vartheta_{z,k}=0$.
Thus, positive observed probability at a category in one instrument stratum cannot coexist with zero probability at that category in another stratum.

Compatibility therefore requires a vector $h$ satisfying $\ell_k\le h_k\le u_k$ for every $k$ and $\sum_kh_k=1$.
Such a vector necessarily satisfies
\[
\ell_k\le u_k\quad\text{for every }k,
\qquad
\sum_k\ell_k\le1\le\sum_ku_k.
\]
Conversely, suppose these inequalities hold.
Initially set $h_k=\ell_k$ for each $k=1,\ldots,K$, where $h_k$ represents the conditional probability $Q(Y^t=y_k\mid X=x)$.
Distribute the remaining probability $1-\sum_k\ell_k$ successively across the categories without increasing any probability above its upper bound $u_k$.
This is possible because $\sum_k(u_k-\ell_k)\ge1-\sum_k\ell_k$.
The resulting vector gives nonnegative normalized missing probabilities in every stratum through the displayed formula.
Using these probabilities in the construction of Proposition~\ref{prop:sharpness_msm_iv}, with treatment relabeled when $t=0$, gives a full-data extension satisfying the required restrictions.
This proves the compatibility characterization.
The constructed probabilities are measurable functions of $x$: the allocation uses only finite sums, differences, minima, and maxima of the measurable bounds $\ell_k(x)$ and $u_k(x)$, and the formula for $\vartheta_{z,k}(x)$ preserves measurability because $1-\pi_z(x)>0$.

We next show that the stated allocations minimize and maximize the conditional mean $\sum_k y_kh_k$.
Consider the increasing-order allocation $h^-$.
If $1-\sum_k\ell_k>0$, let $r$ be the last category receiving additional probability in this allocation.
For every other feasible vector $h$, normalization gives
\[
\sum_k y_k(h_k-h_k^-)
=\sum_k(y_k-y_r)(h_k-h_k^-)\ge0.
\]
Indeed, for $k<r$, we have $h_k^-=u_k$, so both factors in the corresponding summand are nonpositive.
For $k>r$, we have $h_k^-=\ell_k$, so both factors are nonnegative.
The summand for $k=r$ is zero.
If $1-\sum_k\ell_k=0$, the only feasible vector is $h_k=\ell_k$ for every $k$.
Thus $h^-$ minimizes the conditional mean.
The decreasing-order allocation $h^+$ maximizes it by the same argument with the inequalities reversed.

It remains to establish population sharpness.
The two allocations are measurable in $x$ and, by the compatibility construction above, are attained by full-data extensions satisfying the required restrictions.
Averaging their conditional means over $X$ therefore gives the stated sharp population bounds.
Combining the two treatment-arm constructions as in Corollary~\ref{cor:sharp_ate} proves the ATE claim.
\end{proof}

\begin{proof}[Proof of Proposition~\ref{prop:supp_finite_outcome_eif}]
Under the stated regularity assumption, for almost every $x$, sufficiently small perturbations of the conditional probabilities leave unchanged both the arguments attaining the envelope maxima and minima and the last category receiving additional probability.
The probability assigned to that category is one minus the sum assigned to the other categories.
Consequently, sufficiently small changes in the treatment and outcome probabilities change the value of $M_P^{tb}(x)$ without changing the formula used to calculate it.
That formula is differentiable in these probabilities.

Each selected envelope expression has bounded derivatives because the probabilities lie in $[0,1]$ and the sensitivity parameters are fixed and finite.
The allocated probabilities are finite sums and differences of these expressions, and $M_P^{tb}(x)$ is their weighted sum with fixed coefficients $y_1,\ldots,y_K$.
Its derivatives are therefore bounded uniformly in $x$.
Moreover, the finite sums, maxima, and minima defining the
allocation are Lipschitz functions of the probabilities.

Consider a regular parametric submodel through $P$ with score $S$.
Together with positivity and regularity of this submodel, these bounds justify integrating the pointwise derivative over $X$.
The chain rule gives
\[
\dot M_P^{tb}(x)
=\sum_{z=0}^1\dot M_{\pi_z,P}^{tb}(x)\dot\pi_{z,P}^t(x)
+\sum_{z=0}^1\sum_{k=1}^K
\dot M_{a_{z,k},P}^{tb}(x)\dot a_{z,k,P}^t(x).
\]
The conditional-score identities are
\[
\begin{aligned}
\dot\pi_{z,P}^t(x)
&=\E_P[\{I(T=t)-\pi_{z,P}^t(x)\}S(O)\mid X=x,Z=z],\\
\dot a_{z,k,P}^t(x)
&=\E_P[\{I(Y=y_k)-a_{z,k,P}^t(x)\}S(O)
\mid T=t,X=x,Z=z].
\end{aligned}
\]
Inverse weighting embeds these derivatives into the observed-data
score inner product. Perturbing the distribution of $X$ contributes
$M_P^{tb}(X)-\mu^{tb}(P)$. Therefore the inner product of
\eqref{eq:supp_finite_outcome_eif} with $S$ equals the derivative
of $\mu^{tb}(P)$.
The proposed influence function has mean zero by conditional expectation and is square integrable by the boundedness established above and uniform positivity.
Since it represents the derivative along every regular submodel,
it is the canonical gradient in the nonparametric observed-data
model.
Finally, adding the same constant to all coefficients $\dot M_{a_{z,k},P}^{tb}(X)$, $k=1,\ldots,K$, leaves their contribution to the influence function unchanged, since
\[
\sum_{k=1}^K
\{I(Y=y_k)-a_{z,k,P}^t(X)\}=0.
\]
This proves that the gradient is well defined despite the
probability-simplex constraint.
\end{proof}

\begin{proof}[Smoothing the finite-support bounds]
For a feasible box $[\ell,u]$, the allocation argument gives the following finite representations:
\[
\begin{aligned}
\mathcal W^-(\ell,u)
&=\max_{1\le r\le K}
\left\{y_r+\sum_{k<r}(y_k-y_r)u_k
+\sum_{k>r}(y_k-y_r)\ell_k\right\},\\
\mathcal W^+(\ell,u)
&=\min_{1\le r\le K}
\left\{y_r+\sum_{k<r}(y_k-y_r)\ell_k
+\sum_{k>r}(y_k-y_r)u_k\right\}.
\end{aligned}
\]
Indeed, for each $r$, normalization gives $\sum_k y_k h_k=y_r+\sum_k(y_k-y_r)h_k$.
Bounding each summand using $\ell_k\le h_k\le u_k$ gives the corresponding lower and upper candidates.
Equality holds when $r$ is the final receiving category of the extremizing allocation.
If $\sum_k\ell_k=1$, the only feasible vector is $h=\ell$; the lower candidate with $r=1$ and the upper candidate with $r=K$ both equal $\sum_k y_k\ell_k$.
If $\sum_ku_k=1$, the only feasible vector is $h=u$; the lower candidate with $r=K$ and the upper candidate with $r=1$ both equal $\sum_k y_ku_k$.

Replacing envelope maxima by lower smooth approximations decreases $\ell$, while replacing envelope minima by upper smooth approximations increases $u$.
Every candidate in the lower representation consequently decreases, and every candidate in the upper representation increases.
Approximating the final maximum from below and the final minimum from above therefore preserves outward ordering.
A smoothed lower bound may be negative, but it is a constraint on a category probability, not a probability itself. Every originally feasible vector still satisfies the smoothed inequalities, so the resulting lower and upper mean bounds remain valid.

For fixed positive smoothing, the resulting finite compositions are differentiable, with derivatives given by the chain rule and smooth maximum/minimum weights.
The preceding influence-function proof therefore applies with the smoothed conditional mean and its derivatives, without active-branch uniqueness.
The corresponding one-step estimator and its inference follow by the cross-fitted argument under the associated nuisance-rate and variance conditions.
\end{proof}

\begin{proof}[Proof of Proposition~\ref{prop:supp_finite_iv_bounds}]
Fix $t\in\{0,1\}$ and a covariate value $x$.

We first prove necessity of the compatibility conditions.
The IV restrictions imply a common conditional potential-outcome density $h_Q^t$.
In each instrument category, consistency and composition give
\[
h_Q^t(y\mid x)
=f_{j,P}^t(y\mid x)
\{\pi_{j,P}^t(x)+(1-\pi_{j,P}^t(x))
\lambda_{j,Q}^t(y,x)\}.
\]
Dividing by the reference density and applying the sensitivity restrictions gives $L_P^t\le\rho_Q^t\le U_P^t$.
In particular, $L_P^t\le U_P^t$.
Since the common density integrates to one, integrating these inequalities under the reference outcome law gives the two conditional normalization inequalities.

We next prove sufficiency and characterize the feasible common densities.
Suppose~\eqref{eq:supp_finite_iv_compatibility} holds.
The reference candidate gives $L_P^t\ge A_{j_0,P}^t>0$.
Since $U_P^t\le B_{j,P}^t\gamma_{j,P}^t$ for every $j$, the inequality $L_P^t\le U_P^t$ implies $\gamma_{j,P}^t>0$ almost surely under the reference outcome law.
Also, $U_P^t\le B_{j_0,P}^t<\infty$, so both envelopes are bounded.

A normalized function $\rho$ between the envelopes can be constructed explicitly. If their conditional expectations differ, set
\[
\rho(y,x)=L_P^t(x,y)
+\frac{1-\E_P(L_P^t\mid T=t,X=x,Z=z_{j_0})}
{\E_P(U_P^t-L_P^t\mid T=t,X=x,Z=z_{j_0})}
\{U_P^t(x,y)-L_P^t(x,y)\}.
\]
The compatibility inequalities ensure that the coefficient lies in $[0,1]$, and the resulting function has conditional expectation one under the reference outcome law.
If the denominator is zero, both envelope expectations equal one, and we take $\rho=L_P^t=U_P^t$ almost surely.

For any normalized $\rho$ satisfying $L_P^t\le\rho\le U_P^t$, set $h^t=\rho f_{j_0,P}^t$ and define
\[
\lambda_j^t(y,x)
=\frac{\rho(y,x)/\gamma_{j,P}^t(y,x)-\pi_{j,P}^t(x)}
{1-\pi_{j,P}^t(x)}.
\]
The envelope inequalities imply $\Lambda_j^{-1}\le\lambda_j^t\le\Lambda_j$.
Moreover,
\[
\int\lambda_j^t(y,x)f_{j,P}^t(y\mid x)\,dy
=\frac{\int h^t(y\mid x)\,dy-\pi_{j,P}^t(x)}
{1-\pi_{j,P}^t(x)}
=1.
\]
These ratios therefore define valid missing-outcome distributions in every instrument category.

To realize these distributions in a full-data extension, draw $(X,Z,T)$ from $P$ and assign $Y^t$ its observed conditional law when $T=t$ and its constructed missing law when $T=1-t$.
Set $Y^{t,z_j}=Y^t$ for every $j$.
The conditional density of $Y^t$ given $X=x,Z=z_j$ is then $h^t$ for every instrument category.
The construction of Proposition~\ref{prop:sharpness_msm_iv}, applied to these laws, preserves the observed-data distribution and establishes the required restrictions for this potential outcome.
Thus every normalized $\rho$ between the envelopes is attainable, proving sufficiency and the characterization of feasible common densities.

We now establish invariance to the reference category.
Using $z_{j_1}$ instead of $z_{j_0}$ divides every candidate envelope and every feasible $\rho$ by the positive ratio $\gamma_{j_1,P}^t$, while multiplying the reference density by that ratio.
Their product, the common potential-outcome density, is unchanged.
Consequently, the feasible common densities and their mean bounds do not depend on the reference category.

It remains to establish the sharp mean representations and restricted-class outward ordering.
Under the first-moment condition used in the main text, the proof of Theorem~\ref{thm:sharp_bounds_reps} applies to the bounded envelopes $L_P^t,U_P^t$ and the reference outcome law.
For atomless laws, it gives normalized threshold extremizers, with the same boundary conventions when normalization binds.

For any feasible normalized $\rho$ and any finite threshold $q$,
\[
\begin{aligned}
\int y\rho\,f_{j_0,P}^t\,dy
-\theta_{P,q}^{t-}(x)
&=\int(y-q)(\rho-\rho_{P,q}^{t-})
f_{j_0,P}^t\,dy\ge0,\\
\int y\rho\,f_{j_0,P}^t\,dy
-\theta_{P,q}^{t+}(x)
&=\int(y-q)(\rho-\rho_{P,q}^{t+})
f_{j_0,P}^t\,dy\le0.
\end{aligned}
\]
The signs follow from the lower and upper allocations.
Averaging over $X$ and optimizing over the prescribed threshold class therefore gives the restricted-class outward bounds.
Optimizing over all measurable thresholds gives the sharp population means, by the same argument as in Theorem~\ref{thm:sharp_bounds_reps}.
\end{proof}

\begin{proof}[Proof of Proposition~\ref{prop:supp_finite_iv_eif}]
Fix $(t,b,q)$ and a regular submodel with score $S$.
Let $s_j^t(y,x)$ be the conditional score of $Y\mid(T=t,X=x,Z=z_j)$.
For $j\ne j_0$,
\[
\dot\gamma_{j,P}^t
=\gamma_{j,P}^t(s_j^t-s_{j_0}^t).
\]
Hold the threshold and its allocation indicators fixed.
The no-tie condition gives the envelope derivatives specified in the statement.
Differentiating $\theta_{P,q}^{tb}$ and collecting conditional outcome-score terms yields
\[
\dot\theta_{P,q}^{tb}(x)
=\sum_{j=1}^J
\int H_{j,P,q}^{tb}(y,x)s_j^t(y,x)f_{j,P}^t(y\mid x)\,dy
+\sum_{j=1}^J C_{j,P,q}^{tb}(x)\dot\pi_{j,P}^t(x).
\]
For the reference category, differentiation of the integrating density contributes $\rho_{P,q}^{tb}(y,x)\{y-q(x)\}$, and differentiation of the denominators of the other density ratios contributes $-\sum_{j\ne j_0}\gamma_{j,P}^tD_{\gamma_j,P,q}^{tb}$.
Their sum is $H_{j_0,P,q}^{tb}$.
For a nonreference category, changing measure by $f_{j,P}^t=\gamma_{j,P}^t f_{j_0,P}^t$ gives $H_{j,P,q}^{tb}=D_{\gamma_j,P,q}^{tb}$.
The propensity derivative is the reference-law expectation defining $C_{j,P,q}^{tb}$.

By the conditional-score identities, the terms involving $I(T=t)-\pi_{j,P}^t(X)$ represent the treatment-probability derivatives, while those involving
$H_{j,P,q}^{tb}(Y,X)-\overline H_{j,P,q}^{tb}(X)$ represent the conditional outcome-distribution derivatives.
The inverse-probability weights in \eqref{eq:supp_finite_iv_eif} ensure that the inner product of their sum with $S$ equals $\E_P\{\dot\theta_{P,q}^{tb}(X)\}$.
The first term accounts for the distribution of $X$.
The integrated differentiation is justified as in the proof of Theorem~\ref{thm:eif_fixed_q}\textup{(i)}.
Multiplying each envelope by the reference density expresses it as a finite maximum or minimum of weighted outcome densities.
The no-tie condition and positivity give an integrated remainder of order $o(|s|)$ from changes in the determining category.
Because the outcome domain and the fixed threshold function are bounded, multiplication by $y-q(x)$ and the allocation indicators preserves this order.
Finite compatible sensitivity parameters and uniform positivity also bound the intercategory density ratios above and away from zero.
The displayed influence function is centered by conditional expectation and is square integrable by these bounds.
Its inner product with every score equals the derivative of the functional, so it is the canonical gradient.
\end{proof}

\begin{proof}[Remaining finite-instrument extensions]
For the optimizer conclusion, decompose the perturbed optimized
value into the fixed-threshold perturbation and the reoptimization
difference, exactly as in Theorem~\ref{thm:eif_fixed_q}\textup{(ii)}.
Under the corresponding first-order optimizer condition, the latter
is negligible. Cross-fitting uses the same validation-fold expansion
with the finite-instrument gradient; a finite sum over categories
does not change the argument.

For exact envelopes, let the gap be the difference between the
largest and second-largest candidate in the maximum, or the
second-smallest and smallest in the minimum. If all candidate errors
are at most $a_n$, a change in the active category requires the
true gap to be at most $2a_n$. The linearization error is bounded
by a constant times $a_n$ on this set. A margin bound of order
$v^\kappa$ for these gaps therefore gives the same
$O_p(a_n^{1+\kappa})$ switching remainder. For fixed smoothing,
smooth derivative weights replace the active indicators, and the
Hessian bound gives the corresponding quadratic remainder.

If the outcome also has finite support, the feasible common
probability vector is obtained by intersecting the category bounds
over all $J$ instrument strata. The proof of
Proposition~\ref{prop:supp_finite_outcome_bounds} is unchanged.
The gradient proof similarly sums the conditional-score terms over
$j=1,\ldots,J$. This probability representation uses no reference
instrument category, proving the stated reference-free formulation.
Finally, additive smoothing need not be reference invariant:
a smooth maximum at a fixed additive scale is not homogeneous under
the outcome-dependent rescaling used to change reference categories.
\end{proof}

\begin{proof}[Proof of Proposition~\ref{prop:supp_binary_joint_bounds}]
We prove compatibility first and then establish sharpness of
the population bounds. All conditional probabilities below
are taken in measurable versions, and statements concerning
$x$ hold $P_X$-almost surely.

\paragraph{Necessity.}
Let $Q$ be a full-data extension satisfying the stated
restrictions, and define
\[
w_{ztuv}(x)
=
Q(T=t,Y^0=u,Y^1=v\mid X=x,Z=z).
\]
These probabilities are nonnegative.
By consistency,
\[
\sum_{\substack{u,v\in\{0,1\}\\y_t(u,v)=y}}
w_{ztuv}(x)
=
Q(T=t,Y^t=y\mid X=x,Z=z)
=
p_{zty,P}(x),
\]
which proves \eqref{eq:supp_joint_observed}.
Joint exchangeability gives
\[
\sum_{t=0}^1w_{ztuv}(x)
=
Q(Y^0=u,Y^1=v\mid X=x),
\]
independently of $z$, proving
\eqref{eq:supp_joint_common}.

The observed and missing conditional probabilities of $Y^t=y$
are, respectively,
\[
a_{z,P}^t(y,x)
\quad\text{and}\quad
\frac{
\displaystyle
\sum_{\substack{u,v\in\{0,1\}\\y_t(u,v)=y}}
w_{z,1-t,u,v}(x)}
{1-\pi_{z,P}^t(x)}.
\]
The MSM bounds the second probability between
$a_{z,P}^t(y,x)/\Lambda_z$ and
$\Lambda_z a_{z,P}^t(y,x)$.
Multiplication by $1-\pi_{z,P}^t(x)>0$ gives
\eqref{eq:supp_joint_msm}.
Hence $w(x)\in\mathcal W_P(x)$.

\paragraph{Existence of measurable selections.}
Write the sixteen coordinates of $w$ as a vector in
$\mathbb R^{16}$.
The observed-data constraints imply
\[
\sum_{t,u,v}w_{ztuv}(x)=1,\qquad z=0,1.
\]
Thus $\mathcal W_P(x)$ is a closed subset of $[0,1]^{16}$
defined by finitely many linear equalities and inequalities.
Whenever nonempty, it is a compact polytope.

For completeness, measurable feasible and optimizing
selections can be obtained by enumerating its possible
vertices.
Express all constraints as finitely many weak linear
inequalities, replacing each equality by two inequalities.
Their coefficients are measurable functions of $x$.
For every choice of sixteen constraints, solve the associated
linear system wherever its coefficient matrix is nonsingular.
Each resulting candidate vector is measurable on that set,
and the condition that it satisfies all constraints is
measurable.
Every vertex appears among these finitely many candidates,
because the active constraint normals at a vertex span
$\mathbb R^{16}$.

A nonempty compact polytope has a vertex.
Selecting the first feasible candidate in a fixed enumeration
therefore gives a measurable feasible vector.
Likewise, a linear objective attains its minimum and maximum
at vertices. Selecting the first feasible candidate attaining
the smallest or largest objective gives measurable optimizing
vectors.
Selections may be defined arbitrarily on the $P_X$-null set
where feasibility fails.

\paragraph{Sufficiency.}
Suppose $\mathcal W_P(x)$ is nonempty almost surely, and choose
a measurable feasible vector $w(x)$.
Construct a law $Q$ by drawing $(X,Z)$ from its observed
distribution and assigning conditional probabilities
\[
Q(T=t,Y^0=u,Y^1=v\mid X=x,Z=z)=w_{ztuv}(x).
\]
These probabilities define a conditional distribution because
they are nonnegative, measurable, and sum to one.
Set $Y=Y^T$.
Then
\[
Q(T=t,Y=y\mid X=x,Z=z)
=
\sum_{\substack{u,v\in\{0,1\}\\y_t(u,v)=y}}
w_{ztuv}(x)
=
p_{zty,P}(x).
\]
Consequently, the induced law of $(X,Z,T,Y)$ is exactly $P$.

Equation~\eqref{eq:supp_joint_common} gives
\[
Q(Y^0=u,Y^1=v\mid X=x,Z=z)=r_{uv}(x),
\]
independently of $z$, so
$(Y^0,Y^1)\perp Z\mid X$.
Moreover, summing the observed-data constraints over $y$
gives $Q(T=t\mid X=x,Z=z)=\pi_{z,P}^t(x)$.
The missing conditional probabilities are therefore the
quotients displayed in the necessity argument.
They are nonnegative and sum to one, and
\eqref{eq:supp_joint_msm} gives the required sensitivity bounds.
If $a_{z,P}^t(y,x)=0$, its upper sensitivity bound forces the
missing probability to be zero as well.
Thus the required absolute continuity also holds.
Since the marginal law of $Y^t$ is common across instrument
strata and treatment probabilities are positive, these
constraints also imply agreement of the observed outcome
supports across strata.

Finally, define
\[
Y^{t,0}=Y^{t,1}=Y^t,\qquad t=0,1.
\]
Composition and consistency then hold by construction.
Distributional exclusion holds because these variables are
equal, and instrument exchangeability follows from
$(Y^0,Y^1)\perp Z\mid X$.
This constructs a full-data extension satisfying all the
stated restrictions.
Using individual-level exclusion in this construction proves
existence; it does not impose that restriction on every
full-data law in the model.

\paragraph{Sharpness.}
For any compatible extension $Q$, its induced vector
$w_Q(x)$ belongs to $\mathcal W_P(x)$, and
\[
\E_Q\{c(Y^0,Y^1)\mid X=x\}
=
\sum_{u,v}c(u,v)r_{uv,Q}(x).
\]
By the definitions of the conditional extrema,
\[
m_{c,P}^-(x)
\le
\E_Q\{c(Y^0,Y^1)\mid X=x\}
\le
m_{c,P}^+(x).
\]
Because $Q_X=P_X$, integration gives the asserted population
bounds.

Let $w^-(x)$ and $w^+(x)$ be the measurable optimizing
selections established above.
Applying the full-data construction to each selection gives
compatible laws attaining
$\E_P\{m_{c,P}^-(X)\}$ and
$\E_P\{m_{c,P}^+(X)\}$, respectively.
All expectations are finite because $c$ is defined on the
finite set $\{0,1\}^2$.

For any $\alpha\in[0,1]$, the vector
\[
w^\alpha(x)=(1-\alpha)w^-(x)+\alpha w^+(x)
\]
is measurable and feasible, since all constraints are linear.
The corresponding full-data extension has objective value
\[
(1-\alpha)\E_P\{m_{c,P}^-(X)\}
+\alpha\E_P\{m_{c,P}^+(X)\}.
\]
Every value between the bounds is therefore attained, proving
that the displayed interval is the sharp identified set.
\end{proof}

\begin{proof}[Proof of Proposition~\ref{prop:supp_joint_exchangeability_cutoff}]
We first show that, when $1\le\Lambda_0,\Lambda_1\le3$,
every feasible pair of conditional potential-outcome
distributions admits an extension satisfying joint
exchangeability. We then establish sharpness of the
constant three.

\paragraph{Preserving the feasible marginal distributions.}
Fix a covariate value $x$ and any pair of conditional
potential-outcome distributions feasible under the original
IV-MSM. Write
\[
h_{tu}=Q(Y^t=u\mid X=x),
\qquad t,u\in\{0,1\},
\]
and suppress $x$ in the calculations below.
We seek an extension whose common joint distribution is
\[
H_{uv}=h_{0u}h_{1v},
\qquad u,v\in\{0,1\}.
\]
The product distribution is used only to construct an extension;
independence between the potential outcomes is not an
additional model assumption.

Fix an instrument stratum $z$, and write $\pi=P(T=1\mid X=x,Z=z)$,
\[
A_u=P(T=0,Y=u\mid X=x,Z=z),\qquad\text{and}\qquad
B_v=P(T=1,Y=v\mid X=x,Z=z).
\]
Original-model feasibility implies
\[
0\le A_u\le h_{0u},\qquad
0\le B_v\le h_{1v},\qquad
\sum_u A_u=1-\pi,\qquad
\sum_v B_v=\pi.
\]
We must split $H$ into nonnegative untreated and treated
probability tables so that the untreated table has row
totals $A_u$ and the treated table has column totals $B_v$.

For a binary pair, such a split exists if and only if
\begin{equation}
H_{uv}\le A_u+B_v,
\qquad u,v\in\{0,1\}.
\label{eq:supp_joint_product_feasibility}
\end{equation}
To verify this explicitly, put
$R_u=h_{0u}-A_u$ and $C_v=B_v$.
A treated table with these row and column totals necessarily
has the form
\[
W_1=
\begin{pmatrix}
R_0-C_1+\theta & C_1-\theta\\
R_1-\theta & \theta
\end{pmatrix}.
\]
The conditions $0\le W_1\le H$ hold precisely when
\[
\begin{aligned}
\max\{0,R_1-H_{10},C_1-H_{01},C_1-R_0\}
\le\theta\\
\le
\min\{H_{11},R_1,C_1,H_{00}+C_1-R_0\}.
\end{aligned}
\]
Using the marginal feasibility conditions above, comparison
of the lower and upper limits reduces to the four
inequalities in
\eqref{eq:supp_joint_product_feasibility}.
The untreated table is then $W_0=H-W_1$.

\paragraph{The sensitivity restrictions imply the required split.}
We establish a lower bound on an observed category mass.
Suppose the observed treatment arm has probability
$r\in(0,1)$, its conditional category probability is $a$,
and the corresponding missing category probability is $b$.
If the marginal category probability is $s$, then
\[
s=ra+(1-r)b.
\]
The sensitivity restrictions with parameter at most three
imply
\[
b\le3a,\qquad
1-b\ge(1-a)/3.
\]
Consequently,
\[
s\le(3-2r)a,\qquad
1-s\ge\frac{1+2r}{3}(1-a),
\]
and hence
\[
ra\ge F_r(s)
:=
\max\left\{
\frac{rs}{3-2r},
\;
r-\frac{3r(1-s)}{1+2r}
\right\}.
\]
Applying this inequality to $Y^0=u$ in the untreated arm
and $Y^1=v$ in the treated arm gives
\[
A_u+B_v
\ge
F_{1-\pi}(h_{0u})+F_\pi(h_{1v}).
\]

We claim that
\begin{equation}
F_{1-\pi}(a)+F_\pi(b)\ge ab,
\qquad a,b\in[0,1].
\label{eq:supp_joint_scalar_inequality}
\end{equation}
The breakpoints of these two piecewise-linear functions are
\[
a_*=\frac{1+2\pi}{4},
\qquad
b_*=\frac{3-2\pi}{4}.
\]
On each of the four rectangles determined by these
breakpoints, the difference
$F_{1-\pi}(a)+F_\pi(b)-ab$ is bilinear and attains its
minimum at a corner.
Its values at the nine possible corners are
\[
\begin{array}{c|ccc}
& b=0 & b=b_* & b=1\\
\hline
a=0
&0&\pi/4&\pi\\
a=a_*
&(1-\pi)/4&(1-2\pi)^2/16&\pi/4\\
a=1
&1-\pi&(1-\pi)/4&0
\end{array}
\]
and are all nonnegative.
This proves \eqref{eq:supp_joint_scalar_inequality}.
Taking $a=h_{0u}$ and $b=h_{1v}$ therefore gives
\[
A_u+B_v\ge h_{0u}h_{1v}=H_{uv},
\]
which establishes
\eqref{eq:supp_joint_product_feasibility}.

The construction can be performed in every instrument
stratum using the same joint distribution $H$.
It preserves the observed masses $A_u,B_v$ and the chosen
marginal probabilities $h_{tu}$.
Therefore it also preserves the missing category masses
$h_{0u}-A_u$ and $h_{1v}-B_v$.
In particular, it preserves the original sensitivity
restrictions at $\Lambda_z$, not merely their relaxation
to three.

Restoring $x$, all quantities in the construction are
measurable functions of the conditional probabilities.
Choosing $\theta$ as the midpoint of its feasible interval
gives measurable treated and untreated tables.
The full-data construction in
Proposition~\ref{prop:supp_binary_joint_bounds}
then supplies an extension satisfying all the original
restrictions and joint exchangeability.

Thus every feasible pair of conditional marginal
distributions under the original model remains feasible
under the strengthened model.
The reverse inclusion follows immediately because the
strengthened model adds a restriction.
The feasible pairs, compatibility conditions, and sharp
population mean and ATE bounds therefore coincide.

\paragraph{Sharpness of the constant three.}
Fix any common sensitivity parameter $\Lambda>3$.
Consider a population without covariates, with
$P(Z=0)=P(Z=1)=1/2$, and set
\[
d=\frac{1}{2(\Lambda+1)}.
\]
Let the observed conditional probabilities be
\[
\begin{array}{c|cccc}
& T=0,Y=0&T=0,Y=1&T=1,Y=0&T=1,Y=1\\
\hline
Z=0&d&\Lambda d&\Lambda d&d\\
Z=1&\Lambda d&d&\Lambda d&d
\end{array}.
\]
All cells are positive and treatment has probability
$1/2$ in both strata.
The binary sensitivity restrictions give the feasible
marginal means
\[
\mu^0=\frac12,
\qquad
\mu^1\in\left[\frac{1}{2\Lambda},\frac12\right].
\]
For example, the observed success probabilities for $Y^0$
are $\Lambda/(\Lambda+1)$ and $1/(\Lambda+1)$.
Their respective feasible mean intervals have lower and
upper limits equal to $1/2$, forcing the common mean to
equal $1/2$.
The observed success probability for $Y^1$ is
$1/(\Lambda+1)$ in both strata, yielding the displayed
interval for $\mu^1$.
The original-model sharp ATE interval is therefore
\[
\left[-\frac{\Lambda-1}{2\Lambda},\,0\right].
\]

To impose joint exchangeability, write $m=\mu^1$ and
$r=Q(Y^0=1,Y^1=1)$.
The common joint probability table must be
\[
H=
\begin{pmatrix}
1/2-m+r&m-r\\
1/2-r&r
\end{pmatrix}.
\]
For $m$ in its original feasible interval, allocation of
treatment within this table reproduces the observed
probabilities in stratum $Z=0$ if and only if
\[
r\in[\max\{0,m-2d\},m].
\]
In stratum $Z=1$, the corresponding condition is
\[
r\in[0,\min\{m,2d\}].
\]
These conditions follow by applying the binary-table
allocation calculation above to each stratum.
The intervals intersect if and only if $m\le4d$.
Every such intersection supplies a common joint table and
feasible treatment allocations, while preserving the
marginal sensitivity restrictions.

Since $\Lambda>3$, we have
\[
\frac{1}{2\Lambda}<4d=\frac{2}{\Lambda+1}<\frac12.
\]
Thus the strengthened model is compatible and its sharp
ATE interval is
\[
\left[
-\frac{\Lambda-1}{2\Lambda},
\;
\frac{2}{\Lambda+1}-\frac12
\right]
=
\left[
-\frac{\Lambda-1}{2\Lambda},
\;
\frac{3-\Lambda}{2(\Lambda+1)}
\right].
\]
Its upper limit is strictly below zero for every
$\Lambda>3$, whereas the original-model upper limit is
zero. This establishes strict tightening for every
$\Lambda>3$ and proves sharpness of the constant three.
\end{proof}